\pdfoutput=1
\RequirePackage{fixltx2e}
\documentclass[11pt,a4paper]{article}
\usepackage[utf8]{inputenc}
\usepackage{jheppub}
\hypersetup{unicode=true,bookmarksopen=true}

\usepackage{mathtools}
\usepackage{lmodern}
\usepackage[T1]{fontenc}
\usepackage{bbm}
\DeclareMathAlphabet{\mathbfi}{OML}{cmm}{b}{it}

\usepackage{graphicx}
\usepackage{subcaption}

\let\originalleft\left
\let\originalright\right
\renewcommand{\left}{\mathopen{}\mathclose\bgroup\originalleft}
\renewcommand{\right}{\aftergroup\egroup\originalright}

\makeatletter
\newenvironment{equations}[1][]{\subequations\ifx\relax#1\relax\else\label{#1}\fi\align\ignorespaces}{\endalign\ignorespacesafterend\endsubequations}
\def\@spliteq#1{\begin{equation}\begin{split}#1\end{split}\end{equation}}
\def\splitequation{\collect@body\@spliteq}

\makeatother

\renewcommand{\vec}[1]{{\ifnum9<1#1\mathbf{#1}\else\ifcat\noexpand#1\relax\boldsymbol{#1}\else\mathbfi{#1}\fi\fi}}
\newcommand{\mathe}{\mathrm{e}}
\newcommand{\mathi}{\mathrm{i}}
\let\oldre\Re
\let\oldim\Im
\renewcommand{\Re}{\oldre\mathfrak{e}\,}
\renewcommand{\Im}{\oldim\mathfrak{m}\,}
\newcommand{\total}{\mathop{}\!\mathrm{d}}

\newcommand{\abs}[1]{{\left\lvert{#1}\right\rvert}}

\newcommand{\1}{\mathbbm{1}}

\newcommand{\eqend}[1]{\,#1}
\newcommand{\bigo}[1]{\mathcal{O}\left({#1}\right)}

\newcommand{\hypergeom}[2]{\,{}_{#1}\mathrm{F}_{#2}}

\newcommand{\op}{\mathcal{O}}
\newcommand{\Pf}{\mathop{\mathcal{P}\!f}}

\allowdisplaybreaks

\begin{document}

\title{Constructing CFTs from AdS flows}

\author{Markus B. Fr{\"o}b}
\affiliation{Institut f{\"u}r Theoretische Physik, Universit{\"a}t Leipzig, Br{\"u}derstra{\ss}e 16, 04103 Leipzig, Germany}

\emailAdd{mfroeb@itp.uni-leipzig.de}

\abstract{We study the renormalization group flow equations for correlation functions of weakly coupled quantum field theories in AdS. Taking the limit where the external points approach the conformal boundary, we obtain a flow of conformally invariant correlation functions. We solve the flow for one- and two-point functions and show that the corrections to the conformal dimensions can be obtained as an integral over the Mellin amplitude of the four-point function. We also derive the flow of the Mellin amplitude for higher $n$-point functions. We then consider the flows at tree level and one loop (in AdS), and show that one obtains exactly the recursion relations for the corresponding Mellin amplitudes derived earlier by Fitzpatrick et al.~\cite{fitzpatricketal2011} at tree level and Yuan~\cite{yuan2017,yuan2018} at one loop. As an application, we furthermore compute one-loop corrections to the conformal dimensions for some operators in the CFT dual to an $\mathrm{O}(N)$ scalar model in AdS.}


\maketitle

\section{Introduction}

The AdS/CFT correspondence~\cite{maldacena1998,witten1998,gubserklebanovpolyakov1998} is without doubt one of the most important discoveries in theoretical physics, and has found a multitude of applications. In this correspondence, the partition functions of a string theory on an asymptotically AdS background equals the one of a conformal field theory on its boundary~\cite{witten1998,gubserklebanovpolyakov1998,aharonygubsermaldacenaoogurioz2000}, with the sources of the CFT partition function being the appropriately rescaled boundary values of fields in the bulk. The original (and most studied) example if the correspondence between Type IIB string theory on the $\text{AdS}{}_5 \times \text{S}_5$ spacetime (the bulk) and maximally supersymmetric Yang--Mills theory (SYM) on the four-dimensional boundary of $\text{AdS}{}_5$. SYM theory is specified by its gauge group, taken to be $\mathrm{SU}(N)$, and the Yang--Mills coupling $g$, while for string theory we have the string tension $\alpha'$ and the AdS curvature radius $R$. The original conjecture is that in the weak-coupling limit $\alpha' \to 0$ and with $R$ fixed and large, where string theory reduces to classical supergravity, it can be described by SYM in the planar strong-coupling limit: $N \to \infty$ and $g^2 N$ fixed and large. Beyond this limit, $1/N$ non-planar corrections on the CFT side should correspond to perturbative quantum corrections in the bulk, and corrections in $1/(g^2 N)$ should correspond to $\alpha'$ corrections from string theory. Because of the strong-weak duality, the correspondence is difficult to prove, and evidence comes mostly from cases where symmetry places strong restrictions on the form of the result: the symmetry group $\mathrm{SO}(2,d)$ of $(d+1)$-dimensional AdS space acts as conformal group on its $d$-dimensional flat boundary.

Shortly after the discovery of the AdS/CFT correspondence, Bertola, Bros, Moschella and Schaeffer~\cite{bertolaetal1999,bertolaetal2000} have shown that one can in fact obtain quite arbitrary CFTs by a procedure very reminiscent of the correspondence. Assume there exists a scalar QFT on AdS spacetime, whose correlation functions satisfy the usual axioms for QFTs (positivity, hermiticity, causality, spectrum being bounded from below), are invariant under the AdS symmetry group, and have a certain fall-off behaviour as one approaches the conformal boundary. Under these conditions, they have shown that one can restrict the correlation functions to time-like hypersurfaces, that these restrictions define a (possibly non-Lagrangian) QFT, and that in the limit where the hypersurfaces approach the conformal boundary of AdS these QFTs become conformal.\footnote{If gravity is not dynamical, the boundary theory does not possess a stress tensor, and therefore is not a physical CFT. What is meant here is that the correlation functions are covariant under the action of the conformal algebra. For lack of a better short name, we still call these theories ``CFT''.} An advantage over the usual methods that are used to construct QFTs is that since the conformal symmetry of the boundary QFT descends from the AdS symmetry of the bulk QFT, is not broken at intermediate steps if one uses an AdS-invariant regularisation, and is therefore manifestly non-anomalous.

Bertola et al.~\cite{bertolaetal1999,bertolaetal2000} and Bros, Epstein and Moschella~\cite{brosepsteinmoschella2002} have shown that the required assumptions are fulfilled for free scalar fields, and D{\"u}tsch and Rehren~\cite{duetschrehren2002} have argued that one can identify each Feynman diagram in a perturbative expansion of both the AdS QFT and the boundary CFT to which it should restrict (possibly up to renormalisation ambiguities). Therefore, it seems that any perturbatively interacting scalar field theory in AdS can indeed be restricted to the boundary via the approach of Bertola et al. and defines an interacting CFT there. Moreover, for free fields of any spin (and more generally tree-level diagrams, i.e., classically interacting fields) the approach of Bertola et al. corresponds exactly to the method used in the AdS/CFT correspondence to relate propagators of classical fields in AdS to correlation functions of the boundary CFT. Basically, this statement holds because Witten diagrams (which give the correlation functions of the boundary CFT) are boundary limits of the corresponding Feynman diagrams for the AdS fields.

Already some time ago, the structure of tree-level Witten diagrams has been thoroughly clarified for all kinds of quantum fields, and both recursion relations and Feynman-type rules~\cite{fitzpatricketal2011,paulos2011,nandanvolovichwen2012,kharelsiopsis2013,goncalvespenedonestrevisani2015,nizamirudrasarkarverma2017,sleighttaronna2018,albayrakchowdhurykharel2019,zhou2020} have been given to construct the tree-level diagrams in the Mellin representation~\cite{mack2009a,mack2009b}. On the other hand, loop diagrams in AdS are technically much more difficult, and many results are only available at one loop~\cite{mansfieldnolland2000,hoffmannmesrefruehl2000,fitzpatrickkaplan2012,giombiklebanov2013,aharonyaldaybissiperlmutter2017,giombisleighttaronna2018,cardona2017,aldaybissi2017,bertansachs2018,bertansachsskvortsov2019,carmidipietrokomatsu2019,ponomarev2020,albayrakchowdhurykharel2020,carmi2020,carmi2021,carmipenedonessilvazhiboedov2021,albayrakkharel2021,aldaybissizhou2022,costantinofichet2021,fichet2021a}. Only recently, the full recursive structure of the Mellin amplitude for loop diagrams has been uncovered~\cite{yuan2017,yuan2018}, and certain classes of loop diagrams could be resummed~\cite{cornalbacostapenedonesschiappa2007,cornalbacostapenedones2007,carmidipietrokomatsu2019,carmi2020,carmi2021,fichet2021b}. However, most constructions work at the level of individual Witten diagrams, and it is useful to be able to treat all diagrams corresponding to a given order in perturbation theory at once. In particular, this is possible if the theory has enhanced symmetry (such as supersymmetry) and/or the external operators are protected, and then also higher-loop results are available (see for example~\cite{edenpetkouschubertsokatchev2001,uruchurtu2011,rastellizhou2017,arutyunovfrolovklabberssavin2017,rastellizhou2018,apriledrummondhesloppaul2018,rastelliroumpedakiszhou2019,aldayzhou2020,huangyuan2021,drummondpaul2022,aldaychester2022,caronhuotcoronadotrinhzahraee2022} and references therein).

For quantum field theory in flat space, one method that has been used as an alternative to a Feynman diagram expansion is the method of renormalisation group flow equations, which exist in many different incarnations, among others Stueckelberg and Petermann~\cite{stueckelbergpetermann1953}, Wilson~\cite{wilson1971}, Wegner and Houghton~\cite{wegnerhoughton1973}, Polchinski~\cite{polchinski1984}, Keller, Kopper and Salmhofer~\cite{kellerkoppersalmhofer1992}, Wetterich~\cite{wetterich1993} and Morris~\cite{morris1994}. For our purposes, the most useful formulation is the one of~\cite{kellerkoppersalmhofer1992}, in which one introduces both a UV cutoff $\Lambda_0$ and an IR cutoff $\Lambda$, and studies the flow of correlation functions as the IR cutoff $\Lambda$ is varied, holding the UV cutoff $\Lambda_0$ fixed. Using this formulation, perturbative renormalisability of a variety of four-dimensional quantum field theories has been proven both in Euclidean~\cite{kellerkopper1991,kellerkopper1994,mueller2003,koppermueller2009,froebhollandhollands2016,efremovguidakopper2017} and Lorentzian~\cite{kellerkopperschophaus1997} signature, among others. Moreover, even though the cutoff procedure breaks gauge symmetry, adapting the BRST formulation to the flow equations it has been shown that gauge invariance can be restored in the physical limit $\Lambda \to 0$, $\Lambda_0 \to \infty$, and that the relevant Ward identities are fulfilled~\cite{froebhollandhollands2016}.\footnote{However, see also the recent works~\cite{asnafigieszambelli2019,igarashiitohmorris2019} for BRST-invariant formulations of the RG flow equations in special cases.} The RG flow equations have furthermore been used to prove renormalisability of correlation functions with composite operator insertions~\cite{kellerkopper1992,froebhollandhollands2016}, the perturbative existence of the operator product expansion for general quantum field theories~\cite{kellerkopper1993,hollandhollands2013,hollandhollands2015b,froebholland2016} and even its convergence at each order in perturbation theory~\cite{hollandskopper2012,hollandhollandskopper2016}; moreover, it has been used to derive novel formulas to recursively determine the renormalised OPE coefficients order by order in perturbation theory~\cite{hollandhollands2015a,froebholland2016}.\footnote{See also~\cite{bochicchio2017,becchettibochicchio2019,bochicchiopallante2022} for similar derivations.}

In this article, we apply the RG flow equations to a weakly coupled quantum field theory on (Euclidean) AdS. In the formalism of~\cite{kellerkoppersalmhofer1992}, the flow equations relate the $\Lambda$ derivative of an $n$-point correlation function to the integral over an $(n+2)$-point correlation function with two legs connected by the $\Lambda$ derivative of the propagator, and the product of a $(k+1)$-point correlation function and an $(n-k+1)$-point correlation function with one leg each connected by the $\Lambda$ derivative of the propagator; the schematic form of the flow equations is depicted in figure~\ref{fig:flow}. Taking the external legs to the boundary, we obtain the flow of Witten diagrams to arbitrary loop order. However, the internal propagators are still in the bulk, in particular (the $\Lambda$ derivative of) the propagator that appears explicitly in the flow. To obtain a flow purely in terms of boundary correlation functions, we use the split representation~\cite{leonhardtmanvelyanruehl2003,leonhardtruehlmanvelyan2004,penedones2011,paulos2011,balitsky2011,giecold2012,costagoncalvespenedones2014} for the free propagator in AdS, generalised to include the UV and IR cutoffs. Since the cutoff regularisation that is used in this formalism is AdS invariant, the resulting boundary correlation functions are conformally invariant for all values of the cutoffs.\footnote{This observation was already made in~\cite{antunes2021}.} That is, the cutoffs in AdS do not translate to cutoffs in the CFT, but rather one obtains a family of CFTs depending on the AdS cutoffs as parameters. Nevertheless, bulk counterterms are still required (and give the boundary conditions of the flow), but one may of course choose to work in a dimensional renormalisation scheme with minimal subtraction and without explicit counterterms. In this regard, the approach and flow equations presented here are completely different from the holographic renormalisation group~\cite{deboerverlindeverlinde2000,deboer2000,li2000,heemskerkpolchinski2011,sathiapalansonoda2017}, in which a radial cutoff in the bulk corresponds to a UV cutoff in the CFT.

We then use the Mellin representation~\cite{mack2009a,mack2009b} of conformally invariant $n$-point correlation functions with $n \geq 3$ to derive the flow of Mellin amplitudes. On the other hand, for the one- and two-point functions, which determine the conformal dimensions of the operators dual to the AdS quantum fields, we can solve the flow exactly. It turns out that the boundary terms of the flow, the counterterms in AdS, give the contact amplitudes at tree level, that is, Witten diagrams with a single interaction vertex to which all bulk-to-boundary propagators are connected. Witten diagrams with a more complicated topology are recursively constructed by the flow, by either connecting together two diagrams with less external legs (from the second term in the flow equation) or closing two legs with a propagator, forming a loop (from the first term in the flow equation). We thus obtain a systematic way of computing loop corrections, which on the level of individual Witten diagrams reduces to the (by now) known recursion relations. However, in principle the flow equations could also be solved non-perturbatively and numerically, given initial data. For this, one would have to use the OPE to relate the Mellin amplitude of higher $n$-point functions to the three-point amplitude, and obtain in this way a closed (infinite-dimensional) system of ODEs for the conformal data of primary operators in the CFT. Employing the results for the flow and the OPE in flat space~\cite{hollandskopper2012,hollandhollands2015a,hollandhollandskopper2016}, this has been done by Hollands~\cite{hollands2018} to derive the change in conformal data under a marginal perturbation of the CFT under study (and similar results have been obtained by Behan~\cite{behan2018}). However, applying this result in the simplest setting, a perturbation of mean-field (or generalised free-field) theory, results in degenerate boundary conditions, which prohibits the numerical solution of the system of ODEs. In contrast, the natural initial conditions for the flow derived in this article are obtained in the planar limit $N \to \infty$ (or more general large central charge~\cite{heemskerkpenedonespolchinskisully2009,elshowkpapadodimas2012,fitzpatrickkaplan2013,aldaybissi2017}), where one might expect non-degenerate boundary conditions. However, we do not consider the OPE in this article and leave these studies to future work.

The remainder of this article is organised as follows: We first review the embedding formalism for AdS and the split representation for the propagator, and derive the split representation for the propagator with UV and IR cutoffs in section~\ref{sec:embeddingsplit}. In section~\ref{sec:singletrace}, we derive the flow equations for the correlation functions of single-trace operators of the CFT, which are dual to basic fields in the AdS bulk, as well as the flow of the corresponding Mellin amplitudes. These flows are extended to multi-trace operators of the CFT, dual to composite fields in the bulk, in the next section~\ref{sec:multitrace}. In section~\ref{sec:flow_tree_oneloop}, we solve the flow at tree level and one loop to obtain explicit expressions for the one-loop corrections to the conformal dimensions and two-point normalisation factors in the general case, and in section~\ref{sec:oneloop_corrections} we apply these formulas to compute one-loop corrections to the conformal dimensions for some operators in the CFT dual to an $\mathrm{O}(N)$ scalar model in AdS. We conclude in section~\ref{sec:conclusion}, and give details on the computation of various AdS and Mellin--Barnes integrals in the appendix~\ref{sec:appendix}.

\section{Embedding formalism and split representation}
\label{sec:embeddingsplit}

We shortly review both the embedding formalism for AdS and the various propagators, including their split representation; a nice introduction to this topic can be found in~\cite{costagoncalvespenedones2014}. Apart from establishing notations, we also derive the split representation for the propagator with UV and IR cutoffs.

\subsection{Embedding formalism}

We consider Euclidean $(d+1)$-dimensional AdS or hyperbolic space, obtained as the hyperboloid $X \cdot X \equiv X^A X_A = - \ell^2$ in the ambient $(d+2)$-dimensional Minkowski space with metric $\eta_{AB}$ and Cartesian coordinates $X^A$. A useful coordinate system are Poincar{\'e} coordinates
\begin{equation}
\label{eq:coordinates_poincare}
X^0 = \frac{1}{2 z} \left( \ell^2 + z^2 + \vec{x}^2 \right) \eqend{,} \qquad X^a = \frac{\ell}{z} \vec{x}^a \eqend{,} \qquad X^{d+1} = \frac{1}{2 z} \left( \ell^2 - z^2 - \vec{x}^2 \right) \eqend{,}
\end{equation}
where the induced metric reads
\begin{equation}
\total s^2 = \frac{\ell^2}{z^2} \left[ \total z^2 + \delta_{ab} \total \vec{x}^a \total \vec{x}^b \right] \eqend{.}
\end{equation}
In these coordinates, the invariant chordal distance reads
\begin{equation}
u(X,Y) = \frac{1}{z_x z_y} \left[ (z_x-z_y)^2 + ( \vec{x} - \vec{y} )^2 \right] \eqend{,}
\end{equation}
and the integration measure is
\begin{equation}
\total X = \left( \frac{\ell}{z} \right)^{d+1} \total z \total^d \vec{x} \eqend{.}
\end{equation}

The conformal boundary of EAdS is obtained in the limit $\chi \to \infty$ or $z \to 0$, and can be identified with light rays passing through the origin of the embedding space: $P^A P_A = 0$, $P^A \sim \lambda P^A$ for $\lambda \in \mathbb{R}$. In Poincar{\'e} coordinates, we choose the standard representative $P^A = \lim_{z \to 0} \left( z X^A/\ell \right)$, which in components reads
\begin{equation}
P^0 = \frac{1}{2 \ell} \left( \ell^2 + \vec{p}^2 \right) \eqend{,} \qquad P^a = \vec{p}^a \eqend{,} \qquad P^{d+1} = \frac{1}{2 \ell} \left( \ell^2 - \vec{p}^2 \right) \eqend{,}
\end{equation}
and define the induced integration measure on the boundary as $\total P \equiv \total^d \vec{p}$. The invariant distance from a point in the bulk to a boundary one is given by
\begin{equation}
\ell^2 + (X-P) \cdot (X-P) = - 2 P \cdot X = \frac{\ell}{z} \left[ z^2 + ( \vec{x} - \vec{p} )^2 \right] \eqend{,}
\end{equation}
and the one between two boundary points reads
\begin{equation}
(P-Q) \cdot (P-Q) = - 2 P \cdot Q = ( \vec{p} - \vec{q} )^2 \eqend{.}
\end{equation}
For the boundary limits, we note the relations
\begin{equation}
\lim_{z_x \to 0} \left[ \frac{z_x}{\ell} u(X,Y) \right] = - 2 \ell^{-2} P \cdot Y \eqend{,} \qquad \lim_{z_x \to 0} \left[ \frac{z_x}{\ell} ( - 2 \ell^{-2} Q \cdot X ) \right] = - 2 \ell^{-2} P \cdot Q \eqend{.}
\end{equation}

The covariant derivative in hyperbolic space can be obtained by projecting the ambient space partial derivative onto the hyperboloid with the projector $g_A^B \equiv \delta_A^B + \ell^{-2} X_A X^B$, which is the induced metric on the hyperboloid, and we have in particular
\begin{equation}
\label{eq:nabla_projected}
\nabla_A \phi = \left( \delta_A^B + \ell^{-2} X_A X^B \right) \partial_B \phi \eqend{,} \qquad \nabla^2 \phi = \left( \eta^{AB} + \ell^{-2} X^A X^B \right) \partial_A \partial_B \phi + \frac{d+1}{\ell^2} X^A \partial_A \phi \eqend{.}
\end{equation}
Acting on a function of invariant distances, we also obtain
\begin{equations}
\ell^2 \nabla^2 f(u(X,Y)) &= u (u+4) f''(u) + (d+1) (u+2) f'(u) \eqend{,} \label{eq:nabla_bulk} \\
\ell^2 \nabla^2 f(- 2 \ell^{-2} X \cdot P) &= (- 2 \ell^{-2} X \cdot P)^2 f'' + (d+1) (- 2 \ell^{-2} X \cdot P) f' \eqend{.} \label{eq:nabla_boundary}
\end{equations}

\subsection{Propagators}

The scalar bulk propagator is given by~\cite{penedones2011}
\begin{splitequation}
\label{eq:bulk_propagator}
G_\Delta(X,Y) &= \frac{\ell^{1-d} \Gamma(\Delta)}{2 \pi^\rho \Gamma\left( \Delta - \rho + 1 \right)} u^{-\Delta} \hypergeom{2}{1}\left( \Delta, \Delta - \rho + \frac{1}{2}; 2 \Delta - d + 1; - \frac{4}{u(X,Y)} \right) \\
&= \frac{\ell^{1-d}}{2^{d+1} \pi^\frac{d+1}{2}} \int_{-\mathi \infty}^{\mathi \infty} \frac{\Gamma(\Delta-t) \Gamma\left( t - \rho + \frac{1}{2} \right) \Gamma(t)}{\Gamma(\Delta-d+1+t)} \left[ \frac{4}{u(X,Y)} \right]^t \frac{\total t}{2 \pi \mathi} \eqend{,}
\end{splitequation}
where $\Delta$ is related to the mass $m^2 = \ell^{-2} \Delta (\Delta - d)$ and we set $d = 2 \rho$. For $\Delta > \rho - 1/2$, the integration contour is a straight line with $\rho-1/2 < \Re t < \Delta$, while for smaller $\Delta$ it is indented in such a way that the poles of the $\Gamma$ functions of the form $\Gamma(a+t)$ (the left poles) lie to the left of the contour and the poles of the $\Gamma$ functions of the form $\Gamma(a-t)$ (the right poles) lie to its right.  The bulk-to-boundary propagator is obtained as the limit
\begin{splitequation}
\label{eq:bulk_to_boundary_propagator}
G_{\partial \Delta}(X,P) &\equiv \lim_{z_y \to 0} \left( \frac{z_y}{\ell} \right)^{-\Delta} G_\Delta(X,Y) = \frac{\ell^{1-d} \Gamma(\Delta)}{2 \pi^\rho \Gamma\left( \Delta - \rho + 1 \right)} \left( - 2 \ell^{-2} P \cdot X \right)^{-\Delta} \eqend{,}
\end{splitequation}
and the boundary propagator as
\begin{splitequation}
\label{eq:boundary_to_boundary_propagator}
\lim_{z_x \to 0} \left( \frac{z_x}{\ell} \right)^{-\Delta} G_{\partial \Delta}(X,P) = \frac{\ell^{1-d} \Gamma(\Delta)}{2 \pi^\rho \Gamma\left( \Delta - \rho + 1 \right)} \left( - 2 \ell^{-2} P \cdot Q \right)^{-\Delta} \eqend{.}
\end{splitequation}
Using equation~\eqref{eq:nabla_boundary}, one easily checks that
\begin{equation}
\label{eq:nabla2_boundary_propagator}
\left[ \nabla^2 + \frac{\Delta(d-\Delta)}{\ell^2} \right] G_{\partial \Delta}(X,P) = 0 \eqend{.}
\end{equation}
For given mass $m$, there are two solutions $\Delta_\pm = \rho \pm \sqrt{ m^2 \ell^2 + \rho^2 }$ for the dimension parameter $\Delta$, with $\Delta = \Delta_+$ known as standard quantisation (corresponding to Dirichlet boundary conditions for the scalar field) and $\Delta = \Delta_-$ known as alternative quantisation (corresponding to Neumann boundary conditions). For scalar fields satisfying the Breitenlohner--Freedman bound $m^2 \ell^2 > - \rho^2$~\cite{breitenlohnerfreedman1982a,breitenlohnerfreedman1982b}, standard quantisation always leads to a viable theory without ghosts, while alternative quantisation can only be used for fields with mass greater than the BF bound but less than $m^2 \ell^2 < 1-\rho^2$~\cite{balasubramaniankrauslawrence1999,klebanovwitten1999}.\footnote{However, see~\cite{andrademarolf2012} for exceptions from this general picture.} In the following, we will always use standard quantisation wirh $\Delta = \Delta_+ > \rho$, and leave a study of the alternative to future work.

The bulk propagator has a spectral representation~\cite{penedones2011}
\begin{equation}
G_\Delta(X,Y) = \ell^{1-d} \int_{-\infty}^\infty \frac{1}{\nu^2 + (\Delta-\rho)^2} \Omega_\nu(u(X,Y)) \total \nu
\end{equation}
with the harmonic function
\begin{equation}
\Omega_\nu(u) = \ell^{d-1} \frac{\mathi \nu}{2\pi} \left[ G_{\rho+\mathi \nu}(u) - G_{\rho-\mathi \nu}(u) \right]
\end{equation}
satisfying
\begin{equation}
\nabla^2 \Omega_\nu(u) = - \frac{\rho^2+\nu^2}{\ell^2} \Omega_\nu(u) \eqend{.}
\end{equation}
For the harmonic function, we have the Mellin--Barnes representation
\begin{equation}
\Omega_\nu(u) = \frac{\nu \sinh\left( \pi \nu \right)}{2^{d+1} \pi^\frac{d+3}{2}} \int_{0 < \Re t < \rho} \Gamma(\rho-\mathi \nu-t) \Gamma(\rho+\mathi \nu-t) \frac{\Gamma(t)}{\Gamma\left( \frac{1}{2} + \rho - t \right)} \left( \frac{4}{u} \right)^t \frac{\total t}{2 \pi \mathi}
\end{equation}
and the split representation
\begin{equation}
\Omega_\nu(u(X,Y)) = \ell^{d-2} \frac{\nu^2}{\pi} \int G_{\partial (\rho+\mathi \nu)}(X,P) G_{\partial (\rho-\mathi \nu)}(Y,P) \total P \eqend{.}
\end{equation}
Since the propagator fulfills
\begin{equation}
\left[ \nabla^2 - \frac{\Delta (\Delta-d)}{\ell^2} \right] G_\Delta(X,Y) = - \delta(X,Y) \eqend{,}
\end{equation}
we obtain from the split representation that
\begin{equation}
\label{eq:omega_harmonic_integral}
\int_{-\infty}^\infty \Omega_\nu(u(X,Y)) \total \nu = \ell^{1+d} \delta(X,Y) \eqend{.}
\end{equation}

Given the hyperbolic heat kernel $K_{d,\Delta}$ which fulfills the equation
\begin{equation}
\left[ \partial_t - \nabla^2 + \frac{\Delta (\Delta-d)}{\ell^2} \right] K_{d,\Delta}(X,Y,t) = 0
\end{equation}
with boundary condition
\begin{equation}
\label{eq:heatkernel_boundary}
\lim_{t \to 0} K_{d,\Delta}(X,Y,t) = \delta(X,Y) \eqend{,}
\end{equation}
we can obtain the propagator as
\begin{equation}
G_\Delta(X,Y) = \int_0^\infty K_{d,\Delta}(X,Y,t) \total t \eqend{,}
\end{equation}
since then we have as required
\begin{equation}
\left[ \nabla^2 - \frac{\Delta (\Delta-d)}{\ell^2} \right] G_\Delta(X,Y) = \int_0^\infty \partial_t K_{d,\Delta}(X,Y,t) \total t = - \lim_{t \to 0} K_{d,\Delta}(X,Y,t) = - \delta(X,Y) \eqend{.}
\end{equation}
For the heat kernel, we assume the spectral representation
\begin{equation}
K_{d,\Delta}(X,Y,t) = \int_{-\infty}^\infty \tilde{K}_{d,\Delta}(\nu,t) \Omega_\nu(u(X,Y)) \total \nu \eqend{,}
\end{equation}
whence it follows that
\begin{equation}
0 = \int_{-\infty}^\infty \left[ \partial_t \tilde{K}_{d,\Delta}(\nu,t) + \frac{\nu^2 + (\Delta-\rho)^2}{\ell^2} \tilde{K}_{d,\Delta}(\nu,t) \right] \Omega_\nu(u(X,Y)) \total \nu
\end{equation}
with solution
\begin{equation}
\tilde{K}_{d,\Delta}(\nu,t) = C \exp\left[ - \frac{\nu^2 + (\Delta-\rho)^2}{\ell^2} t \right] \eqend{.}
\end{equation}
As $t \to 0$, we use the boundary condition~\eqref{eq:heatkernel_boundary} together with~\eqref{eq:omega_harmonic_integral} to determine $C = \ell^{-d-1}$.

The propagator with IR cutoff $\Lambda$ and UV cutoff $\Lambda_0$ can then be obtained by restricting the $t$ integration to a finite range, and it follows that
\begin{splitequation}
\label{eq:cutoff_propagator_int}
G^{\Lambda,\Lambda_0}_\Delta(X,Y) &= \int_{\Lambda_0^{-2}}^{\Lambda^{-2}} K_{d,\Delta}(X,Y,t) \total t = \int_{-\infty}^\infty \int_{\Lambda_0^{-2}}^{\Lambda^{-2}} \tilde{K}_{d,\Delta}(\nu,t) \total t \, \Omega_\nu(X,Y) \total \nu \\
&= \ell^{-d-1} \int_{-\infty}^\infty \int_{\Lambda_0^{-2}}^{\Lambda^{-2}} \exp\left[ - \frac{\nu^2 + (\Delta-\rho)^2}{\ell^2} t \right] \total t \, \Omega_\nu(X,Y) \total \nu \\
&= \ell^{1-d} \int_{-\infty}^\infty \frac{1}{\nu^2 + (\Delta-\rho)^2} \left[ \mathe^{- \frac{\nu^2 + (\Delta-\rho)^2}{\ell^2 \Lambda_0^2} } - \mathe^{ - \frac{\nu^2 + (\Delta-\rho)^2}{\ell^2 \Lambda^2} } \right] \, \Omega_\nu(X,Y) \total \nu \eqend{.}
\end{splitequation}
Taking a derivative with respect to the IR cutoff $\Lambda$, we finally obtain
\begin{splitequation}
\label{eq:cutoff_propagator}
&\partial_\Lambda G^{\Lambda,\Lambda_0}_\Delta(X,Y) = - \frac{2}{\Lambda^3} \ell^{-1-d} \int_{-\infty}^\infty \exp\left( - \frac{\nu^2 + (\Delta-\rho)^2}{\ell^2 \Lambda^2} \right) \Omega_\nu(X,Y) \total \nu \\
&\quad= - \frac{2}{\ell^3 \Lambda^3} \int_{-\infty}^\infty \exp\left( - \frac{\nu^2 + (\Delta-\rho)^2}{\ell^2 \Lambda^2} \right) \frac{\nu^2}{\pi} \int G_{\partial (\rho+\mathi \nu)}(X,P) G_{\partial (\rho-\mathi \nu)}(Y,P) \total P \total \nu \eqend{.}
\end{splitequation}
Similar considerations apply to propagators of higher-spin fields; see~\cite{costagoncalvespenedones2014} for the bulk propagators and their split representations. The cutoff propagator has a similar structure to the one in flat space~\cite{hollandskopper2012}, and in particular the mass (that is $\Delta$) only appears in the exponentials. For a field with mass above the Breitenlohner--Freedman bound~\cite{breitenlohnerfreedman1982a,breitenlohnerfreedman1982b} (and standard quantisation), we have $\Delta > \rho$ and the IR cutoff is of course not actually necessary to regularise the theory. It instead is used as a technical tool to derive and solve the flow equations. On the other hand, for a field with $\Delta = \rho$, we see from the explicit expression~\eqref{eq:cutoff_propagator_int} that the IR cutoff regulates the theory also at small $\nu \approx 0$.

\section{Single-trace operators}
\label{sec:singletrace}

We first derive the CFT flow equation for single-trace operators $\op_A$, which are the operators dual to basic fields $\phi_A$ in the EAdS bulk. We use a condensed notation, where the index $A$ denotes both the type of field and its spinorial or Lorentz indices, and repeated indices are implicitly summed over. We denote by $\Delta_A$ the conformal dimension of $\op_A$, which for a spin-$J$ field is related to the bulk mass as $\ell^2 m^2 = \Delta (\Delta-d) - J$~\cite{costagoncalvespenedones2014}, generalising the relation for scalar fiels from the last section. The derivation proceeds in various steps: we first derive flow equations for generic quantum field theories, and then specialise to EAdS and correlation functions on the boundary. We then solve the flow explicitly for the one- and two-point boundary correlators, while for three- and higher point boundary correlators we derive the flow equations for the corresponding Mellin amplitudes.

\subsection{Flow equations for generic quantum field theories}

Consider the generating functional of Euclidean correlation functions
\begin{equation}
\label{eq:generating_z1}
Z[J] \equiv N \int \exp\left( - \frac{1}{\hbar} S[\phi] + \int J^A(X) \phi_A(X) \total X \right) \mathcal{D} \phi \eqend{,}
\end{equation}
where the constant $N$ is chosen such that $Z[0] = 1$, and where we have explicitly introduced $\hbar$ as loop counting parameter. We split the action into a free part
\begin{equation}
\label{eq:free_action}
S_0[\phi] = \frac{1}{2} \int \phi_A(X) P^{AB} \phi_B(X) \total X
\end{equation}
and the interaction $S_\text{int}$ (which includes the necessary counterterms), and perform the shift $\phi_A(X) \to \phi_A(X) + \hbar \int G_{AB}(X,Y) J^B(Y) \total Y$, where $G_{AB}$ is the propagator fulfilling
\begin{equation}
P^{AB} G_{BC}(X,Y) = \delta^A_C \, \delta(X,Y) \eqend{.}
\end{equation}
This results in
\begin{equation}
\label{eq:generating_z2}
Z[J] = N \int \exp\left( - \frac{1}{\hbar} S_0[\phi] + \frac{1}{\hbar} S_0[\Phi] - \frac{1}{\hbar} S_\text{int}[\phi + \Phi] \right) \mathcal{D} \phi \eqend{,}
\end{equation}
where we defined the amputated source
\begin{equation}
\label{eq:amputated_source_def}
\Phi_A(X) \equiv \hbar \int G_{AB}(X,Y) J^B(Y) \total Y \eqend{,} \qquad P^{AB} \Phi_B = \hbar J^A \eqend{.}
\end{equation}
Taking the logarithm, we thus obtain the generating functional of connected amputated correlation functions (discarding an irrelevant additive constant)
\begin{equation}
\label{eq:generating_w}
W[\Phi] \equiv - \hbar \ln Z[J] = - S_0[\Phi] - \hbar \ln \int \exp\left( - \frac{1}{\hbar} S_0[\phi] - \frac{1}{\hbar} S_\text{int}[\phi + \Phi] \right) \mathcal{D} \phi \eqend{.}
\end{equation}

\begin{figure}[h]
\includegraphics[width=\textwidth]{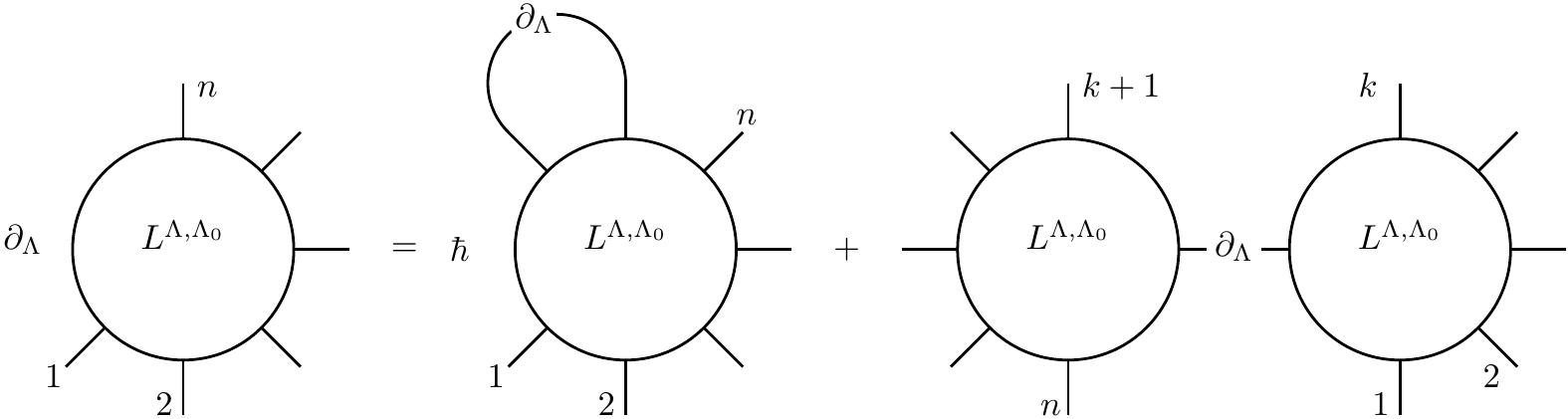}
\caption{Schematic depiction of the flow equations for $n$-point correlation functions.}
\label{fig:flow}
\end{figure}

To derive the flow equation, we regularise the propagator and introduce an IR cutoff $\Lambda$ as well as a UV cutoff $\Lambda_0$. We denote the regularised propagator by $G_{AB}^{\Lambda,\Lambda_0}$, and the physical limit is given by $\Lambda \to 0$, $\Lambda_0 \to \infty$. The inverse of this propagator enters the regularised free action $S_0^{\Lambda, \Lambda_0}$, which we use to define the normalised Gaussian measure
\begin{equation}
\label{eq:gaussian_measure}
\total \mu^{\Lambda, \Lambda_0}(\phi) \equiv \left[ \int \exp\left( - \frac{1}{\hbar} S_0^{\Lambda, \Lambda_0}[\phi] \right) \mathcal{D} \phi \right]^{-1} \exp\left( - \frac{1}{\hbar} S_0^{\Lambda, \Lambda_0}[\phi] \right) \mathcal{D} \phi \eqend{.}
\end{equation}
Subtracting the trivial free part from $W$, we obtain the regularised generating functional of connected amputated correlation functions (where only interactions contribute)
\begin{equation}
\label{eq:generating_l}
L^{\Lambda,\Lambda_0}[\Phi] = - \hbar \ln \int \exp\left( - \frac{1}{\hbar} S_\text{int}[\phi + \Phi] \right) \total \mu^{\Lambda, \Lambda_0}(\phi) \eqend{.}
\end{equation}
The flow equation gives the dependence of $L^{\Lambda,\Lambda_0}$ on the cutoff $\Lambda$, which follows from the general formula for the dependence of Gaussian measures on a parameter~\cite{salmhofer1999}. Completing the square, it follows that
\begin{equation}
\label{eq:gaussian_exponential}
\int \exp\left( \int J^A \phi_A \total X \right) \total \mu^{\Lambda, \Lambda_0}(\phi) = \exp\left( \frac{\hbar}{2} \iint J^A(X) G^{\Lambda, \Lambda_0}_{AB}(X,Y) J_B(Y) \total X \total Y \right) \eqend{,}
\end{equation}
and from this
\begin{splitequation}
&\partial_\Lambda \int \exp\left( \int J^A \phi_A \total X \right) \total \mu^{\Lambda,\Lambda_0}(\phi) \\
&\quad= \exp\left( \frac{\hbar}{2} \iint J^A(X) G^{\Lambda,\Lambda_0}_{AB}(X,Y) J_B(Y) \total X \total Y \right) \frac{\hbar}{2} \iint J^A(X) \partial_\Lambda G^{\Lambda,\Lambda_0}_{AB}(X,Y) J_B(Y) \total X \total Y \\
&\quad= \frac{\hbar}{2} \int \iint \frac{\delta}{\delta \phi_A(X)} \partial_\Lambda G^{\Lambda,\Lambda_0}_{AB}(X,Y) \frac{\delta}{\delta \phi_B(Y)} \total X \total Y \exp\left( \int J^A \phi_A \total X \right) \total \mu^{\Lambda,\Lambda_0}(\phi) \eqend{.}
\end{splitequation}
Expanding in powers of $J$, the above equality holds for polynomials, and thus for $L^{\Lambda,\Lambda_0}$~\eqref{eq:generating_l} by expanding in powers of $S_\text{int}$; it can be shown to also hold more generally~\cite{salmhofer1999}. We obtain
\begin{splitequation}
&\partial_\Lambda \exp\left( - \frac{1}{\hbar} L^{\Lambda,\Lambda_0}[\Phi] \right) = \partial_\Lambda \int \exp\left( - \frac{1}{\hbar} S_\text{int}[\phi + \Phi] \right) \total \mu^{\Lambda, \Lambda_0}(\phi) \\
&\quad= \frac{\hbar}{2} \int \iint \frac{\delta}{\delta \phi_A(X)} \partial_\Lambda G^{\Lambda,\Lambda_0}_{AB}(X,Y) \frac{\delta}{\delta \phi_B(Y)} \total X \total Y \exp\left( - \frac{1}{\hbar} S_\text{int}[\phi + \Phi] \right) \total \mu^{\Lambda, \Lambda_0}(\phi) \\
&\quad= \frac{\hbar}{2} \int \iint \frac{\delta}{\delta \Phi_A(X)} \partial_\Lambda G^{\Lambda,\Lambda_0}_{AB}(X,Y) \frac{\delta}{\delta \Phi_B(Y)} \total X \total Y \exp\left( - \frac{1}{\hbar} S_\text{int}[\phi + \Phi] \right) \total \mu^{\Lambda, \Lambda_0}(\phi) \\
&\quad= \frac{\hbar}{2} \iint \frac{\delta}{\delta \Phi_A(X)} \partial_\Lambda G^{\Lambda,\Lambda_0}_{AB}(X,Y) \frac{\delta}{\delta \Phi_B(Y)} \total X \total Y \exp\left( - \frac{1}{\hbar} L^{\Lambda,\Lambda_0}[\Phi] \right) \eqend{,}
\end{splitequation}
where we could exchange functional derivatives with respect to $\phi$ with the ones with respect to $\Phi$, and then
\begin{splitequation}
\label{eq:l_flow}
\partial_\Lambda L^{\Lambda,\Lambda_0}[\Phi] &= \frac{\hbar}{2} \iint \partial_\Lambda G^{\Lambda,\Lambda_0}_{AB}(X,Y) \frac{\delta^2 L^{\Lambda,\Lambda_0}[\Phi]}{\delta \Phi_A(X) \delta \Phi_B(Y)} \total X \total Y \\
&\quad- \frac{1}{2} \iint \frac{\delta L^{\Lambda,\Lambda_0}[\Phi]}{\delta \Phi_A(X)} \partial_\Lambda G^{\Lambda,\Lambda_0}_{AB}(X,Y) \frac{\delta L^{\Lambda,\Lambda_0}[\Phi]}{\delta \Phi_B(Y)} \total X \total Y \eqend{.}
\end{splitequation}
Expanding in powers of $\Phi$, we obtain the flow equations for the regularised connected and amputated correlation functions:
\begin{splitequation}
\label{eq:l_n_flow}
&\partial_\Lambda L^{\Lambda,\Lambda_0}_{A_1 \cdots A_n}(X_1,\ldots,X_n) = \frac{\hbar}{2} \iint \partial_\Lambda G^{\Lambda,\Lambda_0}_{BC}(Y,Z) L^{\Lambda,\Lambda_0}_{A_1 \cdots A_n BC}(X_1,\ldots,X_n,Y,Z) \total Y \total Z \\
&\qquad- \frac{1}{2} \sum_{I \cup J = \{1,\ldots,n\}} \iint L^{\Lambda,\Lambda_0}_{\{ A_I \} B}(\{ X_I \},Y) \partial_\Lambda G^{\Lambda,\Lambda_0}_{BC}(Y,Z) L^{\Lambda,\Lambda_0}_{\{ A_J \} C}(\{ X_J \},Z) \total Y \total Z \eqend{,}
\end{splitequation}
where the sum in the second line runs over all disjoint subsets $I$ and $J$. The flow equations are depicted in figure~\ref{fig:flow}: the first term on the right-hand side is a loop correction that connects two external legs of a higher $n$-point correlation function with a propagator, while the second term connects two lower $n$-point functions with a propagator.

Finally, let us assume that the regularisation is such that the regularised propagator $G_{AB}^{\Lambda,\Lambda_0}$ vanishes as $\Lambda \to \Lambda_0$, which is the case for the regularised AdS propagator~\eqref{eq:cutoff_propagator_int}. The Gaussian measure $\total \mu^{\Lambda,\Lambda_0}$~\eqref{eq:gaussian_measure} then reduces to a functional $\delta$ measure in this limit, i.e., $\int F[\phi] \total \mu^{\Lambda_0,\Lambda_0}(\phi) = F[0]$, which can be clearly seen from~\eqref{eq:gaussian_exponential} where the right-hand side gives $1$ in this limit. From equation~\eqref{eq:generating_l}, we thus obtain the boundary condition
\begin{equation}
\label{eq:l_boundary}
L^{\Lambda_0,\Lambda_0}[\Phi] = S_\text{int}[\Phi] \eqend{,}
\end{equation}
or expanding in powers of $\Phi$
\begin{equation}
\label{eq:l_n_boundary}
L^{\Lambda_0,\Lambda_0}_{A_1 \cdots A_n}(X_1,\ldots,X_n) = \left. \frac{\delta^n S_\text{int}[\Phi]}{\delta \Phi_{A_1}(X_1) \cdots \delta \Phi_{A_n}(X_n)} \right\rvert_{\Phi = 0} \eqend{.}
\end{equation}
The physical interpretation of the flow is the following: the regulated propagator $G_{AB}^{\Lambda,\Lambda_0}$ describes the propagation of fluctuating modes between $\Lambda$ and $\Lambda_0$. In the generating functional $L^{\Lambda,\Lambda_0}$ these modes are integrated out, but it still contains unintegrated fluctuating modes between $0$ and $\Lambda$; the physical theory is obtained in the limit $\Lambda \to 0$. When setting $\Lambda = \Lambda_0$, no modes at all are integrated out, and we recover the bare action at the scale $\Lambda_0$. In particular, $S_\text{int}$ must contain all the counterterms (depending on the UV cutoff $\Lambda_0$) that are needed to obtain a finite result for $L^{0,\infty}_{A_1 \cdots A_n}$.

The system of flow equations~\eqref{eq:l_n_flow}, together with the boundary conditions~\eqref{eq:l_n_boundary}, have been used to prove perturbative renormalisability for scalar and gauge theories~\cite{kellerkopper1991,kellerkopper1994,kellerkopperschophaus1997,mueller2003,koppermueller2009,froebhollandhollands2016,efremovguidakopper2017}, as well as the existence and properties of the operator product expansion~\cite{kellerkopper1993,hollandskopper2012,hollandhollands2013,hollandhollands2015b,hollandhollandskopper2016,froebholland2016}, and recursive formulas for the construction of OPE coefficients~\cite{hollandhollands2015a,froebholland2016} and CFT data~\cite{hollands2018}. Expanding in powers of $\hbar$, one can easily generate all connected Feynman diagrams.

\subsection{Flow equations in EAdS and on the boundary}

The flow equations~\eqref{eq:l_n_flow} together with the boundary conditions~\eqref{eq:l_n_boundary} for the flow are obviously also valid for quantum field theories in hyperbolic space. Since we derived the flow for amputated correlation functions, it is in fact very easy to obtain boundary correlators: we simply have to convolve the amputated correlation functions with bulk-to-boundary propagators $G_{A \partial B}(X,P)$. To simplify the following discussion, we assume that the free action~\eqref{eq:free_action} and thus the propagator $G_{AB}(X,Y)$ are diagonal in field space, which can always be achieved by adding auxiliary fields, and furthermore restrict to scalar fields.

Let us thus define
\begin{equation}
\label{eq:witten_s}
S^{\Lambda,\Lambda_0}_{A_1 \cdots A_n}(P_1,\ldots,P_n) \equiv \int\dotsi\int L^{\Lambda,\Lambda_0}_{A_1 \cdots A_n}(X_1,\ldots,X_n) \prod_{k=1}^n G_{\partial A_k}(X_k,P_k) \total X_k \eqend{,}
\end{equation}
which are exactly the connected boundary correlation functions described by Witten diagrams (to all loop orders). In order to express the flow equations~\eqref{eq:l_n_flow} purely in terms of these correlators, we need to use the split representation for the bulk propagators appearing there. Since this involves bulk-to-boundary propagators with complex conformal dimensions, we assume that the connected correlators are analytic functions of the respective conformal dimensions --- which, as we will see later, holds by construction for the boundary correlation functions that we construct using flow equations. The relevant split representation reads~\eqref{eq:cutoff_propagator}
\begin{splitequation}
\label{eq:dlambda_g_split}
\partial_\Lambda G^{\Lambda,\Lambda_0}_A(X,Y) &= - \frac{2}{\ell^3 \Lambda^3} \int_{-\infty}^\infty \exp\left( - \frac{\nu^2 + (\Delta_A-\rho)^2}{\ell^2 \Lambda^2} \right) \frac{\nu^2}{\pi} \\
&\qquad\times \int G_{\partial A(\rho+\mathi \nu)}(X,P) G_{\partial A(\rho-\mathi \nu)}(Y,P) \total P \total \nu \eqend{,}
\end{splitequation}
where by the notation $G_{\partial A(\Delta')}$ we denote the bulk-to-boundary propagator of the field $\phi_A$, with the conformal dimension analytically continued to $\Delta'$.

Inserting the split representation~\eqref{eq:dlambda_g_split} into the flow equations~\eqref{eq:l_n_flow} and using the definition~\eqref{eq:witten_s}, we obtain the flow equations for the boundary correlators. This reads
\begin{splitequation}
\label{eq:s_n_flow_pre}
&\partial_\Lambda S^{\Lambda,\Lambda_0}_{A_1 \cdots A_n}(P_1,\ldots,P_n) = - \frac{1}{\ell^3 \Lambda^3} \sum_B \int_{-\infty}^\infty \exp\left( - \frac{\nu^2 + (\Delta_B-\rho)^2}{\ell^2 \Lambda^2} \right) \frac{\nu^2}{\pi} \\
&\hspace{5em}\times \int \Bigg[ \hbar \, S^{\Lambda,\Lambda_0}_{A_1 \cdots A_n B(\rho+\mathi \nu) B(\rho-\mathi \nu)}(P_1,\ldots,P_n,Q,Q) \\
&\hspace{6em}- \sum_{I \cup J = \{1,\ldots,n\}} S^{\Lambda,\Lambda_0}_{\{ A_I \} B(\rho+\mathi \nu)}(\{ P_I \},Q) S^{\Lambda,\Lambda_0}_{\{ A_J \} B(\rho-\mathi \nu)}(\{ P_J \},Q) \Bigg] \total Q \total \nu \eqend{,}
\end{splitequation}
where have now made the sum over all basic fields $\phi_B$ of the theory explicit, and use the same notation as for the bulk-to-boundary propagator: $A(\Delta')$ denotes the boundary correlator of the field $\phi_A$, with the conformal dimension analytically continued to $\Delta'$. The boundary conditions for the flow at $\Lambda = \Lambda_0$ are obtained by inserting the definition~\eqref{eq:witten_s} into the boundary conditions~\eqref{eq:l_n_boundary}, and we obtain
\begin{equation}
\label{eq:s_n_boundary}
S^{\Lambda_0,\Lambda_0}_{A_1 \cdots A_n}(P_1,\ldots,P_n) = \int\dotsi\int \left. \frac{\delta^n S_\text{int}[\Phi]}{\delta \Phi_{A_1}(X_1) \cdots \delta \Phi_{A_n}(X_n)} \right\rvert_{\Phi = 0} \prod_{k=1}^n G_{\partial A_k}(X_k,P_k) \total X_k \eqend{.}
\end{equation}

Since the regularisation and thus the cutoff propagator~\eqref{eq:dlambda_g_split} are EAdS-invariant, the boundary correlators $S^{\Lambda,\Lambda_0}_{A_1 \cdots A_n}$ actually enjoy full conformal symmetry.\footnote{This observation was already made in~\cite{antunes2021}.} That is, the bulk regularisation does \emph{not} regularise the boundary correlators in the traditional sense, which can be seen in the loop contribution (proportional to $\hbar$) in the flow equation~\eqref{eq:s_n_flow_pre}, where a boundary correlator is evaluated at coincident points $Q$. To also regularise this divergence, we would have to place an additional IR cutoff in the bulk, evaluating the boundary correlators not at the boundary $z = 0$, but rather at a finite $z = \epsilon$. One could then extract the divergences as $\epsilon \to 0$, add counterterms to cancel them, and finally take the limit $\epsilon \to 0$ to obtain the finite renormalised result~\cite{skenderis2002}. However, we will use an equivalent but quicker way, displacing one of the two points $Q \to Q'$ and taking the finite part $\Pf$ of the result as $Q' \to Q$. That is, instead of the formal but divergent flow equation~\eqref{eq:s_n_flow_pre}, we take
\begin{splitequation}
\label{eq:s_n_flow}
&\partial_\Lambda S^{\Lambda,\Lambda_0}_{A_1 \cdots A_n}(P_1,\ldots,P_n) = - \frac{1}{\ell^3 \Lambda^3} \sum_B \int_{-\infty}^\infty \exp\left( - \frac{\nu^2 + (\Delta_B-\rho)^2}{\ell^2 \Lambda^2} \right) \frac{\nu^2}{\pi} \\
&\hspace{5em}\times \int \Pf_{Q' \to Q} \Bigg[ \hbar \, S^{\Lambda,\Lambda_0}_{A_1 \cdots A_n B(\rho+\mathi \nu) B(\rho-\mathi \nu)}(P_1,\ldots,P_n,Q,Q') \\
&\hspace{6em}- \sum_{I \cup J = \{1,\ldots,n\}} S^{\Lambda,\Lambda_0}_{\{ A_I \} B(\rho+\mathi \nu)}(\{ P_I \},Q) S^{\Lambda,\Lambda_0}_{\{ A_J \} B(\rho-\mathi \nu)}(\{ P_J \},Q) \Bigg] \total Q \total \nu \eqend{,}
\end{splitequation}
which gives a finite result.

Using the flow equation~\eqref{eq:s_n_flow} together with the boundary conditions~\eqref{eq:s_n_boundary} of the flow, one can now construct boundary correlators to arbitrary order in the loop expansion in $\hbar$. At each order, the integration over the flow parameter $\Lambda$ is trivial since the dependence on $\Lambda$ is only in the exponentials; the difficult integrations are the ones over the boundary point $Q$ and the spectral parameter $\nu$.

\subsection{Solution of the flow equations for one- and two-point correlators}
\label{sec_flow_one_two}

To illustrate the flow equation~\eqref{eq:s_n_flow}, let us solve it explicitly at low orders. We start with $n = 1$, where the left-hand side should vanish if $\phi_A$ does not have a VEV (at least in the physical limit $\Lambda \to 0$). On the right-hand side, the loop correction is a three-point function, which by conformal symmetry has the general form
\begin{splitequation}
\label{eq:s_3_form}
S^{\Lambda,\Lambda_0}_{A_1 A_2 A_3}(P_1,P_2,P_3) &= c^{\Lambda,\Lambda_0}_{A_1 A_2 A_3} \left( - 2 \ell^{-2} P_1 \cdot P_2 \right)^{\Delta_{312}^{\Lambda,\Lambda_0}} \\
&\quad\times \left( - 2 \ell^{-2} P_1 \cdot P_3 \right)^{\Delta_{213}^{\Lambda,\Lambda_0}} \left( - 2 \ell^{-2} P_2 \cdot P_3 \right)^{\Delta_{123}^{\Lambda,\Lambda_0}} \eqend{,}
\end{splitequation}
where we defined
\begin{equation}
\label{eq:delta_3_def}
\Delta_i^{\Lambda,\Lambda_0} \equiv \Delta_i + \delta \Delta_i^{\Lambda,\Lambda_0} \eqend{,} \qquad \Delta_{ijk}^{\Lambda,\Lambda_0} \equiv \frac{1}{2} \left( \Delta_i^{\Lambda,\Lambda_0} - \Delta_j^{\Lambda,\Lambda_0} - \Delta_k^{\Lambda,\Lambda_0} \right)
\end{equation}
and took into account that the normalisation factor $c^{\Lambda,\Lambda_0}_{A_1 A_2 A_3}$ and the perturbative corrections $\delta \Delta_i^{\Lambda,\Lambda_0}$ to the conformal dimensions can depend on the cutoffs $\Lambda$ and $\Lambda_0$. The term that appears in the flow equation is
\begin{splitequation}
&\int \Pf_{Q' \to Q} S^{\Lambda,\Lambda_0}_{A B(\rho+\mathi \nu) B(\rho-\mathi \nu)}(P,Q,Q') \total Q = c^{\Lambda,\Lambda_0}_{A B(\rho+\mathi \nu) B(\rho-\mathi \nu)} \int \left( - 2 \ell^{-2} P \cdot Q \right)^{- \frac{1}{2} \Delta_A^{\Lambda,\Lambda_0} - \mathi \nu} \\
&\hspace{8em}\times \Pf_{Q' \to Q} \left[ \left( - 2 \ell^{-2} P \cdot Q' \right)^{- \frac{1}{2} \Delta_A^{\Lambda,\Lambda_0} + \mathi \nu} \left( - 2 \ell^{-2} Q \cdot Q' \right)^{\frac{1}{2} \Delta_A^{\Lambda,\Lambda_0} - \rho} \right] \total Q \\
&\quad= c^{\Lambda,\Lambda_0}_{A B(\rho+\mathi \nu) B(\rho-\mathi \nu)} I_{\frac{1}{2} \Delta_A^{\Lambda,\Lambda_0} + \mathi\nu, \frac{1}{2} \Delta_A^{\Lambda,\Lambda_0} - \mathi\nu, - \frac{1}{2} \Delta_A^{\Lambda,\Lambda_0} + \rho}(P) \eqend{,} \raisetag{1.8em}
\end{splitequation}
with the integral $I_{\mu\nu\lambda}(P)$ defined in equation~\eqref{eq:integral_boundary1_finitepart_def}. We obtain a non-vanishing result only if $\mu+\nu+\lambda = \rho$~\eqref{eq:integral_boundary1_finitepart_result}, which is the case for $\Delta_A^{\Lambda,\Lambda_0} = 0$, that however is excluded by our condition $\Delta_A \geq \rho$ in standard quantisation. That is, a possible non-vanishing result for the one-point function can only occur in alternative quantisation, and then only for fields $\phi_A$ without perturbative corrections to their conformal dimension $\delta \Delta_A^{\Lambda,\Lambda_0} = 0$, thus dual to protected operators. The result then reads
\begin{equation}
\label{eq:1pf_flow_terms1}
\int \Pf_{Q' \to Q} S^{\Lambda,\Lambda_0}_{A B(\rho+\mathi \nu) B(\rho-\mathi \nu)}(P,Q,Q') \total Q = c^{\Lambda,\Lambda_0}_{A B(\rho+\mathi \nu) B(\rho-\mathi \nu)} \delta_{\Delta_A^{\Lambda,\Lambda_0},0} \pi^\rho \ell^d \frac{\Gamma(\rho-\mathi\nu) \Gamma(\rho+\mathi\nu)}{\Gamma(\mathi\nu) \Gamma(-\mathi\nu) \Gamma(d)} \eqend{.}
\end{equation}
In fact, such fields are exactly the ones whose VEV's can survive the transition from the bulk to the boundary: by EAdS invariance, expectation values of scalar fields must be constants, and thus vanish in the boundary limit $z \to 0$ since they are rescaled by a factor $z^\Delta$ coming from the bulk-to-boundary propagator --- except if $\Delta = 0$. In the following, we only work in standard quantisation with $\Delta > \rho$; see for example~\cite{gubsermitra2003,hartmanrastelli2008,giombiyin2012,giombisleighttaronna2018} for computations in alternative quantisation.

For the terms in the second line of the flow equation~\eqref{eq:s_n_flow}, we note that the integrand is invariant under the exchange $\nu \leftrightarrow -\nu$, and we thus obtain twice the same term
\begin{equation}
\label{eq:1pf_flow_terms2}
- 2 \int S^{\Lambda,\Lambda_0}_{A B(\rho+\mathi \nu)}(P,Q) S^{\Lambda,\Lambda_0}_{B(\rho-\mathi \nu)}(Q) \total Q \eqend{.}
\end{equation}
By the above discussion, a self-consistent assumption is that the one-point functions vanish if $\Delta_B \neq 0$, and therefore do not depend analytically on the conformal dimension. Consequently, the analytically continued one-point functions and thus the terms~\eqref{eq:1pf_flow_terms2} vanish. We must then also assume that the boundary conditions~\eqref{eq:s_n_boundary} of the flow vanish for these one-point functions, which can always be achieved by considering fluctuations around the vacuum $\phi_A = 0$, i.e., shifting the fields to have vanishing bare expectation values. In summary, the flow of the one-point functions is only non-vanishing for fields with $\Delta_A = 0$ and dual to protected operators, and is then sourced by the three-point function according to equation~\eqref{eq:1pf_flow_terms1}. Moreover, it does not back-react on the flow of the higher $n$-point functions since the analytically continued one-point functions must be taken to vanish; in other words, the sum over $I$ and $J$ in the last line of the flow equation~\eqref{eq:s_n_flow} only ranges over non-empty subsets. In the following, we only consider the case where $\Delta_A > \rho$ for all fields, and leave a detailed study of the additional complications for fields $\phi_A$ with $\Delta_A = \rho$ or alternative quantisation with $\Delta_A < \rho$ to future work.

Consider then the flow of the two-point function, which is thus given by
\begin{splitequation}
\label{eq:s_2_flow}
&\partial_\Lambda S^{\Lambda,\Lambda_0}_{A_1 A_2}(P_1,P_2) = - \frac{1}{\ell^3 \Lambda^3} \sum_B \int_{-\infty}^\infty \exp\left( - \frac{\nu^2 + (\Delta_B-\rho)^2}{\ell^2 \Lambda^2} \right) \frac{\nu^2}{\pi} \\
&\hspace{5em}\times \int \Pf_{Q' \to Q} \Bigg[ \hbar \, S^{\Lambda,\Lambda_0}_{A_1 A_2 B(\rho+\mathi \nu) B(\rho-\mathi \nu)}(P_1,P_2,Q,Q') \\
&\hspace{6em}- 2 S^{\Lambda,\Lambda_0}_{A_1 B(\rho+\mathi \nu)}(P_1,Q) S^{\Lambda,\Lambda_0}_{A_2 B(\rho-\mathi \nu)}(P_2,Q) \Bigg] \total Q \total \nu \eqend{.}
\end{splitequation}
By conformal symmetry, the two-point functions have the form
\begin{equation}
\label{eq:s_2_form}
S^{\Lambda,\Lambda_0}_{A B}(P,Q) = \delta_{A,B} c^{\Lambda,\Lambda_0}_{AB} ( - 2 \ell^{-2} P \cdot Q )^{-\Delta_A^{\Lambda,\Lambda_0}} \eqend{,}
\end{equation}
where in the free theory~\eqref{eq:boundary_to_boundary_propagator}
\begin{equation}
\label{eq:s_2_free}
c^{\Lambda,\Lambda_0}_{AA} = \frac{\ell^{1-d} \Gamma(\Delta_A)}{2 \pi^\rho \Gamma\left( \Delta_A - \rho + 1 \right)} \eqend{,} \quad \Delta_A^{\Lambda,\Lambda_0} = \Delta_A \eqend{,}
\end{equation}
and since this does not depend analytically on the conformal dimension of each operator separately, we must again assume that the analytically continued two-point functions vanish. It follows that also the flow of the two-point function is only sourced by higher-order boundary correlators, in this case the four-point function. To compute its contribution, we use the Mellin representation~\cite{mack2009a,mack2009b}
\begin{equation}
S^{\Lambda,\Lambda_0}_{A_1 A_2 A_3 A_4}(P_1,P_2,P_3,P_4) = \int M^{\Lambda,\Lambda_0}_{A_1 A_2 A_3 A_4}(\delta_{ij}) \prod_{i<j} \Gamma(\delta_{ij}) \left( - 2 \ell^{-2} P_i \cdot P_j \right)^{- \delta_{ij}} [ \total \delta_{ij} ] \eqend{,}
\end{equation}
where $M^{\Lambda,\Lambda_0}$ is the Mellin amplitude, a meromorphic function of the conformal dimensions $\Delta_i$ and the integration variables $\delta_{ij}$, and the integration measure $[ \total \delta_{ij} ]$ is defined by equation~\eqref{eq:mellin_integration_measure}, with the $\mu_i$ appearing there equal to the full conformal dimensions $\Delta_i^{\Lambda,\Lambda_0}$. The integral that appears in the flow equation~\eqref{eq:s_2_flow} is defined in equation~\eqref{eq:integral_boundary2_finitepart_def}, and we have
\begin{equation}
\partial_\Lambda S^{\Lambda,\Lambda_0}_{A_1 A_2}(P_1,P_2) = - \frac{\hbar}{\ell^3 \Lambda^3} \sum_B \int_{-\infty}^\infty \exp\left( - \frac{\nu^2 + (\Delta_B-\rho)^2}{\ell^2 \Lambda^2} \right) \frac{\nu^2}{\pi} I^{\Lambda,\Lambda_0,\nu}_{A_1 A_2 B}(P_1,P_2) \total \nu \eqend{.}
\end{equation}

The result for the integral $I^{\Lambda,\Lambda_0,\nu}_{A_1 A_2 B}$ is given in equations~\eqref{eq:integral_boundary2_finitepart_result_0} and~\eqref{eq:integral_boundary2_finitepart_result_n}, and is non-vanishing only if the difference in conformal dimensions between the two operators is an even integer: $\Delta_{A_1}^{\Lambda,\Lambda_0} - \Delta_{A_2}^{\Lambda,\Lambda_0} \in 2 \mathbb{Z}$. The simplest case in which this can happen is when the operators are identical, and inserting the result of~\eqref{eq:integral_boundary2_finitepart_result_0} we have in this case
\begin{splitequation}
\label{eq:2pf_flow_terms1}
\partial_\Lambda S^{\Lambda,\Lambda_0}_{A A}(P_1,P_2) &= - \frac{\hbar}{\ell^3 \Lambda^3} \frac{2 \pi^\rho \ell^d \Gamma\left( \Delta_A^{\Lambda,\Lambda_0} - \rho \right)}{\Gamma(\rho)} \left( - 2 \ell^{-2} P_1 \cdot P_2 \right)^{- \Delta_A^{\Lambda,\Lambda_0}} \\
&\quad\times \sum_B \int_{-\infty}^\infty \exp\left( - \frac{\nu^2 + (\Delta_B-\rho)^2}{\ell^2 \Lambda^2} \right) \frac{\nu^2}{\pi} \int \Gamma(\delta_{14}) \Gamma(\rho - \delta_{14}) \\
&\qquad\times \Gamma(\delta_{14} + \mathi \nu) \Gamma(\rho - \delta_{14} - \mathi \nu) \bigg[ \left. \partial_{\delta_{34}} M^{\Lambda,\Lambda_0}_{A A B(\rho+\mathi \nu) B(\rho-\mathi \nu)}(\delta_{ij}) \right\rvert_{\delta_{34} = 0} \\
&\qquad\quad+ \left. M^{\Lambda,\Lambda_0}_{A A B(\rho+\mathi \nu) B(\rho-\mathi \nu)}(\delta_{ij}) \right\rvert_{\delta_{34} = 0} \Big[ \ln \left( - 2 P_1 \cdot P_2 \right) - \gamma \\
&\qquad\qquad+ \psi\left( \Delta_A^{\Lambda,\Lambda_0} - \rho \right) - \psi(\rho - \delta_{14} - \mathi \nu) - \psi(\rho - \delta_{14}) \Big] \bigg] \frac{\total \delta_{14}}{2 \pi \mathi} \total \nu \eqend{,}
\end{splitequation}
where the $\delta_{ij}$ in the Mellin amplitude are given by~\eqref{eq:integral_boundary2_finitepart_deltaij} with $\Delta = 0$ and $\Delta_1 = \Delta_A^{\Lambda,\Lambda_0}$.

We see that the dependence on the coordinates comes out right (the factor $\left( - 2 \ell^{-2} P_1 \cdot P_2 \right)^{- \Delta_A}$ in the first line), but that there are logarithmic corrections (the term $\ln \left( - 2 P_1 \cdot P_2 \right)$ in the second-to-last line). It is well known, and in fact easy to see from the form~\eqref{eq:s_2_form} of the two-point function, that such terms arise from the perturbative corrections $\delta \Delta_A^{\Lambda,\Lambda_0}$ to the conformal dimensions. Taking a $\Lambda$ derivative of equation~\eqref{eq:s_2_form} and comparing with~\eqref{eq:2pf_flow_terms1}, we then obtain the flow of both the normalisation factor and the conformal dimensions. They read
\begin{splitequation}
\label{eq:2pf_flow_c}
\partial_\Lambda c^{\Lambda,\Lambda_0}_{AA} &= - \frac{\hbar}{\ell^3 \Lambda^3} \frac{2 \pi^\rho \ell^d \Gamma\left( \Delta_A^{\Lambda,\Lambda_0} - \rho \right)}{\Gamma(\rho)} \sum_B \int_{-\infty}^\infty \exp\left( - \frac{\nu^2 + (\Delta_B-\rho)^2}{\ell^2 \Lambda^2} \right) \frac{\nu^2}{\pi} \int \Gamma(\rho - \delta_{14}) \\
&\qquad\times \Gamma(\delta_{14}) \Gamma(\delta_{14} + \mathi \nu) \Gamma(\rho - \delta_{14} - \mathi \nu) \bigg[ \left. \partial_{\delta_{34}} M^{\Lambda,\Lambda_0}_{A A B(\rho+\mathi \nu) B(\rho-\mathi \nu)}(\delta_{ij}) \right\rvert_{\delta_{34} = 0} \\
&\qquad\quad+ \left. M^{\Lambda,\Lambda_0}_{A A B(\rho+\mathi \nu) B(\rho-\mathi \nu)}(\delta_{ij}) \right\rvert_{\delta_{34} = 0} \Big[ - \gamma + \psi\left( \Delta_A^{\Lambda,\Lambda_0} - \rho \right) - \psi(\rho - \delta_{14} - \mathi \nu) \\
&\qquad\qquad- \psi(\rho - \delta_{14}) \Big] \bigg] \frac{\total \delta_{14}}{2 \pi \mathi} \total \nu
\end{splitequation}
and
\begin{splitequation}
\label{eq:2pf_flow_delta}
\partial_\Lambda \delta \Delta_A^{\Lambda,\Lambda_0} &= \frac{\hbar}{\ell^3 \Lambda^3} \frac{2 \pi^\rho \ell^d \Gamma\left( \Delta_A^{\Lambda,\Lambda_0} - \rho \right)}{c^{\Lambda,\Lambda_0}_{AA} \Gamma(\rho)} \sum_B \int_{-\infty}^\infty \exp\left( - \frac{\nu^2 + (\Delta_B-\rho)^2}{\ell^2 \Lambda^2} \right) \frac{\nu^2}{\pi} \int \Gamma(\rho - \delta_{14}) \\
&\qquad\times \Gamma(\delta_{14}) \Gamma(\delta_{14} + \mathi \nu) \Gamma(\rho - \delta_{14} - \mathi \nu) \left. M^{\Lambda,\Lambda_0}_{A A B(\rho+\mathi \nu) B(\rho-\mathi \nu)}(\delta_{ij}) \right\rvert_{\delta_{34} = 0} \frac{\total \delta_{14}}{2 \pi \mathi} \total \nu \eqend{,}
\end{splitequation}
and must be supplanted by boundary conditions, which are obtained from the bare interaction according to~\eqref{eq:s_n_boundary}. Since in the derivation of the general flow equation we had subtracted the free two-point function from the generating functional, these are only counterterms and thus at least of order $\bigo{\hbar}$. Their most general form is
\begin{equation}
\label{eq:2pf_flow_boundaryaction}
\frac{1}{2} (-1)^k \int \Phi_A(X) \left( \overleftrightarrow{\nabla}_\mu \overleftrightarrow{\nabla}^\mu \right)^k \Phi_A(X) \total X
\end{equation}
for some $k \in \mathbb{N}_0$ or a sum of such terms, where $\overleftrightarrow{\nabla}_\mu \equiv ( \nabla_\mu - \overleftarrow{\nabla}_\mu )/2$ and which only differs by total derivatives from the simpler expression $\Phi_A(X) ( \nabla^2 )^k \Phi_A(X)$. However, the present form is advantageous since $\overleftrightarrow{\nabla}$ is orthogonal to total derivatives~\cite{fan2011}, in the sense that
\begin{equation}
A \nabla_\mu B = \frac{1}{2} \nabla_\mu (A B) + A \overleftrightarrow{\nabla}_\mu B \eqend{,}
\end{equation}
and we thus do not need to perform any derivatives by parts in the formula~\eqref{eq:s_n_boundary} to obtain the boundary condition. Inserting the action~\eqref{eq:2pf_flow_boundaryaction} into formula~\eqref{eq:s_n_boundary}, we obtain the boundary condition of the flow
\begin{splitequation}
\label{eq:2pf_flow_boundary}
S^{\Lambda_0,\Lambda_0}_{A A}(P_1,P_2) &= (-1)^k \int G_{\partial A}(X,P_1) \left( \overleftrightarrow{\nabla}_\mu \overleftrightarrow{\nabla}^\mu \right)^k G_{\partial A}(X,P_2) \total X \\
&= \left( \frac{\Delta_A \rho}{\ell^2} \right)^k \int G_{\partial A}(X,P_1) G_{\partial A}(X,P_2) \total X \eqend{,}
\end{splitequation}
where we used that
\begin{equation}
- G_{\partial A}(X,P_1) \overleftrightarrow{\nabla}_\mu \overleftrightarrow{\nabla}^\mu G_{\partial A}(X,P_2) = \frac{\Delta_A \rho}{\ell^2} G_{\partial A}(X,P_1) G_{\partial A}(X,P_2) \eqend{,}
\end{equation}
which follows using the formula for the induced covariant derivative~\eqref{eq:nabla_projected} and the explicit expression~\eqref{eq:bulk_to_boundary_propagator} of the bulk-to-boundary propagator. We can thus restrict to the case $k = 0$, where
\begin{splitequation}
S^{\Lambda_0,\Lambda_0}_{A A}(P_1,P_2) &= \int G_{\partial A}(X,P_1) G_{\partial A}(X,P_2) \total X = \frac{\ell^{2-2d}}{4 \pi^d \Gamma^2\left( \Delta_A - \rho + 1 \right)} I_{\Delta_A \Delta_A}(P_1,P_2) \\
&= \frac{\ell^{3-d} \Gamma(\Delta_A) \Gamma(\Delta_A-\rho)}{4 \pi^\rho \Gamma^2\left( \Delta_A - \rho + 1 \right)} \left( - 2 \ell^{-2} P_1 \cdot P_2 \right)^{-\Delta_A} \\
&\qquad\times \left[ \ln \left( - 2 \ell^{-2} P_1 \cdot P_2 \right) + 2 \ln M - \psi(\Delta_A) + \psi(\Delta_A-\rho) \right] \eqend{,}
\end{splitequation}
with the integral $I_{\mu\nu}$ defined in equation~\eqref{eq:integral_bulk2_def} and the result given by equation~\eqref{eq:integral_bulk2_result_0}, and where $M$ is a renormalisation scale that can in principle depend on $\Delta_A$ and needs to be determined. It follows that the boundary conditions $\delta c^{\Lambda_0,\Lambda_0}_{AA}$ and $\delta \Delta^{\Lambda_0,\Lambda_0}_A$ are related:
\begin{equation}
\label{eq:2pf_boundary_relation}
\delta c^{\Lambda_0,\Lambda_0}_{AA} = - c^{\Lambda_0,\Lambda_0}_{AA} \Big[ \psi(\Delta_A-\rho) - \psi(\Delta_A) + 2 \ln M \Big] \delta \Delta_A^{\Lambda_0,\Lambda_0} \eqend{,}
\end{equation}
but since the renormalisation scale $M$ is arbitrary, this is no restriction in practice.

\subsection{The flow of Mellin amplitudes}

For the three- and higher $n$-point functions, conformal symmetry is again manifest by expressing them using the Mellin representation~\cite{mack2009a,mack2009b}
\begin{equation}
\label{eq:s_n_mellin}
S^{\Lambda,\Lambda_0}_{A_1 \cdots A_n}(P_1,\ldots,P_n) = \int M^{\Lambda,\Lambda_0}_{A_1 \cdots A_n}(\delta_{ij}) \prod_{i<j} \Gamma(\delta_{ij}) \left( - 2 \ell^{-2} P_i \cdot P_j \right)^{- \delta_{ij}} [ \total \delta_{ij} ] \eqend{,}
\end{equation}
where $M^{\Lambda,\Lambda_0}$ is the Mellin amplitude, a meromorphic function of the conformal dimensions $\Delta_i$ and the integration variables $\delta_{ij}$, and $[ \total \delta_{ij} ]$ is the Mellin integration measure defined by equation~\eqref{eq:mellin_integration_measure} with the $\mu_k$ appearing there equal to the full conformal dimensions $\Delta_k^{\Lambda,\Lambda_0}$. Note that since the conformal dimensions $\Delta_i = \Delta_{A_i}^{\Lambda,\Lambda_0}$ depend on the cutoffs, also the $\delta^0_{ij}$ and thus the $\delta_{ij}$ which are constrained to satisfy
\begin{equation}
\sum_{i \neq j \in \{1,\ldots,n\}} \delta_{ij} = \Delta_i^{\Lambda,\Lambda_0} \quad (i \in \{1,\ldots,n\})
\end{equation}
depend on $\Lambda$ and $\Lambda_0$, but the integration measure $\left[ \total \delta_{ij} \right]$ itself does not. For simplicity of notation, we do not display this dependence on the cutoffs explicitly.

For the three-point function, there is no integration involved since there exists a unique solution of the required condition $\sum_{l=1}^{k-1} \delta_{lk} + \sum_{l=k+1}^n \delta_{kl} = \Delta_k^{\Lambda,\Lambda_0}$ for $n = 3$, which is
\begin{equation}
\label{eq:mellin3_sol}
\delta_{12} = - \Delta_{312}^{\Lambda,\Lambda_0} \eqend{,} \quad \delta_{13} = - \Delta_{213}^{\Lambda,\Lambda_0} \eqend{,} \quad \delta_{23} = - \Delta_{123}^{\Lambda,\Lambda_0}
\end{equation}
with the $\Delta_{ijk}^{\Lambda,\Lambda_0}$ defined in equation~\eqref{eq:delta_3_def}, and comparing with the previous form of the three-point function~\eqref{eq:s_3_form}, we identify
\begin{equation}
c^{\Lambda,\Lambda_0}_{A_1 A_2 A_3} = M^{\Lambda,\Lambda_0}_{A_1 A_2 A_3}(\delta_{ij}) \Gamma(\delta_{12}) \Gamma(\delta_{13}) \Gamma(\delta_{23}) \eqend{.}
\end{equation}
In the flow~\eqref{eq:s_n_flow} of the three-point function, the five-point function enters as a loop correction, while one- to four-point functions appear already at tree level. However, since the one-point functions vanish and the two-point functions do not depend analytically on the conformal dimension of each operator, such that the analytically continued two-point functions must be taken to vanish, the second type of terms actually vanishes. Therefore, also the flow of the three-point function is only sourced by higher-order boundary correlators, namely the five-point function which has a Mellin representation. For higher $n$-point functions the flow is non-trivial already at tree-level, starting with the four-point function where a product of two three-point functions contributes to the flow~\eqref{eq:s_n_flow}.

Inserting the Mellin representation~\eqref{eq:s_n_mellin} in the flow equation~\eqref{eq:s_n_flow}, on the right-hand side we have two types of terms: the first one where the $(n+2)$-point correlator is integrated over the last two points and the finite part of the result is taken, and the second one where two lower-point correlators are multiplied together. Let us first consider the contribution of the latter ones. Since the one- and two-point functions do not contribute, each of the two lower-point correlators has a Mellin representation
\begin{splitequation}
S^{\Lambda,\Lambda_0}_{\{ A_I \} B(\rho+\mathi \nu)}(\{ P_I \},Q) &= \int M^{\Lambda,\Lambda_0}_{\{ A_I \} B(\rho+\mathi \nu)}(\tilde{\delta}_{ij},l_i) \prod_{i,j \in I, i<j} \Gamma(\tilde{\delta}_{ij}) \left( - 2 \ell^{-2} P_i \cdot P_j \right)^{- \tilde{\delta}_{ij}} \\
&\quad\times \prod_{i \in I} \Gamma(l_i) \left( - 2 \ell^{-2} P_i \cdot Q \right)^{- l_i} [ \total (\tilde{\delta}_{ij}, l_i) ] \eqend{,}
\end{splitequation}
where the constraints on the integration variables are
\begin{equation}
\label{eq:mellin_product_constraints}
l_i + \sum_{i \neq j \in I} \tilde{\delta}_{ij} = \Delta_i^{\Lambda,\Lambda_0} \quad (i \in I) \eqend{,} \qquad \sum_{i \in I} l_i = \rho + \mathi \nu \eqend{.}
\end{equation}
That is, we have split the Mellin variables in order to single out the position $Q$. A similar expression (with $I \to J$ and $\nu \to -\nu$) is obtained for $S^{\Lambda,\Lambda_0}_{\{ A_J \} B(\rho-\mathi \nu)}$. The integral over $Q$ can then be done using Symanzik's formula~\eqref{eq:integral_symanzik}, and we obtain
\begin{splitequation}
&\int \prod_{i \in I} \Gamma(l_i) \left( - 2 \ell^{-2} P_i \cdot Q \right)^{- l_i} \prod_{j \in J} \Gamma(l_j) \left( - 2 \ell^{-2} P_j \cdot Q \right)^{- l_j} \total Q \\
&= \pi^\rho \ell^{2\rho} \int \prod_{1 \leq i < j \leq k} \Gamma(\delta_{ij}) \left( - 2 \ell^{-2} P_i \cdot P_j \right)^{-\delta_{ij}} [ \total \delta_{ij} ]
\end{splitequation}
with the constraints
\begin{equation}
\sum_{i \neq j \in \{1,\ldots,k\}} \delta_{ij} = l_i \eqend{.}
\end{equation}

We thus obtain
\begin{splitequation}
&\int S^{\Lambda,\Lambda_0}_{\{ A_I \} B(\rho+\mathi \nu)}(\{ P_I \},Q) S^{\Lambda,\Lambda_0}_{\{ A_J \} B(\rho-\mathi \nu)}(\{ P_J \},Q) \total Q \\
&= \pi^\rho \ell^{2\rho} \int \mathcal{F}^{\Lambda,\Lambda_0}_{\{ A_I \} B(\rho + \mathi \nu)}(\delta_{ij}) \mathcal{F}^{\Lambda,\Lambda_0}_{\{ A_J \} B(\rho - \mathi \nu)}(\delta_{ij}) \prod_{1 \leq i < j \leq k} \Gamma(\delta_{ij}) \left( - 2 \ell^{-2} P_i \cdot P_j \right)^{-\delta_{ij}} [ \total \delta_{ij} ]
\end{splitequation}
where we shifted $\delta_{ij} \to \delta_{ij} - \tilde{\delta}_{ij}$, which fulfill the correct constraints
\begin{equation}
\sum_{i \neq j \in \{1,\ldots,k\}} \delta_{ij} = \Delta_i^{\Lambda,\Lambda_0} \eqend{,}
\end{equation}
and where the factors are given by
\begin{equation}
\label{eq:mellin_factor}
\mathcal{F}^{\Lambda,\Lambda_0}_{\{ A_I \} B(\rho + \mathi \nu)}(\delta_{ij}) = \int M^{\Lambda,\Lambda_0}_{\{ A_I \} B(\rho+\mathi \nu)}(\tilde{\delta}_{ij},l_i) \prod_{i,j \in I, i<j} \frac{\Gamma(\tilde{\delta}_{ij}) \Gamma(\delta_{ij} - \tilde{\delta}_{ij})}{\Gamma(\delta_{ij})} [ \total (\tilde{\delta}_{ij}, l_i) ] \eqend{.}
\end{equation}
To obtain this expression, it was important that for each factor only the corresponding $\delta_{ij}$ with both indices belonging to the same factor needed to be shifted. Using the constraints~\eqref{eq:mellin_product_constraints} on the $l_i$, we can solve for them explicitly
\begin{equation}
l_i = \Delta_i^{\Lambda,\Lambda_0} - \sum_{i \neq j \in I} \tilde{\delta}_{ij} \quad (i \in I) \eqend{,}
\end{equation}
and only the single constraint
\begin{equation}
\label{eq:mellin_factor_constraint}
\sum_{i,j \in I, i \neq j} \tilde{\delta}_{ij} = \sum_{i \in I} \Delta_i^{\Lambda,\Lambda_0} - (\rho + \mathi \nu)
\end{equation}
remains. This construction is analogous to the one of~\cite{fitzpatricketal2011}, who considered recursion relations for Mellin amplitudes at tree level.

For the first term, we have to take the finite part of the $(n+2)$-point correlator as the last two points scale together and are integrated over, generalising the results~\eqref{eq:integral_boundary2_finitepart_result_0} and~\eqref{eq:integral_boundary2_finitepart_result_n} for the corresponding integral of the four-point function~\eqref{eq:integral_boundary2_finitepart_def} that entered the flow of the two-point function~\eqref{eq:s_2_flow}. Inserting the Mellin representation~\eqref{eq:s_n_mellin}, that means we need to compute
\begin{splitequation}
&\int \Pf_{Q' \to Q} S^{\Lambda,\Lambda_0}_{A_1 \cdots A_n B(\rho+\mathi \nu) B(\rho-\mathi \nu)}(P_1,\ldots,P_n,Q,Q') \total Q \\
&\quad= \int \Pf_{Q' \to Q} \int M^{\Lambda,\Lambda_0}_{A_1 \cdots A_n (\rho+\mathi \nu) (\rho-\mathi \nu)}(\tilde{\delta}_{ij},k_i,l_i,m) \\
&\qquad\times \prod_{1 \leq i < j \leq n} \Gamma(\tilde{\delta}_{ij}) \left( - 2 \ell^{-2} P_i \cdot P_j \right)^{- \tilde{\delta}_{ij}} \prod_{i=1}^n \Gamma(k_i) \left( - 2 \ell^{-2} P_i \cdot Q \right)^{- k_i} \\
&\qquad\times \prod_{i=1}^n \Gamma(l_i) \left( - 2 \ell^{-2} P_i \cdot Q' \right)^{- l_i} \Gamma(m) \left( - 2 \ell^{-2} Q \cdot Q' \right)^{- m} [ \total (\tilde{\delta}_{ij},k_i,l_i,m) ] \total Q \eqend{,}
\end{splitequation}
where we separated the last two points $Q$ and $Q'$, and the integration measure is consequently constrained by
\begin{equations}
&k_i + l_i + \sum_{i \neq j \in \{1,\ldots,n\}} \tilde{\delta}_{ij} = \Delta_i^{\Lambda,\Lambda_0} \quad (1 \leq i \leq n) \eqend{,} \\
&m + \sum_{i=1}^n k_i = \rho+\mathi \nu \eqend{,} \\
&m + \sum_{i=1}^n l_i = \rho-\mathi \nu \eqend{.}
\end{equations}
To compute the integral, we can directly take the residue of the integrand at $m = 0$ to obtain the finite part, and then perform the integral over $Q$ using the Symanzik formula. This gives
\begin{splitequation}
&\int \Pf_{Q' \to Q} S^{\Lambda,\Lambda_0}_{A_1 \cdots A_n B(\rho+\mathi \nu) B(\rho-\mathi \nu)}(P_1,\ldots,P_n,Q,Q') \total Q \\
&\quad= \iint M^{\Lambda,\Lambda_0}_{A_1 \cdots A_n B(\rho+\mathi \nu) B(\rho-\mathi \nu)}(\tilde{\delta}_{ij},k_i,l_i,0) \prod_{1 \leq i < j \leq n} \Gamma(\tilde{\delta}_{ij}) \left( - 2 \ell^{-2} P_i \cdot P_j \right)^{- \tilde{\delta}_{ij}} \\
&\qquad\times \prod_{i=1}^n \Gamma(k_i) \Gamma(l_i) \left( - 2 \ell^{-2} P_i \cdot Q \right)^{- k_i - l_i} [ \total (\tilde{\delta}_{ij},k_i,l_i) ] \total Q \\
&\quad= \pi^\rho \ell^{2 \rho} \iint M^{\Lambda,\Lambda_0}_{A_1 \cdots A_k (\rho+\mathi \nu) (\rho-\mathi \nu)}(\tilde{\delta}_{ij},k_i,l_i,0) \prod_{i=1}^n \frac{\Gamma(k_i) \Gamma(l_i)}{\Gamma(k_i+l_i)} \\
&\qquad\times \prod_{1 \leq i < j \leq n} \Gamma(\tilde{\delta}_{ij}) \Gamma( \delta_{ij} ) \left( - 2 \ell^{-2} P_i \cdot P_j \right)^{- \tilde{\delta}_{ij} - \delta_{ij}} [ \total \delta_{ij} ] [ \total (\tilde{\delta}_{ij},k_i,l_i) ] \eqend{,}
\end{splitequation}
with the integration measure $[ \total \delta_{ij} ]$ constrained by
\begin{equation}
\sum_{i \neq j \in \{1,\ldots,n\}} \delta_{ij} = k_i+l_i \eqend{,}
\end{equation}
and where we assumed that the Mellin amplitude is regular at $m = 0$. Shifting $\delta_{ij} \to \delta_{ij} + \tilde{\delta}_{ij}$, we recover the Mellin representation for a $n$-point function
\begin{splitequation}
&\int \Pf_{Q' \to Q} S^{\Lambda,\Lambda_0}_{A_1 \cdots A_n B(\rho+\mathi \nu) B(\rho-\mathi \nu)}(P_1,\ldots,P_n,Q,Q') \total Q \\
&\quad= \pi^\rho \ell^{2 \rho} \int \mathcal{G}^{\Lambda,\Lambda_0,\nu}_{A_1 \cdots A_k; B}(\delta_{ij}) \prod_{1 \leq i < j \leq n} \Gamma(\delta_{ij}) \left( - 2 \ell^{-2} P_i \cdot P_j \right)^{- \delta_{ij}} [ \total \delta_{ij} ]
\end{splitequation}
with the standard constraints
\begin{equation}
\label{eq:mellin_constraints}
\sum_{i \neq j \in \{1,\ldots,n\}} \delta_{ij} = \Delta_i^{\Lambda,\Lambda_0} \quad (1 \leq i \leq n) \eqend{,}
\end{equation}
and the new amplitude
\begin{splitequation}
\label{eq:mellin_factor2}
\mathcal{G}^{\Lambda,\Lambda_0,\nu}_{A_1 \cdots A_k; B}(\delta_{ij}) &\equiv \int M^{\Lambda,\Lambda_0}_{A_1 \cdots A_k B(\rho+\mathi \nu) B(\rho-\mathi \nu)}(\tilde{\delta}_{ij},k_i,l_i,0) \prod_{i=1}^n \frac{\Gamma(k_i) \Gamma(l_i)}{\Gamma(k_i+l_i)} \\
&\qquad\times \prod_{1 \leq i < j \leq n} \frac{\Gamma(\tilde{\delta}_{ij}) \Gamma(\delta_{ij} - \tilde{\delta}_{ij})}{\Gamma(\delta_{ij})} [ \total (\tilde{\delta}_{ij},k_i) ]
\end{splitequation}
with
\begin{equation}
l_i = \Delta_i^{\Lambda,\Lambda_0} - k_i - \sum_{i \neq j \in \{1,\ldots,n\}} \tilde{\delta}_{ij} \quad (1 \leq i \leq n) \eqend{,}
\end{equation}
whose integration measure is constrained by
\begin{equations}[eq:mellin_factor2_constraint]
&\sum_{i=1}^n \sum_{i \neq j \in \{1,\ldots,n\}} \tilde{\delta}_{ij} = \sum_{i=1}^n \Delta_i^{\Lambda,\Lambda_0} - 2 \rho \eqend{,} \\
&\sum_{i=1}^n k_i = \rho+\mathi \nu \eqend{.}
\end{equations}
This construction is now analogous to the one of~\cite{yuan2017,yuan2018}, who considered the formation of loops in AdS by sewing propagators.

\begin{figure}[h]
  \begin{subfigure}{0.45\textwidth}
    \centering
    \includegraphics[scale=0.5]{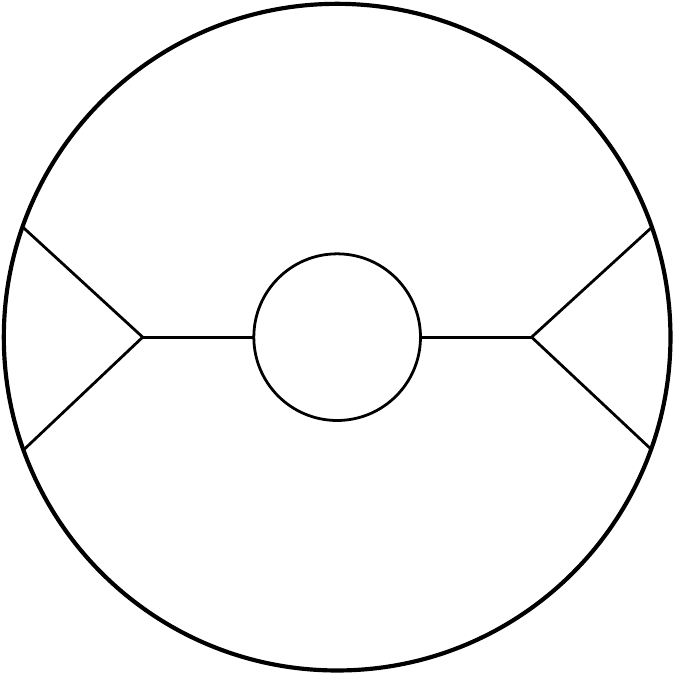}
    \caption{Witten diagram involving a loop correction to the propagator.}
    \label{fig:witten_loop_propagator}
  \end{subfigure}
  \hfill
  \begin{subfigure}{0.45\textwidth}
    \centering
    \includegraphics[scale=0.5]{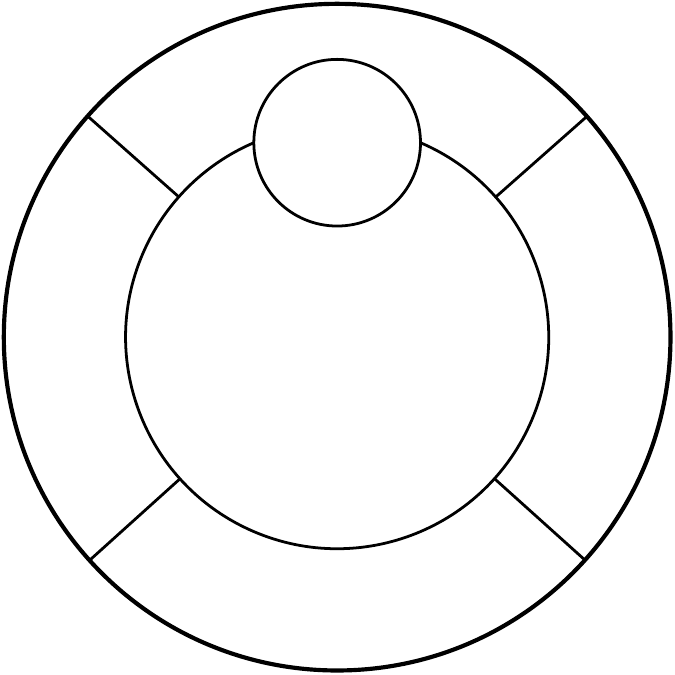}
    \caption{1PI Witten diagram involving a loop correction to the propagator.}
    \label{fig:witten_1pi_propagator}
  \end{subfigure}
  \caption{Witten diagrams involving one-loop corrections to the propagator.}
\end{figure}

It follows that the flow equation~\eqref{eq:s_n_flow} can be written as
\begin{splitequation}
\label{eq:s_n_mellin_flow}
&\partial_\Lambda \int M^{\Lambda,\Lambda_0}_{A_1 \cdots A_n}(\delta_{ij}) \prod_{1 \leq i < j \leq n} \Gamma(\delta_{ij}) \left( - 2 \ell^{-2} P_i \cdot P_j \right)^{- \delta_{ij}} [ \total \delta_{ij} ] \\
&= - \frac{1}{\ell^3 \Lambda^3} \pi^\rho \ell^{2 \rho} \int \sum_B \int_{-\infty}^\infty \exp\left( - \frac{\nu^2 + (\Delta_B-\rho)^2}{\ell^2 \Lambda^2} \right) \frac{\nu^2}{\pi} \\
&\qquad\times \Bigg[ \hbar \mathcal{G}^{\Lambda,\Lambda_0,\nu}_{A_1 \cdots A_n; B}(\delta_{ij}) - \sum_{I \cup J = \{1,\ldots,n\}} \mathcal{F}^{\Lambda,\Lambda_0}_{\{ A_I \} B(\rho + \mathi \nu)}(\delta_{ij}) \mathcal{F}^{\Lambda,\Lambda_0}_{\{ A_J \} B(\rho - \mathi \nu)}(\delta_{ij}) \Bigg] \total \nu \\
&\qquad\times \prod_{1 \leq i < j \leq n} \Gamma(\delta_{ij}) \left( - 2 \ell^{-2} P_i \cdot P_j \right)^{-\delta_{ij}} [ \total \delta_{ij} ] \eqend{,}
\end{splitequation}
where the factors $\mathcal{F}$ are defined by~\eqref{eq:mellin_factor} with the constraint~\eqref{eq:mellin_factor_constraint}, and the factor $\mathcal{G}$ is given by~\eqref{eq:mellin_factor2} with the constraints~\eqref{eq:mellin_factor2_constraint}. On the left-hand side, the $\Lambda$ derivative can act either on the Mellin amplitude $M^{\Lambda,\Lambda_0}$ or on the parameters $\delta_{ij}$, which depend on $\Lambda$ through the constraints which involve the full conformal dimensions $\Delta_i^{\Lambda,\Lambda_0} = \Delta_i + \delta \Delta_i^{\Lambda,\Lambda_0}$. The flow of the conformal dimensions~\eqref{eq:2pf_flow_delta} comes from the flow of the two-point functions~\eqref{eq:2pf_flow_terms1} and is proportional to $\hbar$, that is, it arises from loop corrections to the propagator. It is thus clear that these corrections where the $\Lambda$ derivative acts on the parameters $\delta_{ij}$ corresponds to Witten diagrams where loop corrections to the propagator appear, such as the ones depicted in figure~\ref{fig:witten_loop_propagator}.

The well-known strategy to resum these corrections is to pass to the 1-particle irreducible (1PI) diagrams. However, in our case this is not the full answer, since we also need to exclude diagrams like the one depicted in figure~\ref{fig:witten_1pi_propagator}, which are still 1PI but also include propagator corrections. Instead, we have to consider only those diagrams where all propagators are replaced by exact propagators, i.e., including all loop corrections to the propagator itself. Denoting the Mellin amplitude for these diagrams by $\mathcal{M}$, the flow equation~\eqref{eq:s_n_mellin_flow} results in
\begin{splitequation}
\label{eq:s_n_mellin_flow_exact}
\partial_\Lambda \mathcal{M}^{\Lambda,\Lambda_0}_{A_1 \cdots A_n}(\delta_{ij}) &= - \frac{1}{\ell^3 \Lambda^3} \pi^\rho \ell^{2 \rho} \sum_B \int_{-\infty}^\infty \exp\left( - \frac{\nu^2 + (\Delta_B-\rho)^2}{\ell^2 \Lambda^2} \right) \frac{\nu^2}{\pi} \\
&\times \Bigg[ \hbar \mathcal{G}^{\Lambda,\Lambda_0,\nu}_{A_1 \cdots A_n; B}(\delta_{ij}) - \sum_{I \cup J = \{1,\ldots,n\}} \mathcal{F}^{\Lambda,\Lambda_0}_{\{ A_I \} B(\rho + \mathi \nu)}(\delta_{ij}) \mathcal{F}^{\Lambda,\Lambda_0}_{\{ A_J \} B(\rho - \mathi \nu)}(\delta_{ij}) \Bigg] \total \nu \eqend{,}
\end{splitequation}
to which one adds any $n$-point contact interaction amplitude as boundary condition for $\Lambda = \Lambda_0$. In this equation, the factors $\mathcal{F}$~\eqref{eq:mellin_factor} and $\mathcal{G}$~\eqref{eq:mellin_factor2} are the same as before, with the only difference that $M$ is replaced by $\mathcal{M}$ and the conformal dimensions $\Delta_i$ are now given by the physical limit $\lim_{\Lambda \to 0, \Lambda_0 \to \infty} \Delta_i^{\Lambda,\Lambda_0}$ of the full conformal dimensions, and thus do not depend on $\Lambda$ anymore. In turn, their flow is determined by~\eqref{eq:2pf_flow_delta}, where we conjecture that one can also replace $M$ by $\mathcal{M}$ if simultaneously $\Delta_A^{\Lambda,\Lambda_0}$ is replaced by $\lim_{\Lambda \to 0, \Lambda_0 \to \infty} \Delta_A^{\Lambda,\Lambda_0} = \Delta_A^{0,\infty}$ on the right-hand side. This conjecture is true at least to one-loop order, as we will see in section~\ref{sec:flow_tree_oneloop}.

Lastly, we determine the contact interaction amplitudes. For simplicity, we assume an interaction term
\begin{equation}
\label{eq:contact_interaction}
\frac{\lambda_{A_1 \cdots A_n}}{m_{A_1 \cdots A_n}} \int \phi_{A_1}(X) \cdots \phi_{A_n}(X) \total X
\end{equation}
in AdS, where the multiplicity factor $m_{A_1 \cdots A_n}$ depends on the type of operators: for example, if all $\phi_{A_i}$ are equal, it ensures the standard normalisation $m_{A_1 \cdots A_n} = n!$. The corresponding contribution to the boundary correlator $S^{\Lambda_0,\Lambda_0}$~\eqref{eq:witten_s} is given by the AdS integral
\begin{splitequation}
\label{eq:contact_boundary_correlator}
&S^{\Lambda_0,\Lambda_0}_{A_1 \cdots A_n}(P_1,\ldots,P_n) = \lambda_{A_1 \cdots A_n} \int G_{\partial A_1}(X,P_1) \cdots G_{\partial A_n}(X,P_n) \total X \\
&\quad= \lambda_{A_1 \cdots A_n} \prod_{i=1}^n \frac{\ell^{1-d} \Gamma(\Delta_i)}{2 \pi^\rho \Gamma\left( 1 + \Delta_i - \rho \right)} \int \left( - 2 \ell^{-2} P_1 \cdot X \right)^{-\Delta_1} \cdots \left( - 2 \ell^{-2} P_n \cdot X \right)^{-\Delta_n} \total X \eqend{.}
\end{splitequation}
The integral is easily done using the Symanzik formula~\eqref{eq:integral_symanzik}, and we obtain the contact Mellin amplitude~\cite{penedones2011}
\begin{splitequation}
\label{eq:contact_amplitude}
\mathcal{M}^{\Lambda_0,\Lambda_0}_{A_1 \cdots A_n}(\delta_{ij}) &= \lambda_{A_1 \cdots A_n} \ell^{2\rho+1} \frac{\pi^\rho}{2} \Gamma\left( - \rho + \frac{1}{2} \sum_{i=1}^n \Delta_i \right) \prod_{i=1}^n \frac{\ell^{1-d}}{2 \pi^\rho \Gamma\left( 1 + \Delta_i - \rho \right)} \\
&= \tilde{\lambda}_{A_1 \cdots A_n} \frac{\Gamma\left( - \rho + \frac{1}{2} \sum_{i=1}^n \Delta_i \right)}{\prod_{i=1}^n \Gamma\left( 1 + \Delta_i - \rho \right)} \eqend{.}
\end{splitequation}
As is well known by now, the contact amplitude is independent of the $\delta_{ij}$, while if one considers interactions with derivatives, these give rise to a polynomial in the $\delta_{ij}$.

\section{Multi-trace operators}
\label{sec:multitrace}

Composite operators $\op_\mathcal{A}$ in the EAdS bulk are obtained as normal-ordered products of basic fields $\phi_A$ and their derivatives, and are thus dual to multi-trace operators in the CFT~\cite{chalmersschalm2000,fan2011}. We again use a condensed notation, where the calligraphic index $\mathcal{A}$ denotes all the fields that make up the composite operator, their spinorial or Lorentz indices, and the derivative indices. In the free theory, the conformal dimension $\Delta_\mathcal{A}$ of $\op_\mathcal{A}$ is obtained simply as the sum of the conformal dimensions of the (operators dual to the) constituent basic fields, and for weak coupling there will be perturbative corrections. We will see that in addition to the straightforward generalisation of the flow equations for single-trace operators, the flow equations of multi-trace operators involve additional source terms containing single-trace operators and other multi-trace operators, such that the flow has to be solved inductively. We again split the derivation in parts: first we derive flow equations for generic quantum field theories, and then specialise to EAdS and correlation functions on the boundary.

\subsection{Flow equations for generic quantum field theories}

To derive a flow equation for composite operators, we include corresponding source terms $K^\mathcal{A}$ in the generating functional~\eqref{eq:generating_z1}, which then reads
\begin{equation}
\label{eq:generating_comp_z1}
Z[J,K] \equiv N \int \exp\left( - \frac{1}{\hbar} S[\phi] + \int J^A(X) \phi_A(X) \total X + \int K^\mathcal{A}(X) \mathcal{N}\left[ \op_\mathcal{A}[\phi] \right](X) \total X \right) \mathcal{D} \phi \eqend{.}
\end{equation}
Here, the expression $\mathcal{N}\left[ \op_\mathcal{A}[\phi](X) \right] = \op_\mathcal{A} + \delta \op_\mathcal{A}$ not only includes the classical expression $\op_\mathcal{A}$ for the composite operator, but also the required counterterms $\delta \op_\mathcal{A}$. These arise both at tree level, namely from the normal-ordering of the operator in the free theory, and from loop corrections; $\mathcal{N}$ thus denotes a normal-ordering in the full interacting theory. To derive the flow equations, we now proceed exactly as the previous section. That is, we split the action into a free part and the interaction, perform the shift $\phi_A(X) \to \phi_A(X) + \hbar \int G_{AB}(X,Y) J^B(Y) \total Y$ with the propagator $G_{AB}$, and take the logarithm to obtain the generating functional of connected amputated correlation functions $W[\Phi,K]$, which now also depends on the sources $K^\mathcal{A}$:
\begin{splitequation}
\label{eq:generating_comp_w}
&W[\Phi,K] \equiv - \hbar \ln Z[J,K] \\
&\ = - S_0[\Phi] - \hbar \ln \int \exp\left( - \frac{1}{\hbar} S_0[\phi] - \frac{1}{\hbar} S_\text{int}[\phi + \Phi] + \int K^\mathcal{A}(X) \mathcal{N}\left[ \op_\mathcal{A}[\phi+\Phi] \right](X) \total X \right) \mathcal{D} \phi \eqend{.}
\end{splitequation}

Regularising the propagator with an IR cutoff $\Lambda$ and a UV cutoff $\Lambda_0$, introducing the normalised Gaussian measure~\eqref{eq:gaussian_measure} and subtracting the trivial free part from $W$, we obtain the regularised generating functional of connected amputated correlation functions (where only interactions contribute)
\begin{equation}
\label{eq:generating_comp_l}
L^{\Lambda,\Lambda_0}[\Phi,K] = - \hbar \ln \int \exp\left( - \frac{1}{\hbar} S_\text{int}[\phi + \Phi] + \int K^\mathcal{A}(X) \mathcal{N}\left[ \op_\mathcal{A}[\phi+\Phi] \right](X) \total X \right) \total \mu^{\Lambda, \Lambda_0}(\phi) \eqend{.}
\end{equation}
The flow equation for $L^{\Lambda,\Lambda_0}$ is derived exactly as in the previous section, since also the source terms for the composite operators $\op_\mathcal{A}$ only depend on the sum $\phi+\Phi$, such that functional derivatives with respect to $\phi$ can be exchanged with the ones with respect to $\Phi$. We thus obtain the generalisation of the flow equation~\eqref{eq:l_flow}:
\begin{splitequation}
\label{eq:l_comp_flow}
\partial_\Lambda L^{\Lambda,\Lambda_0}[\Phi,K] &= \frac{\hbar}{2} \iint \partial_\Lambda G^{\Lambda,\Lambda_0}_{AB}(X,Y) \frac{\delta^2 L^{\Lambda,\Lambda_0}[\Phi,K]}{\delta \Phi_A(X) \delta \Phi_B(Y)} \total X \total Y \\
&\quad- \frac{1}{2} \iint \frac{\delta L^{\Lambda,\Lambda_0}[\Phi,K]}{\delta \Phi_A(X)} \partial_\Lambda G^{\Lambda,\Lambda_0}_{AB}(X,Y) \frac{\delta L^{\Lambda,\Lambda_0}[\Phi,K]}{\delta \Phi_B(Y)} \total X \total Y \eqend{.}
\end{splitequation}
From this equation, we derive the flow equation for connected amputated correlation functions with insertions of the composite operators $\op_\mathcal{A}$ by taking functional derivatives with respect to $K^\mathcal{A}$. Let us denote these correlation functions by
\begin{equation}
L^{\Lambda,\Lambda_0}\left( \op_{\mathcal{A}_1}(Z_1) \cdots \op_{\mathcal{A}_k}(Z_k) \right)[\Phi] \equiv - \frac{1}{\hbar} \left. \frac{\delta^k L^{\Lambda,\Lambda_0}[\Phi,K]}{\delta K^{\mathcal{A}_1}(Z_1) \cdots \delta K^{\mathcal{A}_k}(Z_k)} \right\rvert_{K^\mathcal{A} = 0} \eqend{,}
\end{equation}
where we have compensated for the extra factor $-\hbar$ in the definition of $W$~\eqref{eq:generating_comp_w}.\footnote{For the single-trace operators, this compensation was done in the definition of the amputated source~\eqref{eq:amputated_source_def}.}

Using that derivatives with respect to the $\Phi_A$ commute with the ones with respect to the $K^\mathcal{A}$, for a single insertion we obtain
\begin{splitequation}
\label{eq:l_comp1_flow}
\partial_\Lambda L^{\Lambda,\Lambda_0}\left( \op_\mathcal{A}(Z) \right)[\Phi] &= \frac{\hbar}{2} \iint \partial_\Lambda G^{\Lambda,\Lambda_0}_{AB}(X,Y) \frac{\delta^2 L^{\Lambda,\Lambda_0}\left( \op_\mathcal{A}(Z) \right)[\Phi]}{\delta \Phi_A(X) \delta \Phi_B(Y)} \total X \total Y \\
&\quad- \iint \frac{\delta L^{\Lambda,\Lambda_0}[\Phi]}{\delta \Phi_A(X)} \partial_\Lambda G^{\Lambda,\Lambda_0}_{AB}(X,Y) \frac{\delta L^{\Lambda,\Lambda_0}\left( \op_\mathcal{A}(Z) \right)[\Phi]}{\delta \Phi_B(Y)} \total X \total Y \eqend{,}
\end{splitequation}
and we see that the source term of the flow equation (the second line) contains the regularised generating functional for correlation functions without a composite operator insertion. Taking more than one functional derivative, it follows that
\begin{splitequation}
\label{eq:l_compk_flow}
&\partial_\Lambda L^{\Lambda,\Lambda_0}\left( \op_{\mathcal{A}_1}(Z_1) \cdots \op_{\mathcal{A}_k}(Z_k) \right)[\Phi] \\
&= \frac{\hbar}{2} \iint \partial_\Lambda G^{\Lambda,\Lambda_0}_{AB}(X,Y) \frac{\delta^2 L^{\Lambda,\Lambda_0}\left( \op_{\mathcal{A}_1}(Z_1) \cdots \op_{\mathcal{A}_k}(Z_k) \right)[\Phi]}{\delta \Phi_A(X) \delta \Phi_B(Y)} \total X \total Y \\
&- \iint \frac{\delta L^{\Lambda,\Lambda_0}[\Phi]}{\delta \Phi_A(X)} \partial_\Lambda G^{\Lambda,\Lambda_0}_{AB}(X,Y) \frac{\delta L^{\Lambda,\Lambda_0}\left( \op_{\mathcal{A}_1}(Z_1) \cdots \op_{\mathcal{A}_k}(Z_k) \right)[\Phi]}{\delta \Phi_B(Y)} \total X \total Y \\
&+ \frac{\hbar}{2} \sum_{I \cup J = \{1,\ldots,k\}} \iint \frac{\delta L^{\Lambda,\Lambda_0}\left( \{ \op_{\mathcal{A}_I}(Z_I) \} \right)[\Phi]}{\delta \Phi_A(X)} \partial_\Lambda G^{\Lambda,\Lambda_0}_{AB}(X,Y) \frac{\delta L^{\Lambda,\Lambda_0}\left( \{ \op_{\mathcal{A}_J}(Z_J) \} \right)[\Phi]}{\delta \Phi_B(Y)} \total X \total Y \eqend{,}
\end{splitequation}
where the sum in the last line runs over all disjoint non-empty subsets $I$ and $J$, and now the source term also contains the regularised generating functional for correlation functions with a smaller number of composite operator insertions.

To obtain the flow equations for the correlation functions (with operator insertions) themselves, we again expand in powers of $\Phi$:
\begin{splitequation}
\label{eq:l_n_comp1_flow}
&\partial_\Lambda L^{\Lambda,\Lambda_0}_{A_1 \cdots A_n}\left( \op_\mathcal{A}(Z) ; X_1,\ldots,X_n \right) \\
&\quad= \frac{\hbar}{2} \iint \partial_\Lambda G^{\Lambda,\Lambda_0}_{BC}(Y,Y') L^{\Lambda,\Lambda_0}_{A_1 \cdots A_n BC}\left( \op_\mathcal{A}(Z) ; X_1,\ldots,X_n,Y,Y' \right) \total Y \total Y' \\
&\qquad- \sum_{I \cup J = \{1,\ldots,n\}} \iint L^{\Lambda,\Lambda_0}_{\{ A_I \} B}(\{ X_I \},Y) \partial_\Lambda G^{\Lambda,\Lambda_0}_{BC}(Y,Y') L^{\Lambda,\Lambda_0}_{\{ A_J \} C}\left( \op_\mathcal{A}(Z) ; \{ X_J \},Y' \right) \total Y \total Y' \eqend{,}
\end{splitequation}
where the sum in the second line runs over all disjoint subsets $I$ and $J$, and for multiple insertions
\begin{splitequation}
\label{eq:l_n_compk_flow}
&\partial_\Lambda L^{\Lambda,\Lambda_0}_{A_1 \cdots A_n}\left( \op_{\mathcal{A}_1}(Z_1) \cdots \op_{\mathcal{A}_k}(Z_k) ; X_1,\ldots,X_n \right) \\
&\quad= \frac{\hbar}{2} \iint \partial_\Lambda G^{\Lambda,\Lambda_0}_{BC}(Y,Y') L^{\Lambda,\Lambda_0}_{A_1 \cdots A_n BC}\left( \op_{\mathcal{A}_1}(Z_1) \cdots \op_{\mathcal{A}_k}(Z_k); X_1,\ldots,X_n,Y,Y' \right) \total Y \total Y' \\
&\qquad- \sum_{\smash{\substack{\mathstrut \\ I \cup J = \{1,\ldots,n\}}}} \iint L^{\Lambda,\Lambda_0}_{\{ A_I \} B}(\{ X_I \},Y) \partial_\Lambda G^{\Lambda,\Lambda_0}_{BC}(Y,Y') \\
&\hspace{10em}\times L^{\Lambda,\Lambda_0}_{\{ A_J \} C}\left( \op_{\mathcal{A}_1}(Z_1) \cdots \op_{\mathcal{A}_k}(Z_k); \{ X_J \},Y' \right) \total Y \total Y' \\
&\qquad+ \frac{\hbar}{2} \sum_{\smash{\substack{\strut \\ I \cup J = \{1,\ldots,n\} \\ K \cup L = \{1,\ldots,k\}}}} \iint L^{\Lambda,\Lambda_0}_{\{ A_I \} B}\left( \{ \op_{\mathcal{A}_K}(Z_K) \} ; \{ X_I \},Y \right) \partial_\Lambda G^{\Lambda,\Lambda_0}_{BC}(Y,Y') \\
&\hspace{10em}\times L^{\Lambda,\Lambda_0}_{\{ A_J \} C}\left( \{ \op_{\mathcal{A}_L}(Z_L) \} ; \{ X_J \},Y' \right) \total Y \total Y' \eqend{,}
\end{splitequation}
where the sums in the second and third line run over all disjoint subsets $I$ and $J$, and all non-empty disjoint subsets $K$ and $L$. The flow equations are depicted in figure~\ref{fig:flow}: the first term on the right-hand side is a loop correction that connects two external legs of a higher $n$-point correlation function with a composite operator insertion with a propagator, the second term is a source term connecting a lower $n$-point function without composite operator insertions with another one with composite operator insertions with a propagator, and the last term is a source term connecting two lower-order $n$-point functions with a smaller number of composite operator insertions with a propagator.

Finally, let us assume that the regularisation is such that the covariance vanishes as $\Lambda \to \Lambda_0$. The Gaussian measure $\total \mu^{\Lambda,\Lambda_0}$ then reduces to a functional $\delta$ measure in this limit, and from equation~\eqref{eq:generating_comp_l} we obtain the boundary conditions
\begin{equations}
\label{eq:l_comp_boundary}
L^{\Lambda_0,\Lambda_0}\left( \op_\mathcal{A}(Z) \right)[\Phi] &= \mathcal{N}\left[ \op_\mathcal{A}[\Phi] \right](Z) \eqend{,} \\
L^{\Lambda_0,\Lambda_0}\left( \op_{\mathcal{A}_1}(Z_1) \cdots \op_{\mathcal{A}_k}(Z_k) \right)[\Phi] &= 0 \eqend{,}
\end{equations}
or expanding in powers of $\Phi$
\begin{equation}
\label{eq:l_n_comp_boundary}
L^{\Lambda_0,\Lambda_0}_{A_1 \cdots A_n}\left( \op_\mathcal{A}(Z) ; X_1,\ldots,X_n \right) = \left. \frac{\delta^n \mathcal{N}\left[ \op_\mathcal{A}[\Phi] \right](Z)}{\delta \Phi_{A_1}(X_1) \cdots \delta \Phi_{A_n}(X_n)} \right\rvert_{\Phi = 0} \eqend{.}
\end{equation}
The vanishing boundary condition for correlation functions with multiple insertions of composite operators can be changed by including terms in the bare action that are multilinear in the sources $K^\mathcal{A}$. Such terms can then be used to define oversubtracted correlation functions in the sense of Zimmermann, which have improved singular behaviour when two insertions converge to a single point~\cite{wilsonzimmermann1972,zimmermann1973}.

\subsection{Flow equations in EAdS and on the boundary}

To derive the flow of boundary correlators from the bulk flow equations~\eqref{eq:l_n_comp1_flow} and~\eqref{eq:l_n_compk_flow} with boundary conditions~\eqref{eq:l_n_comp_boundary}, we again have to convolve the amputated correlation functions with bulk-to-boundary propagators $G_{A \partial B}(X,P)$. On the other hand, since the composite operator insertions have not been amputated, we simply rescale the corresponding points to the boundary to obtain multi-trace operator insertions there. As for the single-trace operators, to simplify the discussion we assume that the free action~\eqref{eq:free_action} and thus the propagator $G_{AB}(X,Y)$ are diagonal in field space, which can always be achieved by adding auxiliary fields, and restrict to scalar fields.

In analogy to equation~\eqref{eq:witten_s} for single-trace operators, we thus define
\begin{splitequation}
\label{eq:witten_s_comp}
&S^{\Lambda,\Lambda_0}_{A_1 \cdots A_n}\left( \op_{\mathcal{A}_1}(R_1) \cdots \op_{\mathcal{A}_k}(R_k) ; P_1,\ldots,P_n \right) \equiv \lim_{z_i \to 0} \prod_{i=1}^k \left( \frac{z_i}{\ell} \right)^{-\Delta_{\mathcal{A}_i}} \\
&\qquad\times \int\dotsi\int L^{\Lambda,\Lambda_0}_{A_1 \cdots A_n}\left( \op_{\mathcal{A}_1}(Z_1) \cdots \op_{\mathcal{A}_k}(Z_k) ; X_1,\ldots,X_n \right) \prod_{k=1}^n G_{\partial A_k}(X_k,P_k) \total X_k \eqend{,}
\end{splitequation}
where $R_i$ is the boundary point corresponding to the bulk point $Z_i$, and $z_i$ the corresponding radial coordinate. These are the connected boundary correlation functions with insertions of multi-trace operators, which are dual to composite operators in the bulk; by abuse of notation, we denote both by the same symbol $\op_\mathcal{A}$. To derive the flow equation for these correlation functions, we use again the split representation~\eqref{eq:dlambda_g_split} to express the bulk-to-bulk propagator appearing in the bulk flow equations~\eqref{eq:l_n_comp1_flow} and~\eqref{eq:l_n_compk_flow} in terms of bulk-to-boundary propagators with analytically continued dimension. This results in
\begin{splitequation}
\label{eq:s_n_comp1_flow_pre}
&\partial_\Lambda S^{\Lambda,\Lambda_0}_{A_1 \cdots A_n}\left( \op_\mathcal{A}(R) ; P_1,\ldots,P_n \right) = - \frac{1}{\ell^3 \Lambda^3} \sum_B \int_{-\infty}^\infty \exp\left( - \frac{\nu^2 + (\Delta_B-\rho)^2}{\ell^2 \Lambda^2} \right) \frac{\nu^2}{\pi} \\
&\quad\times \int \Bigg[ \hbar S^{\Lambda,\Lambda_0}_{A_1 \cdots A_n B(\rho+\mathi \nu) B(\rho-\mathi \nu)}\left( \op_\mathcal{A}(R) ; P_1,\ldots,P_n,Q,Q \right) \\
&\qquad- 2 \sum_{I \cup J = \{1,\ldots,n\}} S^{\Lambda,\Lambda_0}_{\{ A_I \} B(\rho+\mathi \nu)}(\{ P_I \},Q) S^{\Lambda,\Lambda_0}_{\{ A_J \} B(\rho-\mathi \nu)}\left( \op_\mathcal{A}(R) ; \{ P_J \},Q \right) \Bigg] \total Q \total \nu
\end{splitequation}
and for multiple insertions
\begin{splitequation}
\label{eq:s_n_compk_flow_pre}
&\partial_\Lambda S^{\Lambda,\Lambda_0}_{A_1 \cdots A_n}\left( \op_{\mathcal{A}_1}(R_1) \cdots \op_{\mathcal{A}_k}(R_k) ; P_1,\ldots,P_n \right) = - \frac{1}{\ell^3 \Lambda^3} \sum_B \int_{-\infty}^\infty \exp\left( - \frac{\nu^2 + (\Delta_B-\rho)^2}{\ell^2 \Lambda^2} \right) \frac{\nu^2}{\pi} \\
&\quad\times \int \Bigg[ \hbar S^{\Lambda,\Lambda_0}_{A_1 \cdots A_n B(\rho+\mathi \nu) B(\rho-\mathi \nu)}\left( \op_{\mathcal{A}_1}(R_1) \cdots \op_{\mathcal{A}_k}(R_k); P_1,\ldots,P_n,Q,Q \right) \\
&\qquad- 2 \sum_{I \cup J = \{1,\ldots,n\}} S^{\Lambda,\Lambda_0}_{\{ A_I \} B(\rho+\mathi \nu)}(\{ P_I \},Q) S^{\Lambda,\Lambda_0}_{\{ A_J \} B(\rho-\mathi \nu)}\left( \op_{\mathcal{A}_1}(R_1) \cdots \op_{\mathcal{A}_k}(R_k); \{ P_J \},Q \right) \\
&\qquad+ \hbar \sum_{\smash{\substack{\strut \\ I \cup J = \{1,\ldots,n\} \\ K \cup L = \{1,\ldots,k\}}}} S^{\Lambda,\Lambda_0}_{\{ A_I \} B(\rho+\mathi \nu)}\left( \{ \op_{\mathcal{A}_K}(R_K) \} ; \{ P_I \},Q \right) \\
&\hspace{10em}\times S^{\Lambda,\Lambda_0}_{\{ A_J \} B(\rho-\mathi \nu)}\left( \{ \op_{\mathcal{A}_L}(R_L) \} ; \{ P_J \},Q \right) \Bigg] \total Q \total \nu \eqend{,}
\end{splitequation}
where we made the sum over all basic fields $\phi_B$ of the theory explicit, and use the same notation as for the bulk-to-boundary propagator: $A(\Delta')$ denotes the boundary correlator of the field $\phi_A$, with the conformal dimension analytically continued to $\Delta'$.

The boundary conditions for the flow at $\Lambda = \Lambda_0$ are obtained by inserting the definition~\eqref{eq:witten_s_comp} into the boundary conditions~\eqref{eq:l_n_comp_boundary}, and we obtain
\begin{splitequation}
\label{eq:s_n_comp_boundary}
&S^{\Lambda_0,\Lambda_0}_{A_1 \cdots A_n}\left( \op_\mathcal{A}(R) ; P_1,\ldots,P_n \right) = \lim_{z \to 0} \left( \frac{z}{\ell} \right)^{-\Delta_\mathcal{A}} \\
&\qquad\times \int\dotsi\int \left. \frac{\delta^n \mathcal{N}\left[ \op_\mathcal{A}[\Phi] \right](Z)}{\delta \Phi_{A_1}(X_1) \cdots \delta \Phi_{A_n}(X_n)} \right\rvert_{\Phi = 0} \prod_{k=1}^n G_{\partial A_k}(X_k,P_k) \total X_k \eqend{,}
\end{splitequation}
while correlation functions with multiple insertions have vanishing boundary conditions:
\begin{equation}
\label{eq:s_n_comp_boundary2}
S^{\Lambda_0,\Lambda_0}_{A_1 \cdots A_n}\left( \op_{\mathcal{A}_1}(R_1) \cdots \op_{\mathcal{A}_k}(R_k) ; P_1,\ldots,P_n \right) = 0 \eqend{.}
\end{equation}
As in the case of single-trace operators, since the regularisation and thus the cutoff propagator~\eqref{eq:dlambda_g_split} are EAdS-invariant, the full conformal symmetry is manifest for the boundary correlators $S^{\Lambda,\Lambda_0}_{A_1 \cdots A_n}\left( \op_{\mathcal{A}_1}(R_1) \cdots \op_{\mathcal{A}_k}(R_k) ; P_1,\ldots,P_n \right)$, and the bulk regularisation does \emph{not} regularise the boundary correlators. Again, this can be seen clearly in the first term on the right-hand sides of the flow equations~\eqref{eq:s_n_comp1_flow_pre} and~\eqref{eq:s_n_compk_flow_pre}, where to compute a loop contribution we need to evaluate the boundary correlator at coincident points $Q$. In turn, there are again two possibilities for the regularisation and renormalisation of this boundary divergence: either placing an additional IR cutoff in the bulk, evaluating the boundary correlators at some finite $z = \epsilon$ instead of the boundary $z = 0$ and using holographic renormalisation to cancel the divergences~\cite{skenderis2002}, or displacing one of the two points $Q \to Q'$ and taking the finite part $\Pf$ of the result as $Q' \to Q$. As for the single-trace operators, we prefer the second posssibility, and thus replace the formal but divergent flow equations~\eqref{eq:s_n_comp1_flow_pre} and~\eqref{eq:s_n_compk_flow_pre} by the finite versions
\begin{splitequation}
\label{eq:s_n_comp1_flow}
&\partial_\Lambda S^{\Lambda,\Lambda_0}_{A_1 \cdots A_n}\left( \op_\mathcal{A}(R) ; P_1,\ldots,P_n \right) = - \frac{1}{\ell^3 \Lambda^3} \sum_B \int_{-\infty}^\infty \exp\left( - \frac{\nu^2 + (\Delta_B-\rho)^2}{\ell^2 \Lambda^2} \right) \frac{\nu^2}{\pi} \\
&\quad\times \int \Pf_{Q' \to Q} \Bigg[ \hbar S^{\Lambda,\Lambda_0}_{A_1 \cdots A_n B(\rho+\mathi \nu) B(\rho-\mathi \nu)}\left( \op_\mathcal{A}(R) ; P_1,\ldots,P_n,Q,Q' \right) \\
&\qquad- 2 \sum_{I \cup J = \{1,\ldots,n\}} S^{\Lambda,\Lambda_0}_{\{ A_I \} B(\rho+\mathi \nu)}(\{ P_I \},Q) S^{\Lambda,\Lambda_0}_{\{ A_J \} B(\rho-\mathi \nu)}\left( \op_\mathcal{A}(R) ; \{ P_J \},Q \right) \Bigg] \total Q \total \nu
\end{splitequation}
and
\begin{splitequation}
\label{eq:s_n_compk_flow}
&\partial_\Lambda S^{\Lambda,\Lambda_0}_{A_1 \cdots A_n}\left( \op_{\mathcal{A}_1}(R_1) \cdots \op_{\mathcal{A}_k}(R_k) ; P_1,\ldots,P_n \right) = - \frac{1}{\ell^3 \Lambda^3} \sum_B \int_{-\infty}^\infty \exp\left( - \frac{\nu^2 + (\Delta_B-\rho)^2}{\ell^2 \Lambda^2} \right) \frac{\nu^2}{\pi} \\
&\quad\times \int \Pf_{Q' \to Q} \Bigg[ \hbar S^{\Lambda,\Lambda_0}_{A_1 \cdots A_n B(\rho+\mathi \nu) B(\rho-\mathi \nu)}\left( \op_{\mathcal{A}_1}(R_1) \cdots \op_{\mathcal{A}_k}(R_k); P_1,\ldots,P_n,Q,Q' \right) \\
&\qquad- 2 \sum_{I \cup J = \{1,\ldots,n\}} S^{\Lambda,\Lambda_0}_{\{ A_I \} B(\rho+\mathi \nu)}(\{ P_I \},Q) S^{\Lambda,\Lambda_0}_{\{ A_J \} B(\rho-\mathi \nu)}\left( \op_{\mathcal{A}_1}(R_1) \cdots \op_{\mathcal{A}_k}(R_k); \{ P_J \},Q \right) \\
&\qquad+ \hbar \sum_{\smash{\substack{\strut \\ I \cup J = \{1,\ldots,n\} \\ K \cup L = \{1,\ldots,k\}}}} S^{\Lambda,\Lambda_0}_{\{ A_I \} B(\rho+\mathi \nu)}\left( \{ \op_{\mathcal{A}_K}(R_K) \} ; \{ P_I \},Q \right) \\
&\hspace{10em}\times S^{\Lambda,\Lambda_0}_{\{ A_J \} B(\rho-\mathi \nu)}\left( \{ \op_{\mathcal{A}_L}(R_L) \} ; \{ P_J \},Q \right) \Bigg] \total Q \total \nu \eqend{,}
\end{splitequation}
where we recall that the sums run over all disjoint subsets $I$ and $J$ and all non-empty disjoint subsets $K$ and $L$. Expanding the correlation functions in a loop expansion in $\hbar$, one can now construct boundary correlators including insertions of multi-trace operators to arbitrary order in the loop expansion, analogously to the case of single-trace operators.

Conformal symmetry of the boundary correlators is again manifest in the Mellin representation. However, since the flow equation~\eqref{eq:s_n_compk_flow} is almost identical to the one for single-trace operators~\eqref{eq:s_n_flow} (with the differences being the notation and factors of $\hbar$), we refrain from giving the corresponding (lengthy) equations for the Mellin amplitude.

The major difference to the single-trace case are the boundary conditions for the flow, the contact amplitudes. For simplicity, we assume a composite operator without derivatives
\begin{equation}
\op_\mathcal{A} = \frac{1}{m_{A_1 \cdots A_n}} \phi_{A_1} \cdots \phi_{A_n}
\end{equation}
with the same multiplicity factor $m_{A_1 \cdots A_n}$ as for the contact interactions~\eqref{eq:contact_interaction}. The corresponding contribution to the boundary correlator $S^{\Lambda_0,\Lambda_0}$~\eqref{eq:witten_s_comp} is given by
\begin{splitequation}
S^{\Lambda_0,\Lambda_0}_{A_1 \cdots A_n}\left( \op_\mathcal{A}(R) ; P_1,\ldots,P_n \right) &= \lim_{z \to 0} \left( \frac{z}{\ell} \right)^{-\Delta_\mathcal{A}} \prod_{k=1}^n G_{\partial A_k}(Z,P_k) \\
&= \prod_{k=1}^n \left[ \frac{\ell^{1-2\rho} \Gamma(\Delta_{A_k})}{2 \pi^\rho \Gamma\left( \Delta_{A_k} - \rho + 1 \right)} \left( - 2 \ell^{-2} P_k \cdot R \right)^{-\Delta_{A_k}} \right] \eqend{,}
\end{splitequation}
which in contrast to the boundary condition~\eqref{eq:contact_boundary_correlator} for single-trace operators is just the product of boundary correlators, without bulk integration. For a composite operator bilinear in the basic fields, which is dual to a double-trace operator in the CFT, this is already the correct representation: comparing with the general form of the Mellin amplitude~\eqref{eq:s_n_mellin} (with the solution~\eqref{eq:mellin3_sol} of the constraints), one sees that it is just a constant. The exact form of the constant is somewhat ambiguous; identifying $P_3 = R$, we have
\begin{splitequation}
\label{eq:contact_amplitude_compositeboundary}
M^{\Lambda_0,\Lambda_0}_{A_1 A_2 \op_\mathcal{A}}(\delta_{ij}) &= \frac{1}{\Gamma\left( \frac{\Delta_{A_1} + \Delta_{A_2} - \Delta_\mathcal{A}}{2} \right) \Gamma\left( \frac{\Delta_\mathcal{A} + \Delta_{A_1} - \Delta_{A_2}}{2} \right) \Gamma\left( \frac{\Delta_\mathcal{A} + \Delta_{A_2} - \Delta_{A_1}}{2} \right)} \\
&\quad\times \frac{\ell^{2-4\rho} \Gamma(\Delta_{A_1}) \Gamma(\Delta_{A_2})}{4 \pi^{2 \rho} \Gamma\left( \Delta_{A_1} - \rho + 1 \right) \Gamma\left( \Delta_{A_2} - \rho + 1 \right)} \eqend{,}
\end{splitequation}
but since $\Delta_\mathcal{A} = \Delta_{A_1} + \Delta_{A_2}$ (in the free theory), the second and third $\Gamma$ functions in the denominator cancel the ones in the numerator, and the first $\Gamma$ function becomes a $\Gamma(0)$, such that the amplitude formally vanishes. However, this is offset by a diverging $\Gamma(\delta_{12}) = \Gamma(0)$ in the definition of the Mellin amplitude~\eqref{eq:s_n_mellin}. On the other hand, interactions will change the relation between $\Delta_\mathcal{A}$ and the $\Delta_{A_i}$, such that the contact amplitude~\eqref{eq:contact_amplitude_compositeboundary} becomes well-defined. Therefore, the formal vanishing of the contact amplitude in the free theory seems just an artifact of the perturbative expansion around a generalised free theory, i.e., mean-field theory.

On the other hand, for trilinear and higher composite operators, it is not yet clear to me how to derive the corresponding contact Mellin amplitude: it seems that they must be proportional to some Dirac $\delta$ functions in the (independent subset of) Mellin variables $\delta_{ij}$.

\section{Flows at tree level and one loop}
\label{sec:flow_tree_oneloop}

To illustrate the flow equations, we solve the flow at tree level and one loop for low-order $n$-point functions.

\subsection{Tree level}

To obtain the flows at tree level, we simply ignore the first term in the flow equation~\eqref{eq:s_n_flow} which gives the loop corrections. Let us denote the tree-level correlation functions by a superscript $(0)$; the general flow equation~\eqref{eq:s_n_flow} then reduces to
\begin{splitequation}
\label{eq:s_n_flow_tree}
&\partial_\Lambda S^{(0),\Lambda,\Lambda_0}_{A_1 \cdots A_n}(P_1,\ldots,P_n) = \frac{1}{\ell^3 \Lambda^3} \sum_B \int_{-\infty}^\infty \exp\left( - \frac{\nu^2 + (\Delta_B-\rho)^2}{\ell^2 \Lambda^2} \right) \frac{\nu^2}{\pi} \\
&\hspace{5em}\times \int \Pf_{Q' \to Q} \sum_{I \cup J = \{1,\ldots,n\}} S^{(0),\Lambda,\Lambda_0}_{\{ A_I \} B(\rho+\mathi \nu)}(\{ P_I \},Q) S^{\Lambda,\Lambda_0}_{\{ A_J \} B(\rho-\mathi \nu)}(\{ P_J \},Q) \total Q \total \nu \eqend{.}
\end{splitequation}
In section~\ref{sec_flow_one_two} we have seen that the terms on the right-hand side of equation~\eqref{eq:s_n_flow_tree} must be taken to vanish for the one- and two-point function, since they do not depend analytically on the conformal dimensions $\Delta_A$ of the operators. Hence, as computed there, only loop corrections to the one- and two-point functions and consequently to the conformal dimensions $\Delta_A$ exist, and they are constant at tree level.

On the other hand, the Mellin amplitudes of higher $n$-point functions have a non-vanishing flow at tree level, which is given by equation~\eqref{eq:s_n_mellin_flow_exact} but without the term proportional to $\hbar$, and which we repeat for convenience:
\begin{splitequation}
\label{eq:s_n_mellin_flow_exact_tree}
\partial_\Lambda \mathcal{M}^{(0),\Lambda,\Lambda_0}_{A_1 \cdots A_n}(\delta_{ij}) &= \frac{1}{\ell^3 \Lambda^3} \pi^\rho \ell^{2 \rho} \sum_B \int_{-\infty}^\infty \exp\left( - \frac{\nu^2 + (\Delta_B-\rho)^2}{\ell^2 \Lambda^2} \right) \frac{\nu^2}{\pi} \\
&\times \sum_{I \cup J = \{1,\ldots,n\}} \mathcal{F}^{(0),\Lambda,\Lambda_0}_{\{ A_I \} B(\rho + \mathi \nu)}(\delta_{ij}) \mathcal{F}^{(0),\Lambda,\Lambda_0}_{\{ A_J \} B(\rho - \mathi \nu)}(\delta_{ij}) \total \nu \eqend{,}
\end{splitequation}
where~\eqref{eq:mellin_factor}
\begin{equation}
\label{eq:mellin_factor_tree}
\mathcal{F}^{(0),\Lambda,\Lambda_0}_{\{ A_I \} B(\rho + \mathi \nu)}(\delta_{ij}) = \int \mathcal{M}^{(0),\Lambda,\Lambda_0}_{\{ A_I \} B(\rho+\mathi \nu)}(\tilde{\delta}_{ij},l_i) \prod_{i,j \in I, i<j} \frac{\Gamma(\tilde{\delta}_{ij}) \Gamma(\delta_{ij} - \tilde{\delta}_{ij})}{\Gamma(\delta_{ij})} [ \total (\tilde{\delta}_{ij}, l_i) ] \eqend{,}
\end{equation}
and the integration measure is constrained by~\eqref{eq:mellin_factor_constraint}
\begin{equations}[eq:mellin_factor_constraint_tree]
l_i &= \Delta_i - \sum_{i \neq j \in I} \tilde{\delta}_{ij} \quad (i \in I) \eqend{,} \\
\sum_{i,j \in I, i \neq j} \tilde{\delta}_{ij} &= \sum_{i \in I} \Delta_i - (\rho + \mathi \nu) \eqend{.}
\end{equations}

To illustrate the flow~\eqref{eq:s_n_mellin_flow_exact_tree} and compare with existing results at tree level, we will solve the flow explicitly for small $n$. For $n = 3$, at least one of the two $\mathcal{F}$ terms on the right-hand side does not involve at least three operators and thus vanishes. It follows that the Mellin amplitude for $n = 3$ does not flow and is equal to the contact amplitude at $\Lambda = \Lambda_0$:
\begin{equation}
\label{eq:mellin_flow_tree_3}
\mathcal{M}^{(0),\Lambda,\Lambda_0}_{A_1 A_2 A_3}(\delta_{ij}) = \mathcal{M}^{(0),\Lambda_0,\Lambda_0}_{A_1 A_2 A_3}(\delta_{ij}) \eqend{,}
\end{equation}
which is a polynomial in the $\delta_{ij}$~\eqref{eq:contact_amplitude}.

\begin{figure}[h]
\includegraphics[width=\textwidth]{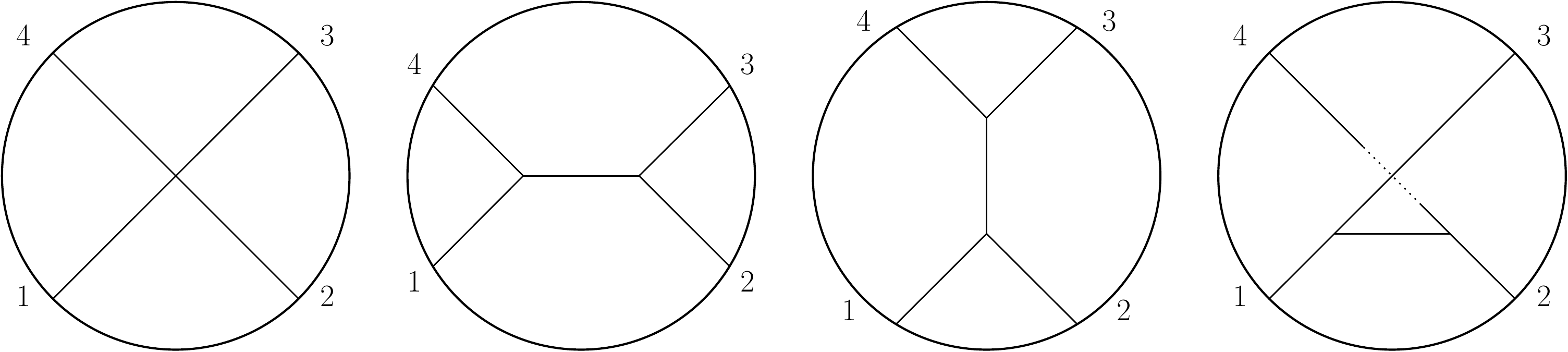}
\caption{Witten diagrams contributing to the four-point function at tree level.}
\label{fig:witten_m4}
\end{figure}

For $n = 4$, there are non-vanishing $\mathcal{F}$ terms on the right-hand side, which both involve three operators. We compute one of them explicitly, say the one involving $A_1$, $A_2$ and $B(\rho+\mathi \nu)$, and the remaining ones follow by a relabeling. Since the constraints~\eqref{eq:mellin_factor_constraint_tree} admit the unique solution
\begin{equations}[eq:f3_constraints_tree]
l_1 &= \frac{\Delta_{A_1} - \Delta_{A_2} + \rho + \mathi \nu}{2} \eqend{,} \\
l_2 &= \frac{- \Delta_{A_1} + \Delta_{A_2} + \rho + \mathi \nu}{2} \eqend{,} \\
\tilde{\delta}_{12} &= \frac{\Delta_{A_1} + \Delta_{A_2} - \rho - \mathi \nu}{2} \eqend{,}
\end{equations}
there is no integration in equation~\eqref{eq:mellin_factor_tree} and we simply obtain
\begin{equation}
\label{eq:f3_tree}
\mathcal{F}^{(0),\Lambda,\Lambda_0}_{A_1 A_2 B(\rho + \mathi \nu)}(\delta_{12}) = \mathcal{M}^{(0),\Lambda,\Lambda_0}_{A_1 A_2 B(\rho+\mathi \nu)}(\tilde{\delta}_{12},l_1,l_2) \frac{\Gamma(\tilde{\delta}_{12}) \Gamma(\delta_{12} - \tilde{\delta}_{12})}{\Gamma(\delta_{12})} \eqend{,}
\end{equation}
which is independent of $\Lambda$ since $\mathcal{M}^{(0),\Lambda,\Lambda_0}_{A_1 A_2 B(\rho+\mathi \nu)}$ is~\eqref{eq:mellin_flow_tree_3}. It follows that the only dependence on $\Lambda$ in the flow equation~\eqref{eq:s_n_mellin_flow_exact_tree} for $\mathcal{M}^{(0),\Lambda,\Lambda_0}_{A_1 A_2 A_3 A_4}$ comes from the explicit exponential, which can be trivially integrated. We thus obtain
\begin{splitequation}
\label{eq:mellin_flow_tree_4}
&\mathcal{M}^{(0),\Lambda,\Lambda_0}_{A_1 A_2 A_3 A_4}(\delta_{ij}) = \mathcal{M}^{(0),\Lambda_0,\Lambda_0}_{A_1 A_2 A_3 A_4}(\delta_{ij}) - \pi^{\rho-1} \ell^{2 \rho-1} \sum_B \int_{-\infty}^\infty \frac{\nu^2}{\nu^2 + (\Delta_B-\rho)^2} \\
&\qquad\times \left[ \exp\left( - \frac{\nu^2 + (\Delta_B-\rho)^2}{\ell^2 \Lambda_0^2} \right) - \exp\left( - \frac{\nu^2 + (\Delta_B-\rho)^2}{\ell^2 \Lambda^2} \right) \right] \\
&\qquad\times \bigg[ \mathcal{F}^{(0),\Lambda_0,\Lambda_0}_{A_1 A_2 B(\rho + \mathi \nu)}(\delta_{12}) \mathcal{F}^{(0),\Lambda_0,\Lambda_0}_{A_3 A_4 B(\rho - \mathi \nu)}(\delta_{34}) + \mathcal{F}^{(0),\Lambda_0,\Lambda_0}_{A_1 A_3 B(\rho + \mathi \nu)}(\delta_{13}) \mathcal{F}^{(0),\Lambda_0,\Lambda_0}_{A_2 A_4 B(\rho - \mathi \nu)}(\delta_{24}) \\
&\qquad\qquad+ \mathcal{F}^{(0),\Lambda_0,\Lambda_0}_{A_1 A_4 B(\rho + \mathi \nu)}(\delta_{14}) \mathcal{F}^{(0),\Lambda_0,\Lambda_0}_{A_2 A_3 B(\rho - \mathi \nu)}(\delta_{23}) \bigg] \total \nu \eqend{,}
\end{splitequation}
where we also used the symmetry under the exchange $\nu \to -\nu$ to simplify the result. Since at tree level there are no UV divergences, we can directly take the limit $\Lambda_0 \to \infty$, as well as $\Lambda \to 0$. In this limit, the exponentials in the second line of equation~\eqref{eq:mellin_flow_tree_4} simplify to $1$, and we recover the result of~\cite{fitzpatricketal2011}. In fact, each term in~\eqref{eq:mellin_flow_tree_4} directly corresponds to a Witten diagram, which are depicted in figure~\ref{fig:witten_m4}: the boundary condition is the four-point contact interaction, while the terms involving the $\mathcal{F}$ correspond to exchange diagrams in different channels.

For $n = 5$, the non-vanishing $\mathcal{F}$ terms on the right-hand side involve three and four operators. For the ones with four operators, the constraints~\eqref{eq:mellin_factor_constraint_tree} do not have a unique solution, such that two integrations remain. If we order the operators such that $\Delta_{A_1} \geq \Delta_{A_2} \geq \Delta_{A_3}$ and assume that $\rho < \Delta_{A_i} < 2 \rho$ as well as $2 \Delta_{A_1} < \Delta_{A_2} + \Delta_{A_3} + \rho$, we can take the solution
\begin{equations}
l_1 &= 2 \rho - \Delta_{A_1} + s \eqend{,} \\
l_2 &= \frac{\Delta_{A_1} - \Delta_{A_2} + \Delta_{A_3} - \rho + \mathi \nu}{2} - t \eqend{,} \\
l_3 &= \frac{\Delta_{A_1} + \Delta_{A_2} - \Delta_{A_3} - \rho + \mathi \nu}{2} - s + t \eqend{,} \\
\tilde{\delta}_{12} &= \Delta_{A_1} + \Delta_{A_2} - \Delta_{A_3} - \rho - s + t \eqend{,} \\
\tilde{\delta}_{13} &= \Delta_{A_1} - \Delta_{A_2} + \Delta_{A_3} - \rho - t \eqend{,} \\
\tilde{\delta}_{23} &= \frac{(\Delta_{A_2} + \Delta_{A_3} + \rho - 2 \Delta_{A_1}) + 2 \rho - \Delta_{A_1} - \mathi \nu}{2} + s
\end{equations}
in terms of the unconstrained variables $s$ and $t$, which with the above assumptions is such that every variable has positive real part as required. Inserting this solution into the definition of $\mathcal{F}$~\eqref{eq:mellin_factor_tree}, together with the four-point Mellin amplitude~\eqref{eq:mellin_flow_tree_4}, we obtain
\begin{splitequation}
&\mathcal{F}^{(0),\Lambda,\Lambda_0}_{A_1 A_2 A_3 B(\rho + \mathi \nu)}(\delta_{ij}) = \int \Bigg[ \mathcal{M}^{(0),\Lambda_0,\Lambda_0}_{A_1 A_2 A_3 B(\rho + \mathi \nu)}(\tilde{\delta}_{ij},l_i) - \pi^{\rho-1} \ell^{2 \rho-1} \\
&\qquad\times \sum_C \int_{-\infty}^\infty \frac{\mu^2}{\mu^2 + (\Delta_C-\rho)^2} \left[ \exp\left( - \frac{\mu^2 + (\Delta_C-\rho)^2}{\ell^2 \Lambda_0^2} \right) - \exp\left( - \frac{\mu^2 + (\Delta_C-\rho)^2}{\ell^2 \Lambda^2} \right) \right] \\
&\qquad\times \bigg[ \mathcal{F}^{(0),\Lambda_0,\Lambda_0}_{A_1 A_2 C(\rho + \mathi \mu)}(\tilde{\delta}_{12}) \mathcal{F}^{(0),\Lambda_0,\Lambda_0}_{A_3 B(\rho + \mathi \nu) C(\rho - \mathi \mu)}(l_3) + \mathcal{F}^{(0),\Lambda_0,\Lambda_0}_{A_1 A_3 C(\rho + \mathi \mu)}(\tilde{\delta}_{13}) \mathcal{F}^{(0),\Lambda_0,\Lambda_0}_{A_2 B(\rho + \mathi \nu) C(\rho - \mathi \mu)}(l_2) \\
&\qquad\qquad+ \mathcal{F}^{(0),\Lambda_0,\Lambda_0}_{A_1 B(\rho + \mathi \nu) C(\rho + \mathi \mu)}(l_1) \mathcal{F}^{(0),\Lambda_0,\Lambda_0}_{A_2 A_3 C(\rho - \mathi \mu)}(\tilde{\delta}_{23}) \bigg] \total \mu \Bigg] \prod_{1 \leq i < j \leq 3} \frac{\Gamma(\tilde{\delta}_{ij}) \Gamma(\delta_{ij} - \tilde{\delta}_{ij})}{\Gamma(\delta_{ij})} \frac{\total s}{2 \pi \mathi} \frac{\total t}{2 \pi \mathi} \eqend{.}
\end{splitequation}
Finally, the flow of the five-point Mellin amplitude is obtained as
\begin{splitequation}
\label{eq:mellin_flow_tree_5}
\partial_\Lambda \mathcal{M}^{(0),\Lambda,\Lambda_0}_{A_1 A_2 A_3 A_4 A_5}(\delta_{ij}) &= \frac{1}{\ell^3 \Lambda^3} \pi^\rho \ell^{2 \rho} \sum_B \int_{-\infty}^\infty \exp\left( - \frac{\nu^2 + (\Delta_B-\rho)^2}{\ell^2 \Lambda^2} \right) \frac{\nu^2}{\pi} \\
&\times \sum_{I \cup J = \{1,\ldots,5\}} \mathcal{F}^{(0),\Lambda,\Lambda_0}_{\{ A_I \} B(\rho + \mathi \nu)}(\delta_{ij}) \mathcal{F}^{(0),\Lambda,\Lambda_0}_{\{ A_J \} B(\rho - \mathi \nu)}(\delta_{ij}) \total \nu \eqend{,}
\end{splitequation}
where only three- and four-point $\mathcal{F}$ factors are needed on the right-hand side, and the $\Lambda$ integration can again be done easily. We refrain from giving an explicit expression in the general case, which would be quite lengthy.

It is thus clear that the determination of the tree-level Mellin amplitudes of an arbitrary number of operators can be done, and that the integrals over $\Lambda$ are always straightforward. Unfortunately, the construction quickly becomes cumbersome.

\subsection{One loop}

At one loop, we obtain the first corrections to one- and two-point functions and thus to the conformal dimensions (and the normalisation constants). These follow from the result for the flow of the two-point function~\eqref{eq:2pf_flow_terms1}, and are given by equations~\eqref{eq:2pf_flow_c} and~\eqref{eq:2pf_flow_delta}, where the $\delta_{ij}$ in the Mellin amplitude are given by~\eqref{eq:integral_boundary2_finitepart_deltaij} with $\Delta = 0$ and $\Delta_1 = \Delta_A^{\Lambda,\Lambda_0}$. Restricting to one loop amounts to taking tree-level amplitudes and tree-level conformal data on the right-hand side.

We first use the tree-level result~\eqref{eq:mellin_flow_tree_4} for the four-point Mellin amplitude $\mathcal{M}^{(0),\Lambda,\Lambda_0}_{A_1 A_2 A_3 A_4}(\delta_{ij})$ (with the $\delta_{ij}$ in the Mellin amplitude are given by~\eqref{eq:integral_boundary2_finitepart_deltaij} with $\Delta = 0$ and $\Delta_1 = \Delta_A$) together with the tree-level result~\eqref{eq:f3_constraints_tree} and~\eqref{eq:f3_tree} for $\mathcal{F}^{(0),\Lambda,\Lambda_0}$ to obtain
\begin{splitequation}
&\mathcal{M}^{(0),\Lambda,\Lambda_0}_{A A B(\rho+\mathi \nu) B(\rho-\mathi \nu)}(\delta_{ij}) = \mathcal{M}^{(0),\Lambda_0,\Lambda_0}_{A A B(\rho+\mathi \nu) B(\rho-\mathi \nu)}(\delta_{ij}) - \pi^{\rho-1} \ell^{2 \rho-1} \sum_C \int_{-\infty}^\infty \frac{\mu^2}{\mu^2 + (\Delta_C-\rho)^2} \\
&\quad\times \left[ \exp\left( - \frac{\mu^2 + (\Delta_C-\rho)^2}{\ell^2 \Lambda_0^2} \right) - \exp\left( - \frac{\mu^2 + (\Delta_C-\rho)^2}{\ell^2 \Lambda^2} \right) \right] \\
&\quad\times \Bigg[ \mathcal{M}^{(0),\Lambda_0,\Lambda_0}_{A A C(\rho+\mathi \mu)}\left( \frac{2 \Delta_A - \rho - \mathi \mu}{2}, \frac{\rho + \mathi \mu}{2}, \frac{\rho + \mathi \mu}{2} \right) \\
&\qquad\quad\times \mathcal{M}^{(0),\Lambda_0,\Lambda_0}_{B(\rho+\mathi \nu) B(\rho-\mathi \nu) C(\rho-\mathi \mu)}\left( \frac{\rho + \mathi \mu}{2}, \frac{\rho + 2 \mathi \nu - \mathi \mu}{2}, \frac{\rho - 2 \mathi \nu - \mathi \mu}{2} \right) \\
&\qquad\quad\times \frac{\Gamma\left( \frac{2 \Delta_A - \rho - \mathi \mu}{2} \right) \Gamma\left( \delta_{34} - \frac{\rho - \mathi \mu}{2} \right)}{\Gamma(\delta_{34} - \rho + \Delta_A)} \frac{\Gamma\left( \frac{\rho + \mathi \mu}{2} \right) \Gamma\left( \delta_{34} - \frac{\rho + \mathi \mu}{2} \right)}{\Gamma(\delta_{34})} \\
&\qquad+ \mathcal{M}^{(0),\Lambda_0,\Lambda_0}_{A B(\rho+\mathi \nu) C(\rho+\mathi \mu)}\left( \frac{\Delta_A - \mathi (\mu-\nu)}{2}, \frac{\Delta_A + \mathi (\mu-\nu)}{2}, \frac{2 \rho - \Delta_A + \mathi (\mu+\nu)}{2} \right) \\
&\qquad\quad\times \mathcal{M}^{(0),\Lambda_0,\Lambda_0}_{A B(\rho-\mathi \nu) C(\rho-\mathi \mu)}\left( \frac{\Delta_A + \mathi (\mu-\nu)}{2}, \frac{\Delta_A - \mathi (\mu-\nu)}{2}, \frac{2 \rho - \Delta_A - \mathi (\mu+\nu)}{2} \right) \\
&\qquad\quad\times \frac{\Gamma\left( \frac{\Delta_A - \mathi (\mu-\nu)}{2} \right) \Gamma\left( - \delta_{14} - \delta_{34} + \rho - \frac{\Delta_A - \mathi (\mu-\nu)}{2} \right)}{\Gamma(- \delta_{14} - \delta_{34} + \rho)} \frac{\Gamma\left( \frac{\Delta_A + \mathi (\mu-\nu)}{2} \right) \Gamma\left( - \delta_{14} - \delta_{34} + \rho - \frac{\Delta_A + \mathi (\mu+\nu)}{2} \right)}{\Gamma(- \delta_{14} - \delta_{34} + \rho - \mathi \nu)} \\
&\qquad+ \mathcal{M}^{(0),\Lambda_0,\Lambda_0}_{A B(\rho-\mathi \nu) C(\rho+\mathi \mu)}\left( \frac{\Delta_A - \mathi (\mu+\nu)}{2}, \frac{\Delta_A + \mathi (\mu+\nu)}{2}, \frac{2 \rho - \Delta_A + \mathi (\mu-\nu)}{2} \right) \\
&\qquad\quad\times \mathcal{M}^{(0),\Lambda_0,\Lambda_0}_{A B(\rho+\mathi \nu) C(\rho-\mathi \mu)}\left( \frac{\Delta_A + \mathi (\mu+\nu)}{2}, \frac{\Delta_A - \mathi (\mu+\nu)}{2}, \frac{2 \rho - \Delta_A - \mathi (\mu-\nu)}{2} \right) \\
&\qquad\quad\times \frac{\Gamma\left( \frac{\Delta_A - \mathi (\mu+\nu)}{2} \right) \Gamma\left( \delta_{14} - \frac{\Delta_A - \mathi (\mu+\nu)}{2} \right)}{\Gamma(\delta_{14})} \frac{\Gamma\left( \frac{\Delta_A + \mathi (\mu+\nu)}{2} \right) \Gamma\left( \delta_{14} - \frac{\Delta_A + \mathi (\mu-\nu)}{2} \right)}{\Gamma(\delta_{14} + \mathi \nu)} \Bigg] \total \mu \eqend{.}
\end{splitequation}
For the flow~\eqref{eq:2pf_flow_delta} of the conformal dimension, we need to evaluate the amplitude at $\delta_{34} = 0$ and integrate over $\delta_{14}$, while for the flow of the normalisation constant~\eqref{eq:2pf_flow_c} we also need the first derivative with respect to $\delta_{34}$ at $\delta_{34} = 0$. We use the first Barnes lemma~\cite{jantzen2013}
\begin{splitequation}
\label{eq:barneslemma1}
&\int \Gamma(s + \alpha_1) \Gamma(s + \alpha_2) \Gamma(- s + \beta_1) \Gamma(- s + \beta_2) \frac{\total s}{2 \pi \mathi} \\
&\quad= \frac{\Gamma(\alpha_1+\beta_1) \Gamma(\alpha_1+\beta_2) \Gamma(\alpha_2+\beta_1) \Gamma(\alpha_2+\beta_2)}{\Gamma(\alpha_1+\alpha_2+\beta_1+\beta_2)}
\end{splitequation}
to perform the integrals over $\delta_{14}$, as well as its corollary
\begin{splitequation}
\label{eq:barneslemmapoly}
&\int \Gamma(s + \alpha_1) \Gamma(s + \alpha_2) \Gamma(- s + \beta_1) \Gamma(- s + \beta_2) \psi(- s + \beta_2) \frac{\total s}{2 \pi \mathi} \\
&= \frac{\Gamma(\alpha_1+\beta_1) \Gamma(\alpha_1+\beta_2) \Gamma(\alpha_2+\beta_1) \Gamma(\alpha_2+\beta_2)}{\Gamma(\alpha_1+\alpha_2+\beta_1+\beta_2)} \\
&\quad\times \Bigg[ \psi(\alpha_1+\beta_2) + \psi(\alpha_2+\beta_2) - \psi(\alpha_1+\alpha_2+\beta_1+\beta_2) \Bigg] \eqend{,}
\end{splitequation}
which is obtained by taking a derivative w.r.t. $\beta_2$ of~\eqref{eq:barneslemma1}.

This results in
\begin{splitequation}
&\int \Gamma(\delta_{14}) \Gamma(\rho - \delta_{14}) \Gamma(\delta_{14} + \mathi \nu) \Gamma(\rho - \delta_{14} - \mathi \nu) \left. \mathcal{M}^{(0),\Lambda,\Lambda_0}_{A A B(\rho+\mathi \nu) B(\rho-\mathi \nu)}(\delta_{ij}) \right\rvert_{\delta_{34} = 0} \frac{\total \delta_{14}}{2 \pi \mathi} \\
&= \int \Gamma(\delta_{14}) \Gamma(\rho - \delta_{14}) \Gamma(\delta_{14} + \mathi \nu) \Gamma(\rho - \delta_{14} - \mathi \nu) \left. \mathcal{M}^{(0),\Lambda_0,\Lambda_0}_{A A B(\rho+\mathi \nu) B(\rho-\mathi \nu)}(\delta_{ij}) \right\rvert_{\delta_{34} = 0} \frac{\total \delta_{14}}{2 \pi \mathi} \\
&- 2 \pi^{\rho-1} \ell^{2 \rho-1} \sum_C \int_{-\infty}^\infty \frac{\mu^2}{\mu^2 + (\Delta_C-\rho)^2} \left[ \exp\left( - \frac{\mu^2 + (\Delta_C-\rho)^2}{\ell^2 \Lambda_0^2} \right) - \exp\left( - \frac{\mu^2 + (\Delta_C-\rho)^2}{\ell^2 \Lambda^2} \right) \right] \\
&\qquad\times \frac{\Gamma\left( \frac{\Delta_A - \mathi (\mu-\nu)}{2} \right) \Gamma\left( \frac{\Delta_A + \mathi (\mu-\nu)}{2} \right)}{\Gamma\left( 2 \rho - \Delta_A \right)} \Gamma\left( \rho - \frac{\Delta_A - \mathi (\mu-\nu)}{2} \right) \Gamma\left( \rho - \frac{\Delta_A + \mathi (\mu-\nu)}{2} \right) \\
&\qquad\times \Gamma\left( \rho - \frac{\Delta_A + \mathi (\mu+\nu)}{2} \right) \Gamma\left( \rho - \frac{\Delta_A - \mathi (\mu+\nu)}{2} \right) \\
&\qquad\times \mathcal{M}^{(0),\Lambda_0,\Lambda_0}_{A B(\rho+\mathi \nu) C(\rho+\mathi \mu)}\left( \frac{\Delta_A - \mathi (\mu-\nu)}{2}, \frac{\Delta_A + \mathi (\mu-\nu)}{2}, \frac{2 \rho - \Delta_A + \mathi (\mu+\nu)}{2} \right) \\
&\qquad\times \mathcal{M}^{(0),\Lambda_0,\Lambda_0}_{A B(\rho-\mathi \nu) C(\rho-\mathi \mu)}\left( \frac{\Delta_A + \mathi (\mu-\nu)}{2}, \frac{\Delta_A - \mathi (\mu-\nu)}{2}, \frac{2 \rho - \Delta_A - \mathi (\mu+\nu)}{2} \right) \total \mu
\end{splitequation}
and
\begin{splitequation}
&\int \Gamma(\delta_{14}) \Gamma(\rho - \delta_{14}) \Gamma(\delta_{14} + \mathi \nu) \Gamma(\rho - \delta_{14} - \mathi \nu) \bigg[ \left. \partial_{\delta_{34}} \mathcal{M}^{(0),\Lambda,\Lambda_0}_{A A B(\rho+\mathi \nu) B(\rho-\mathi \nu)}(\delta_{ij}) \right\rvert_{\delta_{34} = 0} \\
&\quad+ \left. \mathcal{M}^{\Lambda,\Lambda_0}_{A A B(\rho+\mathi \nu) B(\rho-\mathi \nu)}(\delta_{ij}) \right\rvert_{\delta_{34} = 0} \Big[ - \gamma + \psi\left( \Delta_A - \rho \right) - \psi(\rho - \delta_{14} - \mathi \nu) - \psi(\rho - \delta_{14}) \Big] \bigg] \frac{\total \delta_{14}}{2 \pi \mathi} \\
&= \int \Gamma(\delta_{14}) \Gamma(\rho - \delta_{14}) \Gamma(\delta_{14} + \mathi \nu) \Gamma(\rho - \delta_{14} - \mathi \nu) \bigg[ \left. \partial_{\delta_{34}} \mathcal{M}^{(0),\Lambda_0,\Lambda_0}_{A A B(\rho+\mathi \nu) B(\rho-\mathi \nu)}(\delta_{ij}) \right\rvert_{\delta_{34} = 0} \\
&\quad+ \left. \mathcal{M}^{\Lambda_0,\Lambda_0}_{A A B(\rho+\mathi \nu) B(\rho-\mathi \nu)}(\delta_{ij}) \right\rvert_{\delta_{34} = 0} \Big[ - \gamma + \psi\left( \Delta_A - \rho \right) - \psi(\rho - \delta_{14} - \mathi \nu) - \psi(\rho - \delta_{14}) \Big] \bigg] \frac{\total \delta_{14}}{2 \pi \mathi} \\
&+ \pi^{\rho-1} \ell^{2 \rho-1} \sum_C \int_{-\infty}^\infty \frac{\mu^2}{\mu^2 + (\Delta_C-\rho)^2} \left[ \exp\left( - \frac{\mu^2 + (\Delta_C-\rho)^2}{\ell^2 \Lambda_0^2} \right) - \exp\left( - \frac{\mu^2 + (\Delta_C-\rho)^2}{\ell^2 \Lambda^2} \right) \right] \\
&\quad\times \Bigg[ 2 \mathcal{M}^{(0),\Lambda_0,\Lambda_0}_{A B(\rho+\mathi \nu) C(\rho+\mathi \mu)}\left( \frac{\Delta_A - \mathi (\mu-\nu)}{2}, \frac{\Delta_A + \mathi (\mu-\nu)}{2}, \frac{2 \rho - \Delta_A + \mathi (\mu+\nu)}{2} \right) \\
&\qquad\quad\times \mathcal{M}^{(0),\Lambda_0,\Lambda_0}_{A B(\rho-\mathi \nu) C(\rho-\mathi \mu)}\left( \frac{\Delta_A + \mathi (\mu-\nu)}{2}, \frac{\Delta_A - \mathi (\mu-\nu)}{2}, \frac{2 \rho - \Delta_A - \mathi (\mu+\nu)}{2} \right) \\
&\qquad\quad\times \Gamma\left( \frac{\Delta_A + \mathi (\mu-\nu)}{2} \right) \Gamma\left( \frac{\Delta_A - \mathi (\mu-\nu)}{2} \right) \\
&\qquad\quad\times \frac{\Gamma\left( \rho - \frac{\Delta_A + \mathi (\mu+\nu)}{2} \right) \Gamma\left( \rho - \frac{\Delta_A - \mathi (\mu-\nu)}{2} \right) \Gamma\left( \rho - \frac{\Delta_A + \mathi (\mu-\nu)}{2} \right) \Gamma\left( \rho - \frac{\Delta_A - \mathi (\mu+\nu)}{2} \right)}{\Gamma\left( 2 \rho - \Delta_A \right)} \\
&\qquad\quad\times \Bigg[ \psi\left( \rho - \frac{\Delta_A + \mathi (\mu+\nu)}{2} \right) + \psi\left( \rho - \frac{\Delta_A - \mathi (\mu-\nu)}{2} \right) + \gamma - \psi\left( \Delta_A - \rho \right) \\
&\qquad\qquad+ \psi\left( \rho - \frac{\Delta_A + \mathi (\mu-\nu)}{2} \right) + \psi\left( \rho - \frac{\Delta_A - \mathi (\mu+\nu)}{2} \right) - 2 \psi(2 \rho - \Delta_A) \Bigg] \\
&\qquad- \mathcal{M}^{(0),\Lambda_0,\Lambda_0}_{A A C(\rho+\mathi \mu)}\left( \frac{2 \Delta_A - \rho - \mathi \mu}{2}, \frac{\rho + \mathi \mu}{2}, \frac{\rho + \mathi \mu}{2} \right) \\
&\qquad\quad\times \mathcal{M}^{(0),\Lambda_0,\Lambda_0}_{B(\rho+\mathi \nu) B(\rho-\mathi \nu) C(\rho-\mathi \mu)}\left( \frac{\rho + \mathi \mu}{2}, \frac{\rho + 2 \mathi \nu - \mathi \mu}{2}, \frac{\rho - 2 \mathi \nu - \mathi \mu}{2} \right) \\
&\qquad\quad\times \Gamma\left( \frac{2 \Delta_A - \rho - \mathi \mu}{2} \right) \Gamma\left( \frac{\rho + \mathi \mu}{2} \right) \frac{\Gamma\left( - \frac{\rho - \mathi \mu}{2} \right) \Gamma\left( - \frac{\rho + \mathi \mu}{2} \right)}{\Gamma(\Delta_A - \rho)} \frac{\Gamma^2(\rho) \Gamma(\rho - \mathi \nu) \Gamma(\rho + \mathi \nu)}{\Gamma(2\rho)} \Bigg] \total \mu \eqend{,}
\end{splitequation}
where we used the symmetry of the integrand under the exchange $\mu \to -\mu$ to simplify the expressions. While these results look quite complicated, their structure is simple, and in particular the $\Lambda$ dependence is only contained in the exponentials. Inserting these results into the formulas for the flow of the conformal dimension~\eqref{eq:2pf_flow_delta} and the normalisation factor~\eqref{eq:2pf_flow_c} we can thus perform the integrals over $\Lambda$ easily. In the physical limit $\Lambda \to 0, \Lambda_0 \to \infty$ we obtain
\begin{splitequation}
\label{eq:oneloop_c}
c^{(1),0,\infty}_{AA} &= c^{(1),\infty,\infty}_{AA} + \frac{\pi^{\rho-1} \ell^{2\rho-1} \Gamma\left( \Delta_A - \rho \right)}{\Gamma(\rho)} \sum_B \int_{-\infty}^\infty \frac{\nu^2}{\nu^2 + (\Delta_B-\rho)^2} \\
&\quad\times \int \Gamma(\delta_{14}) \Gamma(\rho - \delta_{14}) \Gamma(\delta_{14} + \mathi \nu) \Gamma(\rho - \delta_{14} - \mathi \nu) \\
&\quad\times \bigg[ \left. \partial_{\delta_{34}} \mathcal{M}^{(0),\infty,\infty}_{A A B(\rho+\mathi \nu) B(\rho-\mathi \nu)}(\delta_{ij}) \right\rvert_{\delta_{34} = 0} + \left. \mathcal{M}^{\infty,\infty}_{A A B(\rho+\mathi \nu) B(\rho-\mathi \nu)}(\delta_{ij}) \right\rvert_{\delta_{34} = 0} \\
&\qquad\times \Big[ - \gamma + \psi\left( \Delta_A - \rho \right) - \psi(\rho - \delta_{14} - \mathi \nu) - \psi(\rho - \delta_{14}) \Big] \bigg] \frac{\total \delta_{14}}{2 \pi \mathi} \total \nu \\
&+ \frac{\pi^{2\rho-2} \ell^{4\rho-2} \Gamma\left( \Delta_A - \rho \right)}{\Gamma(\rho)} \sum_{B,C} \int_{-\infty}^\infty \int_{-\infty}^\infty \\
&\quad\times \frac{\mu^2 \nu^2}{[ \nu^2 + (\Delta_B-\rho)^2 ] [ \mu^2 + (\Delta_C-\rho)^2 ]} \\
&\quad\times \Bigg[ 2 \mathcal{M}^{(0),\infty,\infty}_{A B(\rho+\mathi \nu) C(\rho+\mathi \mu)}\left( \frac{\Delta_A - \mathi (\mu-\nu)}{2}, \frac{\Delta_A + \mathi (\mu-\nu)}{2}, \frac{2 \rho - \Delta_A + \mathi (\mu+\nu)}{2} \right) \\
&\qquad\times \mathcal{M}^{(0),\infty,\infty}_{A B(\rho-\mathi \nu) C(\rho-\mathi \mu)}\left( \frac{\Delta_A + \mathi (\mu-\nu)}{2}, \frac{\Delta_A - \mathi (\mu-\nu)}{2}, \frac{2 \rho - \Delta_A - \mathi (\mu+\nu)}{2} \right) \\
&\qquad\times \Gamma\left( \frac{\Delta_A + \mathi (\mu-\nu)}{2} \right) \Gamma\left( \frac{\Delta_A - \mathi (\mu-\nu)}{2} \right) \\
&\qquad\times \frac{\Gamma\left( \rho - \frac{\Delta_A + \mathi (\mu+\nu)}{2} \right) \Gamma\left( \rho - \frac{\Delta_A - \mathi (\mu-\nu)}{2} \right) \Gamma\left( \rho - \frac{\Delta_A + \mathi (\mu-\nu)}{2} \right) \Gamma\left( \rho - \frac{\Delta_A - \mathi (\mu+\nu)}{2} \right)}{\Gamma\left( 2 \rho - \Delta_A \right)} \\
&\qquad\times \Bigg[ \psi\left( \rho - \frac{\Delta_A + \mathi (\mu+\nu)}{2} \right) + \psi\left( \rho - \frac{\Delta_A - \mathi (\mu-\nu)}{2} \right) + \gamma - \psi\left( \Delta_A - \rho \right) \\
&\qquad\qquad+ \psi\left( \rho - \frac{\Delta_A + \mathi (\mu-\nu)}{2} \right) + \psi\left( \rho - \frac{\Delta_A - \mathi (\mu+\nu)}{2} \right) - 2 \psi(2 \rho - \Delta_A) \Bigg] \\
&\qquad- \mathcal{M}^{(0),\infty,\infty}_{A A C(\rho+\mathi \mu)}\left( \frac{2 \Delta_A - \rho - \mathi \mu}{2}, \frac{\rho + \mathi \mu}{2}, \frac{\rho + \mathi \mu}{2} \right) \Gamma\left( \frac{2 \Delta_A - \rho - \mathi \mu}{2} \right) \\
&\qquad\quad\times \mathcal{M}^{(0),\infty,\infty}_{B(\rho+\mathi \nu) B(\rho-\mathi \nu) C(\rho-\mathi \mu)}\left( \frac{\rho + \mathi \mu}{2}, \frac{\rho + 2 \mathi \nu - \mathi \mu}{2}, \frac{\rho - 2 \mathi \nu - \mathi \mu}{2} \right) \\
&\qquad\quad\times \Gamma\left( \frac{\rho + \mathi \mu}{2} \right) \frac{\Gamma\left( - \frac{\rho - \mathi \mu}{2} \right) \Gamma\left( - \frac{\rho + \mathi \mu}{2} \right)}{\Gamma(\Delta_A - \rho)} \frac{\Gamma^2(\rho) \Gamma(\rho - \mathi \nu) \Gamma(\rho + \mathi \nu)}{\Gamma(2\rho)} \Bigg] \total \mu \total \nu
\end{splitequation}
and
\begin{splitequation}
\label{eq:oneloop_delta}
\Delta_A^{(1),0,\infty} &= \Delta_A^{(1),\infty,\infty} - \frac{\pi^{\rho-1} \ell^{2\rho-1} \Gamma\left( \Delta_A - \rho \right)}{c_{AA} \Gamma(\rho)} \sum_B \int_{-\infty}^\infty \frac{\nu^2}{\nu^2 + (\Delta_B-\rho)^2} \int \Gamma(\delta_{14}) \Gamma(\rho - \delta_{14}) \\
&\qquad\times \Gamma(\delta_{14} + \mathi \nu) \Gamma(\rho - \delta_{14} - \mathi \nu) \left. \mathcal{M}^{(0),\infty,\infty}_{A A B(\rho+\mathi \nu) B(\rho-\mathi \nu)}(\delta_{ij}) \right\rvert_{\delta_{34} = 0} \frac{\total \delta_{14}}{2 \pi \mathi} \total \nu \\
&+ \frac{2 \pi^{2\rho-2} \ell^{4 \rho - 2} \Gamma\left( \Delta_A - \rho \right)}{c_{AA} \Gamma(\rho)} \sum_{B,C} \int_{-\infty}^\infty \int_{-\infty}^\infty \\
&\quad\times \frac{\Gamma\left( \frac{\Delta_A - \mathi (\mu-\nu)}{2} \right) \Gamma\left( \frac{\Delta_A + \mathi (\mu-\nu)}{2} \right)}{\Gamma\left( 2 \rho - \Delta_A \right)} \Gamma\left( \rho - \frac{\Delta_A - \mathi (\mu-\nu)}{2} \right) \Gamma\left( \rho - \frac{\Delta_A + \mathi (\mu-\nu)}{2} \right) \\
&\quad\times \Gamma\left( \rho - \frac{\Delta_A + \mathi (\mu+\nu)}{2} \right) \Gamma\left( \rho - \frac{\Delta_A - \mathi (\mu+\nu)}{2} \right) \\
&\quad\times \mathcal{M}^{(0),\Lambda_0,\Lambda_0}_{A B(\rho+\mathi \nu) C(\rho+\mathi \mu)}\left( \frac{\Delta_A - \mathi (\mu-\nu)}{2}, \frac{\Delta_A + \mathi (\mu-\nu)}{2}, \frac{2 \rho - \Delta_A + \mathi (\mu+\nu)}{2} \right) \\
&\quad\times \mathcal{M}^{(0),\Lambda_0,\Lambda_0}_{A B(\rho-\mathi \nu) C(\rho-\mathi \mu)}\left( \frac{\Delta_A + \mathi (\mu-\nu)}{2}, \frac{\Delta_A - \mathi (\mu-\nu)}{2}, \frac{2 \rho - \Delta_A - \mathi (\mu+\nu)}{2} \right) \\
&\quad\times \frac{\mu^2 \nu^2}{[ \nu^2 + (\Delta_B-\rho)^2 ] [ \mu^2 + (\Delta_C-\rho)^2 ]} \total \mu \total \nu \eqend{.}
\end{splitequation}

For the loop corrections to the three-point coefficients $c^{\Lambda,\Lambda_0}_{A_1 A_2 A_3}$, one would need the explicit solution of the flow of the tree-level five-point Mellin amplitude~\eqref{eq:mellin_flow_tree_5}. While this can be written down in the general case, the expression is very long and not illuminating, such that we refrain from giving it explicitly. Let us remark again that as input to the flow, we only have the tree-level boundary terms/contact amplitudes $\mathcal{M}^{(0),\infty,\infty}$ and the tree-level conformal dimensions $\Delta_A$ and normalisation factors $c_{AA}$.

\section{One-loop corrections to the conformal dimensions}
\label{sec:oneloop_corrections}

We now consider a simple O($N$) model of a scalar field $\phi_i$ with a cubic coupling $\lambda_{ijk}$ and a quartic one $\lambda_{ijkl}$, and would like to derive the one-loop corrections to the conformal data of some operators. In this case, there is only a single fundamental field $\phi_i$ in AdS, dual to a single-trace operator in the CFT, and the sums appearing in the flow equations reduce to a single term. On the other hand, we have a hierarchy of multi-trace operators in the CFT, corresponding to composite operators in AdS such as $\phi^2 \equiv \phi_i \phi_i$. For the dual of this operator, we will determine one-loop corrections to the conformal dimension.

\subsection{Free theory}

We start with the free theory, which does not flow. We first consider the single-trace operator in the CFT, whose two-point function is given by the boundary-to-boundary propagator~\eqref{eq:boundary_to_boundary_propagator}, and we have the tree-level normalisation~\eqref{eq:s_2_free}
\begin{equation}
\label{eq:free_theory_cphi1}
c_{\phi\phi} = \frac{\ell^{1-2\rho} \Gamma(\Delta_\phi)}{2 \pi^\rho \Gamma\left( \Delta_\phi - \rho + 1 \right)} \eqend{,} \quad \Delta_{\phi_1} = \Delta_\phi \eqend{.}
\end{equation}
The four-point function factorises by Wick's theorem, and we obtain
\begin{splitequation}
S_{\phi_i \phi_j \phi_k \phi_l}(P_1,P_2,P_3,P_4) &= c_{\phi\phi}^2 \Bigg[ \delta_{ij} \delta_{kl} \left( - 2 \ell^{-2} P_1 \cdot P_2 \right)^{-\Delta_\phi} \left( - 2 \ell^{-2} P_3 \cdot P_4 \right)^{-\Delta_\phi} \\
&\qquad+ \delta_{ik} \delta_{jl} \left( - 2 \ell^{-2} P_1 \cdot P_3 \right)^{-\Delta_\phi} \left( - 2 \ell^{-2} P_2 \cdot P_4 \right)^{-\Delta_\phi} \\
&\qquad+ \delta_{il} \delta_{jk} \left( - 2 \ell^{-2} P_1 \cdot P_4 \right)^{-\Delta_\phi} \left( - 2 \ell^{-2} P_2 \cdot P_3 \right)^{-\Delta_\phi} \Bigg] \eqend{.}
\end{splitequation}
Using the free-field operator product expansion
\begin{equation}
\label{eq:free_field_ope_1}
\phi_k(P_3) \phi_l(P_4) = S_{\phi_k \phi_l}(P_3,P_4) \1 + \phi_k(P_4) \phi_l(P_4) + \ldots \eqend{,}
\end{equation}
we obtain the three-point function of two $\phi$ and $\phi^2$ by subtracting the first term, taking the limit as $P_3,P_4 \to R$ and summing over $k$ and $l$:
\begin{splitequation}
\label{eq:free_field_3pf_limit}
S_{\phi_i \phi_j \phi^2}(P_1,P_2,R) &= \lim_{P_3,P_4 \to R} \left[ S_{\phi_i \phi_j \phi_k \phi_l}(P_1,P_2,P_3,P_4) - S_{\phi_i \phi_j}(P_1,P_2) S_{\phi_k \phi_l}(P_3,P_4) \right] \\
&= 2 c_{\phi\phi}^2 \delta_{ij} \left( - 2 \ell^{-2} P_1 \cdot R \right)^{-\Delta_\phi} \left( - 2 \ell^{-2} P_2 \cdot R \right)^{-\Delta_\phi} \eqend{,}
\end{splitequation}
which has the conformally invariant form~\eqref{eq:s_3_form} with $\Delta_{\phi^2} = 2 \Delta_\phi$ and
\begin{equation}
c_{\phi_i \phi_j \phi^2} = 2 c_{\phi\phi}^2 \delta_{ij} \eqend{.}
\end{equation}
However, the corresponding Mellin amplitude vanishes because of the relation $\Delta_{\phi^2} = \Delta_{\phi_i} + \Delta_{\phi_j}$, as for the boundary contribution~\eqref{eq:contact_amplitude_compositeboundary}. Taking moreover the limit of this result as $P_1,P_2 \to Q$ and summing over $i$ and $j$, we obtain the two-point function of $\phi^2$:
\begin{equation}
S_{\phi^2 \phi^2}(Q,R) = \lim_{P_1,P_2 \to Q} S_{\phi_i \phi_j \phi^2}(P_1,P_2,R) = 2 c_{\phi\phi}^2 \delta_{ii} \left( - 2 \ell^{-2} Q \cdot R \right)^{-2 \Delta_\phi} \eqend{,}
\end{equation}
which again affirmes $\Delta_{\phi^2} = 2 \Delta_\phi$ and moreover
\begin{equation}
\label{eq:free_theory_cphi2}
c_{\phi^2 \phi^2} = 2 c_{\phi\phi}^2 \delta_{ii} \eqend{.}
\end{equation}
It follows that the normalised three-point function of two $\phi$ and $\phi^2$ is given by
\begin{equation}
\label{eq:free_theory_gammaphiphiphi2}
\gamma_{\phi_i \phi_j \phi^2} \equiv \frac{c_{\phi_i \phi_j \phi^2}}{\sqrt{ c_{\phi\phi}^2 c_{\phi^2 \phi^2} }} = \sqrt{ \frac{2}{\delta_{kk}} } \, \delta_{ij} \eqend{.}
\end{equation}
Note that instead of subtracting the first term in the OPE and then taking the limit as $P_3,P_4 \to R$, we can also take the finite part $\mathcal{P}\!f$ of the limit (assuming generic conformal dimension $\Delta_\phi$), which we will do in the following.

\subsection{Tree level for the single-trace operator}

Since we need to derive the boundary conditions of the flow for composite operators, let us start with tree level. The three-point amplitude does not flow~\eqref{eq:mellin_flow_tree_3} and is thus equal to the contact amplitude~\eqref{eq:contact_amplitude}
\begin{equations}
\mathcal{M}^{(0),\Lambda,\Lambda_0}_{\phi_i \phi_j \phi_k}(\delta_{ij}) &= \tilde{\lambda}_{ijk} \frac{\Gamma\left( - \rho + \frac{3}{2} \Delta_\phi \right)}{\Gamma^3\left( 1 + \Delta_\phi - \rho \right)} \eqend{,} \\
\mathcal{M}^{(0),\Lambda,\Lambda_0}_{\phi_i(D_1) \phi_j(D_2) \phi_k(D_3)}(\delta_{ij}) &= \tilde{\lambda}_{ijk} \frac{\Gamma\left( - \rho + \frac{D_1 + D_2 + D_3}{2} \right)}{\Gamma\left( 1 + D_1 - \rho \right) \Gamma\left( 1 + D_2 - \rho \right) \Gamma\left( 1 + D_3 - \rho \right)} \eqend{,}
\end{equations}
where we recall that the notation $\phi_i(D)$ means that the conformal dimension of $\phi_i$ is analytically continued to $D$.

From this, we determine the $\mathcal{F}$ factors~\eqref{eq:f3_tree}, \eqref{eq:f3_constraints_tree} that are needed in the four-point amplitude:
\begin{equations}
\mathcal{F}^{(0),\Lambda_0,\Lambda_0}_{\phi_i \phi_j \phi_m(\rho + \mathi \nu)}(\delta_{12}) &= \tilde{\lambda}_{ijm} \frac{\Gamma\left( \Delta_\phi - \frac{\rho - \mathi \nu}{2} \right) \Gamma\left( \Delta_\phi - \frac{\rho + \mathi \nu}{2} \right) \Gamma\left( \delta - \Delta_\phi + \frac{\rho + \mathi \nu}{2} \right)}{\Gamma^2\left( 1 + \Delta_\phi - \rho \right) \Gamma\left( 1 + \mathi \nu \right) \Gamma(\delta)} \eqend{,} \\
\mathcal{F}^{(0),\Lambda_0,\Lambda_0}_{\phi_i(D_1) \phi_j(D_2) \phi_m(\rho + \mathi \nu)}(\delta) &= \tilde{\lambda}_{ijm} \frac{\Gamma\left( \frac{D_1 + D_2 - \rho + \mathi \nu}{2} \right) \Gamma\left( \frac{D_1 + D_2 - \rho - \mathi \nu}{2} \right) \Gamma\left( \delta - \frac{D_1 + D_2 - \rho - \mathi \nu}{2} \right)}{\Gamma\left( 1 + D_1 - \rho \right) \Gamma\left( 1 + D_2 - \rho \right) \Gamma\left( 1 + \mathi \nu \right) \Gamma(\delta)} \eqend{.} \label{eq:f3_tree_analytical}
\end{equations}
The four-point amplitude itself is given by~\eqref{eq:mellin_flow_tree_4}:
\begin{splitequation}
\label{eq:tree_mellin_4}
\mathcal{M}^{(0),0,\infty}_{\phi_i \phi_j \phi_k \phi_l}(\delta_{ij}) &= \tilde{\lambda}_{ijkl} \frac{\Gamma\left( 2 \Delta_\phi - \rho \right)}{\Gamma^4\left( 1 + \Delta_\phi - \rho \right)} \\
&\quad- \frac{\pi^{\rho-1} \ell^{2 \rho-1}}{\Gamma^4\left( 1 + \Delta_\phi - \rho \right)} \int_{-\infty}^\infty \frac{1}{\nu^2 + (\Delta_\phi-\rho)^2} \frac{\Gamma^2\left( \Delta_\phi - \frac{\rho - \mathi \nu}{2} \right) \Gamma^2\left( \Delta_\phi - \frac{\rho + \mathi \nu}{2} \right)}{\Gamma\left( \mathi \nu \right) \Gamma\left( - \mathi \nu \right)} \\
&\qquad\times \bigg[ \tilde{\lambda}_{ijm} \tilde{\lambda}_{klm} \frac{\Gamma\left( \delta_{12} - \Delta_\phi + \frac{\rho + \mathi \nu}{2} \right)}{\Gamma(\delta_{12})} \frac{\Gamma\left( \delta_{34} - \Delta_\phi + \frac{\rho - \mathi \nu}{2} \right)}{\Gamma(\delta_{34})} \\
&\qquad\quad+ \tilde{\lambda}_{ikm} \tilde{\lambda}_{jlm} \frac{\Gamma\left( \delta_{13} - \Delta_\phi + \frac{\rho + \mathi \nu}{2} \right)}{\Gamma(\delta_{13})} \frac{\Gamma\left( \delta_{24} - \Delta_\phi + \frac{\rho - \mathi \nu}{2} \right)}{\Gamma(\delta_{24})} \\
&\qquad\quad+ \tilde{\lambda}_{ilm} \tilde{\lambda}_{jkm} \frac{\Gamma\left( \delta_{14} - \Delta_\phi + \frac{\rho + \mathi \nu}{2} \right)}{\Gamma(\delta_{14})} \frac{\Gamma\left( \delta_{23} - \Delta_\phi + \frac{\rho - \mathi \nu}{2} \right)}{\Gamma(\delta_{23})} \bigg] \total \nu \eqend{,}
\end{splitequation}
where we already took the physical limit $\Lambda \to 0$, $\Lambda_0 \to \infty$. Since we only consider a single fundamental field in AdS, the only remnant of the sum over $B$ is the sum over the O($N$) index $m$.

The integral over $\nu$ is absolutely convergent, and analytically solvable in special cases. Instead of the $\delta_{ij}$, we can also express the amplitude in terms of variables $s_{ij}$, which are defined as follows: choose $n$ auxiliary vectors $k_i$ such that
\begin{equation}
- k_i^2 = \Delta_i \quad (1 \leq i \leq n) \eqend{,} \quad \sum_{i=1}^n k_i = 0 \eqend{,}
\end{equation}
and set $\delta_{ij} = k_i \cdot k_j$, which solves the Mellin integration measure constraints~\eqref{eq:mellin_integration_measure}, and define
\begin{equation}
s_{ij} = - (k_i+k_j)^2 = \Delta_i + \Delta_j - 2 \delta_{ij} \eqend{.}
\end{equation}
For these variables, the constraints read
\begin{equation}
\sum_{j=1}^{i-1} s_{ji} + \sum_{j=i+1}^n s_{ij} = (n-4) \Delta_i + \sum_{j=1}^n \Delta_j \eqend{,}
\end{equation}
and they are especially useful for $n = 4$, where they satisfy in addition $s_{12} = s_{34}$, $s_{13} = s_{24}$ and $s_{14} = s_{23}$. It follows that also
\begin{equations}[eq:deltaij_4_relations]
\delta_{34} &= \delta_{12} + \frac{\Delta_3 + \Delta_4 - \Delta_1 - \Delta_2}{2} \eqend{,} \\
\delta_{24} &= \delta_{13} + \frac{\Delta_2 + \Delta_4 - \Delta_1 - \Delta_3}{2} \eqend{,} \\
\delta_{14} &= \delta_{23} + \frac{\Delta_1 + \Delta_4 - \Delta_2 - \Delta_3}{2} \eqend{,}
\end{equations}
and the four-point amplitude~\eqref{eq:tree_mellin_4} simplifies even further. In these variables and after the change of integration variable $\nu \to - \mathi c$, it is easily seen to agree with previous and by now well-established results~\cite{penedones2011}, taking into account the difference in overall normalisation. Moreover and by construction, the four-point amplitude has crossing symmetry.

\subsection{One loop for the single-trace operator}

For the one-loop corrections, we need the contact amplitudes~\eqref{eq:contact_amplitude}
\begin{equations}
\mathcal{M}^{(0),\infty,\infty}_{\phi_1 \phi_m(\rho+\mathi \nu) \phi_n(\rho+\mathi \mu)}(\delta_{ij}) &= \tilde{\lambda}_{1mn} \frac{\Gamma\left( \frac{\Delta_\phi + \mathi (\mu+\nu)}{2} \right)}{\Gamma\left( 1 + \Delta_\phi - \rho \right) \Gamma\left( 1 + \mathi \nu \right) \Gamma\left( 1 + \mathi \mu \right)} \eqend{,} \\
\mathcal{M}^{(0),\infty,\infty}_{\phi_1 \phi_1 \phi_n(\rho+\mathi \mu)}(\delta_{ij}) &= \tilde{\lambda}_{11n} \frac{\Gamma\left( \Delta_\phi - \frac{\rho-\mathi \mu}{2} \right)}{\Gamma^2\left( 1 + \Delta_\phi - \rho \right) \Gamma\left( 1 + \mathi \mu \right)} \eqend{,} \\
\mathcal{M}^{(0),\infty,\infty}_{\phi_m(\rho+\mathi \nu) \phi_m(\rho-\mathi \nu) \phi_n(\rho-\mathi \mu)}(\delta_{ij}) &= \tilde{\lambda}_{mmn} \frac{\Gamma\left( \frac{\rho-\mathi \mu}{2} \right)}{\Gamma\left( 1 + \mathi \nu \right) \Gamma\left( 1 - \mathi \nu \right) \Gamma\left( 1 - \mathi \mu \right)} \eqend{,} \\
\mathcal{M}^{(0),\infty,\infty}_{\phi_1 \phi_1 \phi_m(\rho+\mathi \nu) \phi_m(\rho-\mathi \nu)}(\delta_{ij}) &= \tilde{\lambda}_{11mm} \frac{\Gamma\left( \Delta_\phi \right)}{\Gamma^2\left( 1 + \Delta_\phi - \rho \right) \Gamma\left( 1 + \mathi \nu \right) \Gamma\left( 1 - \mathi \nu \right)} \eqend{.}
\end{equations}
Pluggin them into the general results~\eqref{eq:oneloop_c} and~\eqref{eq:oneloop_delta}, we obtain
\begin{splitequation}
\label{eq:oneloop_c_phi}
c^{(1),0,\infty}_{\phi_1 \phi_1} &= c^{(1),\infty,\infty}_{\phi_1 \phi_1} - \tilde{\lambda}_{11mm} \frac{\pi^{\rho-1} \ell^{2\rho-1} \Gamma(\rho) \Gamma\left( \Delta_\phi \right)}{\Gamma(2\rho) \left( \Delta_\phi - \rho \right) \Gamma\left( 1 + \Delta_\phi - \rho \right)} \int_{-\infty}^\infty \frac{1}{\nu^2 + (\Delta_\phi-\rho)^2} \\
&\quad\times \frac{\Gamma(\rho - \mathi \nu) \Gamma(\rho+\mathi \nu)}{\Gamma\left( \mathi \nu \right) \Gamma\left( - \mathi \nu \right)} \Big[ \gamma - \psi\left( \Delta_\phi - \rho \right) + 2 \psi(\rho) - 2 \psi(2\rho) + \psi(\rho+\mathi \nu) + \psi(\rho-\mathi \nu) \Big] \total \nu \\
&+ \frac{\pi^{2\rho-2} \ell^{4\rho-2}}{\Gamma(\rho) \left( \Delta_\phi - \rho \right) \Gamma\left( 1 + \Delta_\phi - \rho \right)} \int_{-\infty}^\infty \int_{-\infty}^\infty \frac{1}{\Gamma\left( \mathi \nu \right) \Gamma\left( - \mathi \nu \right) \Gamma\left( - \mathi \mu \right) \Gamma\left( \mathi \mu \right)} \\
&\quad\times \frac{1}{[ \nu^2 + (\Delta_\phi-\rho)^2 ] [ \mu^2 + (\Delta_\phi-\rho)^2 ]} \\
&\quad\times \Bigg[ 2 \tilde{\lambda}_{1mn} \tilde{\lambda}_{1mn} \Gamma\left( \frac{\Delta_\phi + \mathi (\mu+\nu)}{2} \right) \Gamma\left( \frac{\Delta_\phi - \mathi (\mu+\nu)}{2} \right) \\
&\qquad\times \Gamma\left( \frac{\Delta_\phi + \mathi (\mu-\nu)}{2} \right) \Gamma\left( \frac{\Delta_\phi - \mathi (\mu-\nu)}{2} \right) \\
&\qquad\times \frac{\Gamma\left( \rho - \frac{\Delta_\phi + \mathi (\mu+\nu)}{2} \right) \Gamma\left( \rho - \frac{\Delta_\phi - \mathi (\mu-\nu)}{2} \right) \Gamma\left( \rho - \frac{\Delta_\phi + \mathi (\mu-\nu)}{2} \right) \Gamma\left( \rho - \frac{\Delta_\phi - \mathi (\mu+\nu)}{2} \right)}{\Gamma\left( 2 \rho - \Delta_\phi \right)} \\
&\qquad\times \Bigg[ \psi\left( \rho - \frac{\Delta_\phi + \mathi (\mu+\nu)}{2} \right) + \psi\left( \rho - \frac{\Delta_\phi - \mathi (\mu-\nu)}{2} \right) + \gamma - \psi\left( \Delta_\phi - \rho \right) \\
&\qquad\qquad+ \psi\left( \rho - \frac{\Delta_\phi + \mathi (\mu-\nu)}{2} \right) + \psi\left( \rho - \frac{\Delta_\phi - \mathi (\mu+\nu)}{2} \right) - 2 \psi(2 \rho - \Delta_\phi) \Bigg] \\
&\qquad- \tilde{\lambda}_{11n} \tilde{\lambda}_{mmn} \Gamma\left( \Delta_\phi - \frac{\rho-\mathi \mu}{2} \right) \Gamma\left( \frac{\rho-\mathi \mu}{2} \right) \Gamma\left( \frac{\rho + \mathi \mu}{2} \right) \\
&\qquad\quad\times \frac{\Gamma\left( - \frac{\rho - \mathi \mu}{2} \right) \Gamma\left( - \frac{\rho + \mathi \mu}{2} \right)}{\Gamma(\Delta_\phi - \rho)} \frac{\Gamma^2(\rho) \Gamma(\rho - \mathi \nu) \Gamma(\rho + \mathi \nu)}{\Gamma(2\rho)} \Bigg] \total \mu \total \nu
\end{splitequation}
and by O($N$) invariance the same result for all $\phi_i$, as well as
\begin{splitequation}
\label{eq:oneloop_delta_phi}
\Delta_{\phi_1}^{(1),0,\infty} &= \Delta_{\phi_1}^{(1),\infty,\infty} - \tilde{\lambda}_{11mm} \frac{2 \pi^{2\rho-1} \ell^{4\rho-2} \Gamma(\rho)}{\Gamma(2\rho) ( \Delta_\phi - \rho )} \int_{-\infty}^\infty \frac{1}{\nu^2 + (\Delta_\phi-\rho)^2} \frac{\Gamma(\rho - \mathi \nu) \Gamma(\rho + \mathi \nu)}{\Gamma\left( \mathi \nu \right) \Gamma\left( - \mathi \nu \right)} \total \nu \\
&+ \tilde{\lambda}_{1mn} \tilde{\lambda}_{1mn} \frac{4 \pi^{3\rho-2} \ell^{6\rho-3}}{\Gamma(\rho) ( \Delta_\phi - \rho ) \Gamma(\Delta_\phi) \Gamma\left( 2 \rho - \Delta_\phi \right)} \int_{-\infty}^\infty \int_{-\infty}^\infty \\
&\quad\times \Gamma\left( \frac{\Delta_\phi - \mathi (\mu-\nu)}{2} \right) \Gamma\left( \frac{\Delta_\phi + \mathi (\mu-\nu)}{2} \right) \Gamma\left( \rho - \frac{\Delta_\phi - \mathi (\mu-\nu)}{2} \right) \\
&\quad\times \Gamma\left( \rho - \frac{\Delta_\phi + \mathi (\mu-\nu)}{2} \right) \Gamma\left( \rho - \frac{\Delta_\phi + \mathi (\mu+\nu)}{2} \right) \Gamma\left( \rho - \frac{\Delta_\phi - \mathi (\mu+\nu)}{2} \right) \\
&\quad\times \frac{\Gamma\left( \frac{\Delta_\phi + \mathi (\mu+\nu)}{2} \right)}{\Gamma\left( \mathi \nu \right) \Gamma\left( \mathi \mu \right)} \frac{\Gamma\left( \frac{\Delta_\phi - \mathi (\mu+\nu)}{2} \right)}{\Gamma\left( - \mathi \nu \right) \Gamma\left( - \mathi \mu \right)} \frac{1}{[ \nu^2 + (\Delta_\phi-\rho)^2 ] [ \mu^2 + (\Delta_\phi-\rho)^2 ]} \total \mu \total \nu \eqend{,}
\end{splitequation}
where we used again the first Barnes lemma~\eqref{eq:barneslemma1} and its corollary~\eqref{eq:barneslemmapoly} to perform the integrals of the contact amplitudes over $\delta_{14}$, and the free-theory normalisation factor~\eqref{eq:s_2_free}.

\begin{figure}[h]
  \begin{subfigure}{0.45\textwidth}
    \centering
    \includegraphics[scale=0.5]{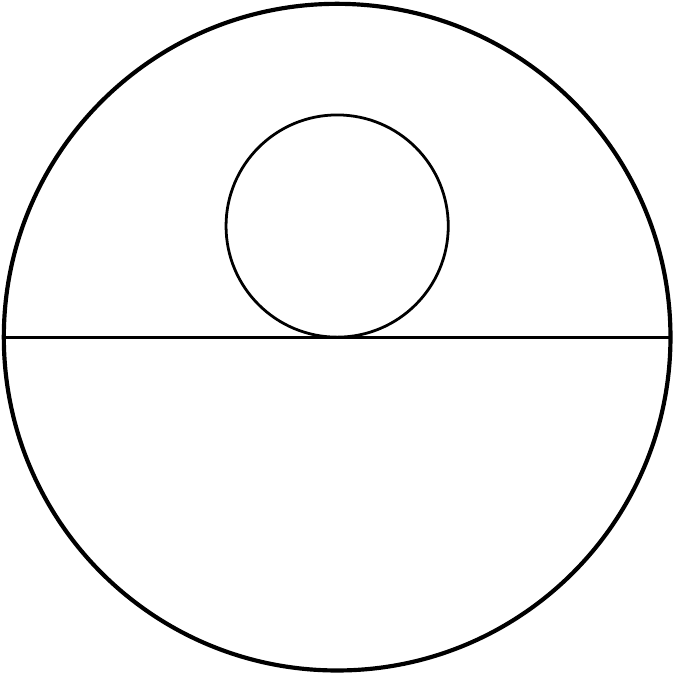}
    \caption{Tadpole Witten diagram for quartic coupling.}
    \label{fig:witten_tadpole}
  \end{subfigure}
  \hfill
  \begin{subfigure}{0.45\textwidth}
    \centering
    \includegraphics[scale=0.5]{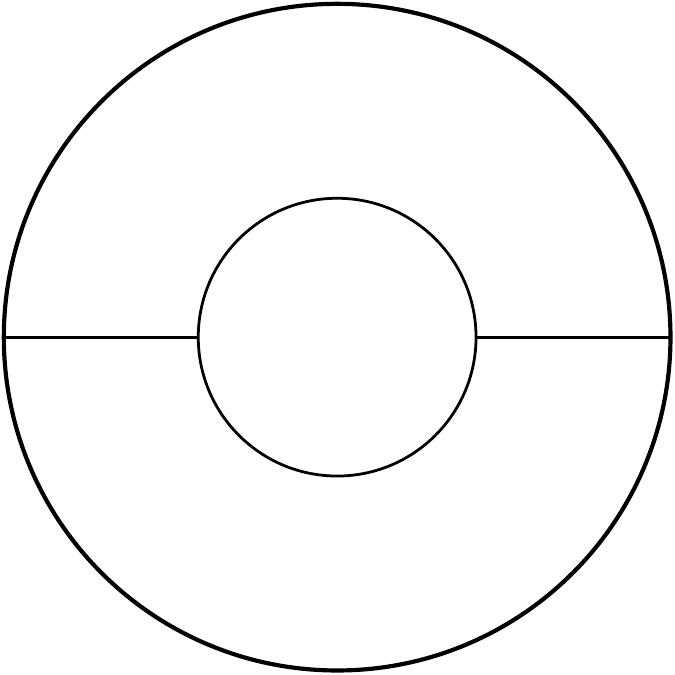}
    \caption{One loop Witten diagram for cubic coupling.}
    \label{fig:witten_loop}
  \end{subfigure}
  \caption{One-loop Witten diagrams contributing to the two-point function.}
\end{figure}

We note various features of the result: the corrections proportional to the quartic coupling $\tilde{\lambda}_{ijkl}$ are clearly divergent for half-integer $\rho$, corresponding to integer dimension of AdS, since for large $\nu$ the integrand grows $\sim \abs{\nu}^{2\rho}$. For the corrections quadratic in the cubic coupling $\tilde{\lambda}_{ijk}$, the integrand decays exponentially in either $\mu$ or $\nu$ if the other variable is held fixed, but grows $\sim \abs{\nu}^{2\rho}$ if $\mu \approx \pm \nu$. These divergences stem from the UV divergences of the AdS field theory, and result from the fact that we have naively taken the limit $\Lambda_0 \to \infty$ of infinite bulk UV cutoff, and thus must work in dimensional regularisation with an analytically continued $\rho$. In fact, the one-loop correction proportional to the quartic coupling results from the tadpole diagram depicted in figure~\ref{fig:witten_tadpole}, which is clearly UV divergent, while the corrections quadratic in the cubic coupling arise from the loop diagram depicted in figure~\ref{fig:witten_loop}. This UV divergence must be cancelled by a counterterm in the boundary term $\Delta_{\phi_1}^{(1),\infty,\infty}$, which is of course possible. However, this makes the loop corrections to the conformal dimension of the operator dual to $\phi_i$ ambiguous, since we can also always add finite counterterms.

The $\nu$ integrals for the contributions proportional to the quartic coupling $\tilde{\lambda}_{ijkl}$ can in fact be evaluated exactly by summing residues in the upper half-plane if the integral converges. This is the case for $\Re \{ a, b \} > 0$ and $\Re (a+b) < 1$, where we compute
\begin{splitequation}
&\int_{-\infty}^\infty \frac{1}{\nu^2 + (\Delta_\phi-\rho)^2} \frac{\Gamma(a + \mathi \nu) \Gamma(b - \mathi \nu)}{\Gamma\left( \mathi \nu \right) \Gamma\left( - \mathi \nu \right)} \total \nu \\
&= - \pi \frac{\Gamma(a + \Delta_\phi-\rho) \Gamma(b - \Delta_\phi+\rho)}{\Gamma\left( 1 + \rho - \Delta_\phi \right) \Gamma\left( \Delta_\phi-\rho \right)} - 2 \sin(b \pi) \sum_{k=0}^\infty \frac{1}{k!} \frac{(b+k) \Gamma(a + b + k)}{- (b+k)^2 + (\Delta_\phi-\rho)^2} \\
&= - \sin[ \pi (\Delta_\phi-\rho) ] \Gamma(a + \Delta_\phi-\rho) \Gamma(b - \Delta_\phi+\rho) \\
&\quad+ \frac{\pi \sin(b \pi)}{\sin[ (a+b) \pi ]} \left[ \frac{\Gamma(b+\Delta_\phi-\rho)}{\Gamma(1-a+\Delta_\phi-\rho)} + \frac{\Gamma(b-\Delta_\phi+\rho)}{\Gamma(1-a-\Delta_\phi+\rho)} \right] \eqend{,}
\end{splitequation}
and thus after analytical continuation $a,b \to \rho$
\begin{splitequation}
&\int_{-\infty}^\infty \frac{1}{\nu^2 + (\Delta_\phi-\rho)^2} \frac{\Gamma(\rho + \mathi \nu) \Gamma(\rho - \mathi \nu)}{\Gamma\left( \mathi \nu \right) \Gamma\left( - \mathi \nu \right)} \total \nu = - \Gamma(\Delta_\phi) \Gamma(2 \rho - \Delta_\phi) \sin[ \pi (\Delta_\phi - \rho) ] \\
&\hspace{6em}+ \frac{\Gamma(\Delta_\phi) \Gamma(2\rho-\Delta_\phi)}{2 \cos( \rho \pi )} \left[ \sin[ \pi (2 \rho - \Delta_\phi) ] + \sin( \pi \Delta_\phi ) \right] \eqend{,}
\end{splitequation}
and after taking a derivative with respect to $a$ or $b$ and then analytically continuing to $a,b \to \rho$
\begin{splitequation}
&\int_{-\infty}^\infty \frac{1}{\nu^2 + (\Delta_\phi-\rho)^2} \frac{\Gamma(\rho + \mathi \nu) \Gamma(\rho - \mathi \nu)}{\Gamma\left( \mathi \nu \right) \Gamma\left( - \mathi \nu \right)} \left[ \psi(\rho + \mathi \nu) + \psi(\rho - \mathi \nu) \right] \total \nu \\
&= \Gamma(\Delta_\phi) \Gamma(2\rho-\Delta_\phi) \left[ \frac{\sin[ \pi (2\rho-\Delta_\phi) ]}{\cos(\pi \rho)} \Big[ \psi(\Delta_\phi) + \psi(2 \rho - \Delta_\phi) \Big] + \frac{\pi}{\cos^2(\pi \rho)} \cos[ \pi (\Delta_\phi - \rho) ] \right] \eqend{.}
\end{splitequation}
We see that there are poles whenever $\rho = k + 1/2$ with $k \in \mathbb{N}_0$, i.e., even-dimensional AdS and an odd-dimensional CFT. Since the counterterms $\Delta_{\phi_1}^{(1),\infty,\infty}$ and $c^{(1),\infty,\infty}_{\phi_1 \phi_1}$ are arbitrary, we can choose them to not only cancel the divergent, but also the finite part, which may be argued to be the most sensible renormalisation scheme.

On the other hand, for an even-dimensional CFT and odd-dimensional AdS with $\rho = k \in \mathbb{N}_0$, we can take $c^{(1),\infty,\infty}_{\phi_1 \phi_1} = 0 = \Delta_{\phi_1}^{(1),\infty,\infty}$ and have the finite results
\begin{splitequation}
c^{(1),0,\infty}_{\phi_1 \phi_1} &= \tilde{\lambda}_{11mm} \frac{\pi^{\rho-1} \ell^{2\rho-1} \Gamma(\rho) \Gamma\left( \Delta_\phi \right)}{\Gamma(2\rho) \left( \Delta_\phi - \rho \right) \Gamma\left( 1 + \Delta_\phi - \rho \right)} \Gamma(\Delta_\phi) \Gamma(2\rho-\Delta_\phi) \Bigg[ - \pi \cos[\pi (\Delta_\phi-\rho)] \\
&\quad+ \sin[\pi (\Delta_\phi-\rho)] \Big[ 2 \gamma - 2 \psi\left( \Delta_\phi - \rho \right) + 4 \psi(\rho) - 4 \psi(2\rho) + \psi(\Delta_\phi) + \psi(2 \rho - \Delta_\phi) \Big] \Bigg] \\
&\quad+ \bigo{\tilde{\lambda}_{ijk}^2}
\end{splitequation}
and
\begin{equation}
\label{eq:one_loop_deltaphi}
\Delta_{\phi_1}^{(1),0,\infty} = \tilde{\lambda}_{11mm} \frac{2 \pi^{2\rho-1} \ell^{4\rho-2} \Gamma(\rho)}{\Gamma(2\rho) ( \Delta_\phi - \rho)} \Gamma(\Delta_\phi) \Gamma(2 \rho - \Delta_\phi) \sin[\pi (\Delta_\phi-\rho)] + \bigo{\tilde{\lambda}_{ijk}^2} \eqend{,}
\end{equation}
where the terms proportional to $\tilde{\lambda}_{ijk}^2$ are the same ones as in equations~\eqref{eq:oneloop_c_phi} and~\eqref{eq:oneloop_delta_phi}, which are finite for any $\rho$. The result~\eqref{eq:one_loop_deltaphi} agrees with \cite[Eq.~(2.59)]{giombisleighttaronna2018} upon taking into account the relative normalisation of the coupling constant~\eqref{eq:contact_amplitude} and using the reflection and duplication formulae for the $\Gamma$ function.

The contributions proportional to the square of the cubic coupling $\tilde{\lambda}_{ijk}$ can be evaluated as an infinite series by summing residues in the upper half-plane. In particular, for $\rho = 1$ the result for the corrections to the conformal dimension~\eqref{eq:oneloop_delta_phi} hugely simplifies and the infinite sums can be evaluated to give $\psi$ functions~\cite{giombisleighttaronna2018,carmidipietrokomatsu2019}. The relevant integral~\eqref{eq:int_loop_rho} is computed in the appendix with the result~\eqref{eq:int_loop_rho1_result} for $\rho = 1$ (taking $\Delta = \Delta_1 = \Delta_2 = \Delta_\phi$), such that the one-loop corrections to the conformal dimension in this case read
\begin{splitequation}
\label{eq:oneloop_delta_phi_rho1}
&\Delta_{\phi_1}^{(1),0,\infty}(\rho = 1) = \tilde{\lambda}_{11mm} 2 \pi^2 \ell^2 + \tilde{\lambda}_{1mn} \tilde{\lambda}_{1mn} \frac{16 \pi^4 \ell^3}{( \Delta_\phi - 1 )^2} \frac{\cos\left( \pi \frac{\Delta_\phi}{2} \right)}{\sin\left( \pi \frac{3 \Delta_\phi}{2} \right)} \\
&\quad+ \tilde{\lambda}_{1mn} \tilde{\lambda}_{1mn} \frac{8 \pi^3 \ell^3}{( \Delta_\phi - 1 )^2} \Bigg[ \psi\left( 2 - \frac{3 \Delta_\phi}{2} \right) - \psi\left( 1 - \frac{\Delta_\phi}{2} \right) - \psi\left( \frac{\Delta_\phi}{2} \right) + \psi\left( \frac{3 \Delta_\phi}{2} - 1 \right) \Bigg] \eqend{,}
\end{splitequation}
which agree with~\cite[Eq.~(2.45)]{giombisleighttaronna2018} after taking into account the relative normalisation of the coupling constant~\eqref{eq:contact_amplitude}.

\subsection{Tree level for the double-trace operator}

For three-point amplitudes involving the composite operator $\phi^2 = \phi_i \phi_i$, the contact amplitude~\eqref{eq:contact_amplitude_compositeboundary} is formally undefined, and we take it to be vanishing. Instead, we consider the limit where two points of the four-point function coincide, which will give the corrections to the three-point OPE coefficient of two $\phi$'s and one operator $\phi^2$. Using the $\mathcal{F}$ factors with analytically continued dimensions~\eqref{eq:f3_tree_analytical}, the four-point amplitude for analytically continued dimensions reads
\begin{splitequation}
\label{eq:mellin_tree_phiphiphiphi}
&\mathcal{M}^{(0),0,\infty}_{\phi_i(D_1) \phi_j(D_2) \phi_k(D_3) \phi_l(D_4)}(\delta_{ij}) = \tilde{\lambda}_{ijkl} \frac{\Gamma\left( \frac{D_1+D_2+D_3+D_4}{2} - \rho \right)}{\prod_{k=1}^4 \Gamma\left( 1 + D_i - \rho \right)} \\
&\quad- \frac{\pi^{\rho-1} \ell^{2 \rho-1}}{\prod_{k=1}^4 \Gamma\left( 1 + D_i - \rho \right)} \int_{-\infty}^\infty \frac{1}{\nu^2 + (\Delta_\phi-\rho)^2} \frac{1}{\Gamma\left( \mathi \nu \right) \Gamma\left( - \mathi \nu \right)} \\
&\qquad\times \bigg[ \tilde{\lambda}_{ijm} \tilde{\lambda}_{klm} \Gamma_{1234}(\delta_{12},\nu) + \tilde{\lambda}_{ikm} \tilde{\lambda}_{jlm} \Gamma_{1324}(\delta_{13},\nu) + \tilde{\lambda}_{ilm} \tilde{\lambda}_{jkm} \Gamma_{1423}(\delta_{14},\nu) \bigg] \total \nu \eqend{,}
\end{splitequation}
with
\begin{splitequation}
\label{eq:mellin_tree_phiphiphiphi_gamma}
\Gamma_{ijkl}(\delta,\nu) &\equiv \Gamma\left( \frac{D_i + D_j - \rho + \mathi \nu}{2} \right) \Gamma\left( \frac{D_i + D_j - \rho - \mathi \nu}{2} \right) \Gamma\left( \frac{D_k + D_l - \rho - \mathi \nu}{2} \right) \\
&\quad\times \Gamma\left( \frac{D_k + D_l - \rho + \mathi \nu}{2} \right) \frac{\Gamma\left( \delta - \frac{D_i + D_j - \rho - \mathi \nu}{2} \right) \Gamma\left( \delta - \frac{D_i + D_j - \rho + \mathi \nu}{2} \right)}{\Gamma(\delta) \Gamma\left( \delta + \frac{D_k + D_l - D_i - D_j}{2} \right)} \eqend{,}
\end{splitequation}
where we took the physical limit $\Lambda \to 0$, $\Lambda_0 \to \infty$ and used the relations~\eqref{eq:deltaij_4_relations}. The four-point function itself is given by
\begin{splitequation}
\label{eq:s_4_tree}
&S^{(0),0,\infty}_{\phi_i(D_1) \phi_j(D_2) \phi_k(D_3) \phi_l(D_4)}(P_1,P_2,P_3,P_4) \\
&= \int \mathcal{M}^{(0),0,\infty}_{\phi_i(D_1) \phi_j(D_2) \phi_k(D_3) \phi_l(D_4)}(\delta_{ij}(s,t)) \Gamma\left( D_{13} + t \right) \Gamma\left( D_{24} + t \right) \Gamma\left( D_{12} - D_{34} + s \right) \\
&\quad\times \Gamma\left( s \right) \Gamma\left( D_{14} + D_{34} - s - t \right) \Gamma\left( D_{23} + D_{34} - s - t \right) \\
&\quad\times \left( - 2 \ell^{-2} P_1 \cdot P_2 \right)^{- \delta_{12}(s,t)} \left( - 2 \ell^{-2} P_1 \cdot P_3 \right)^{- \delta_{13}(s,t)} \left( - 2 \ell^{-2} P_1 \cdot P_4 \right)^{- \delta_{14}(s,t)} \\
&\quad\times \left( - 2 \ell^{-2} P_2 \cdot P_3 \right)^{- \delta_{23}(s,t)} \left( - 2 \ell^{-2} P_2 \cdot P_4 \right)^{- \delta_{24}(s,t)} \left( - 2 \ell^{-2} P_3 \cdot P_4 \right)^{- s} \frac{\total s}{2 \pi \mathi} \frac{\total t}{2 \pi \mathi} \eqend{,}
\end{splitequation}
where we choose the solution
\begin{splitequation}
&\delta_{12}(s,t) = D_{12} - D_{34} + s \eqend{,} \quad \delta_{13}(s,t) = D_{13} + t \eqend{,} \quad \delta_{14}(s,t) = D_{14} + D_{34} - s - t \eqend{,} \\
&\delta_{23}(s,t) = D_{23} + D_{34} - s - t \eqend{,} \quad \delta_{24}(s,t) = D_{24} + t \eqend{,} \quad \delta_{34}(s,t) = s \eqend{,} \\
&D_{ij} = \frac{D_i + D_j}{2} - \frac{D_1+D_2+D_3+D_4}{6}
\end{splitequation}
of the Mellin integration measure constraints~\eqref{eq:mellin_integration_measure}, which are compatible with the relations~\eqref{eq:deltaij_4_relations}.

To obtain the two-point function of $\phi^2$, we set $D_1 = D_2 = D_3 = D_4 = \Delta_\phi$ in the four-point function~\eqref{eq:s_4_tree}, sum over $i$,$j$,$k$,$l$ and compute the finite part as $P_1, P_2 \to P$ and $P_3, P_4 \to Q$. Since for equal dimensions $\delta_{12} = s$, this finite part is quite obviously given by the residue of the integrand at $s = 0$, which reads
\begin{splitequation}
\label{eq:2pf_tree_phi2}
&S^{(0),0,\infty}_{\phi^2 \phi^2}(P,Q) = \mathcal{P}\!f_{P_1, P_2 \to P, P_3, P_4 \to Q} S^{(0),0,\infty}_{\phi_i \phi_i \phi_k \phi_k}(P_1,P_2,P_3,P_4) \\
&\quad= \left( - 2 \ell^{-2} P \cdot Q \right)^{- 2 \Delta_\phi} \int \Gamma^2\left( \frac{\Delta_\phi}{3} + t \right) \Gamma^2\left( \frac{2 \Delta_\phi}{3} - t \right) \Bigg[ \left. \partial_s \mathcal{M}^{(0),0,\infty}_{\phi_i \phi_i \phi_k \phi_k}(\delta_{ij}(s,t)) \right\rvert_{s = 0} \\
&\qquad\quad- \left[ 2 \gamma - 2 \ln \left( - 2 \ell^{-2} P \cdot Q \right) + 2 \psi\left( \frac{2 \Delta_\phi}{3} - t \right) \right] \mathcal{M}^{(0),0,\infty}_{\phi_i \phi_i \phi_k \phi_k}(\delta_{ij}(s=0,t)) \Bigg] \frac{\total t}{2 \pi \mathi} \eqend{,}
\end{splitequation}
such that the corrections to the conformal dimension of $\phi^2$ and the normalisation constant $c_{\phi^2 \phi^2}$ are given by
\begin{splitequation}
\label{eq:delta_deltaphi2}
\delta \Delta_{\phi^2} = - \frac{2}{c_{\phi^2 \phi^2}} \int \Gamma^2\left( \frac{\Delta_\phi}{3} + t \right) \Gamma^2\left( \frac{2 \Delta_\phi}{3} - t \right) \mathcal{M}^{(0),0,\infty}_{\phi_i \phi_i \phi_k \phi_k}(\delta_{ij}(s=0,t)) \frac{\total t}{2 \pi \mathi}
\end{splitequation}
where the free-theory normalisation constant reads~\eqref{eq:free_theory_cphi1}, \eqref{eq:free_theory_cphi2}
\begin{equation}
c_{\phi^2 \phi^2} = \delta_{jj} \frac{\ell^{2-4\rho} \Gamma^2(\Delta_\phi)}{2 \pi^{2\rho} \Gamma^2\left( \Delta_\phi - \rho + 1 \right)} \eqend{,}
\end{equation}
and
\begin{splitequation}
\delta c_{\phi^2 \phi^2} &= \int \Gamma^2\left( \frac{\Delta_\phi}{3} + t \right) \Gamma^2\left( \frac{2 \Delta_\phi}{3} - t \right) \Bigg[ \left. \partial_s \mathcal{M}^{(0),0,\infty}_{\phi_i \phi_i \phi_k \phi_k}(\delta_{ij}(s,t)) \right\rvert_{s = 0} \\
&\qquad\quad- \left[ 2 \gamma + 2 \psi\left( \frac{2 \Delta_\phi}{3} - t \right) \right] \mathcal{M}^{(0),0,\infty}_{\phi_i \phi_i \phi_k \phi_k}(\delta_{ij}(s=0,t)) \frac{\total t}{2 \pi \mathi} \eqend{.}
\end{splitequation}
Both can be calculated using that~\eqref{eq:mellin_tree_phiphiphiphi}
\begin{splitequation}
\label{eq:mellin_tree_phiphiphiphi_summed}
&\mathcal{M}^{(0),0,\infty}_{\phi_i \phi_i \phi_k \phi_k}(\delta_{ij}(s,t)) = \tilde{\lambda}_{iikk} \frac{\Gamma\left( 2 \Delta_\phi - \rho \right)}{\Gamma^4\left( 1 + \Delta_\phi - \rho \right)} \\
&\quad- \frac{\pi^{\rho-1} \ell^{2 \rho-1}}{\Gamma^4\left( 1 + \Delta_\phi - \rho \right)} \int_{-\infty}^\infty \frac{1}{\nu^2 + (\Delta_\phi-\rho)^2} \frac{1}{\Gamma\left( \mathi \nu \right) \Gamma\left( - \mathi \nu \right)} \\
&\qquad\times \bigg[ \tilde{\lambda}_{iim} \tilde{\lambda}_{kkm} \Gamma_{1234}(s,\nu) + \tilde{\lambda}_{ikm} \tilde{\lambda}_{ikm} \Big[ \Gamma_{1234}\left( \frac{\Delta_\phi}{3} + t, \nu \right) + \Gamma_{1234}\left( \frac{2 \Delta_\phi}{3} - s - t, \nu \right) \Big] \bigg] \total \nu \eqend{,} \raisetag{6em}
\end{splitequation}
with~\eqref{eq:mellin_tree_phiphiphiphi_gamma}
\begin{equation}
\Gamma_{1234}(\delta,\nu) = \Gamma^2\left( \Delta_\phi - \frac{\rho + \mathi \nu}{2} \right) \Gamma^2\left( \Delta_\phi - \frac{\rho - \mathi \nu}{2} \right) \frac{\Gamma\left( \delta - \Delta_\phi + \frac{\rho + \mathi \nu}{2} \right) \Gamma\left( \delta - \Delta_\phi + \frac{\rho - \mathi \nu}{2} \right)}{\Gamma^2(\delta)} \eqend{.}
\end{equation}
Note that the term proportional to $\tilde{\lambda}_{iim} \tilde{\lambda}_{kkm}$ in~\eqref{eq:mellin_tree_phiphiphiphi_summed} does not contribute since it and its first derivative vanish at $s = 0$; this is the analogue of the subtracted term in the free-theory OPE~\eqref{eq:free_field_3pf_limit}, which also does not contribute when taking the finite part of the limit. Using Barnes first lemma~\eqref{eq:barneslemma1} and its corollary~\eqref{eq:barneslemmapoly} to perform the integrals over $t$, we obtain
\begin{splitequation}
\delta \Delta_{\phi^2} &= - \frac{4 \pi^{2\rho}}{\ell^{2-4\rho} \Gamma^2\left( 1 + \Delta_\phi - \rho \right) \delta_{jj}} \Bigg[ \tilde{\lambda}_{iikk} \frac{\Gamma\left( 2 \Delta_\phi - \rho \right) \Gamma^2(\Delta_\phi)}{\Gamma(2 \Delta_\phi)} \\
&\qquad\quad- 2 \tilde{\lambda}_{ikm} \tilde{\lambda}_{ikm} \frac{\pi^{\rho-1} \ell^{2 \rho-1}}{\Gamma(\rho) \Gamma^2(\Delta_\phi)} \int_{-\infty}^\infty \frac{1}{\nu^2 + (\Delta_\phi-\rho)^2} \frac{1}{\Gamma\left( \mathi \nu \right) \Gamma\left( - \mathi \nu \right)} \\
&\qquad\qquad\times \Gamma^2\left( \Delta_\phi - \frac{\rho + \mathi \nu}{2} \right) \Gamma^2\left( \Delta_\phi - \frac{\rho - \mathi \nu}{2} \right) \Gamma^2\left( \frac{\rho + \mathi \nu}{2} \right) \Gamma^2\left( \frac{\rho - \mathi \nu}{2} \right) \total \nu \Bigg] \eqend{,}
\end{splitequation}
where the contribution due to the quartic interaction $\tilde{\lambda}_{iikk}$ agrees with~\cite[Eq.~(3.25)]{fitzpatricketal2011}, taking again into account the relative normalisation of the coupling constant~\eqref{eq:contact_amplitude}, and
\begin{splitequation}
\label{eq:delta_c_phi2phi2}
\delta c_{\phi^2 \phi^2} &= - 2 \tilde{\lambda}_{iikk} \frac{\Gamma^4(\Delta_\phi) \Gamma\left( 2 \Delta_\phi - \rho \right)}{\Gamma^4\left( 1 + \Delta_\phi - \rho \right) \Gamma(2 \Delta_\phi)} \Big[ \gamma + 2 \psi(\Delta_\phi) - \psi(2 \Delta_\phi) \Big] \\
&\quad+ \tilde{\lambda}_{ikm} \tilde{\lambda}_{ikm} \frac{4 \pi^{\rho-1} \ell^{2 \rho-1}}{\Gamma(\rho) \Gamma^4\left( 1 + \Delta_\phi - \rho \right)} \int_{-\infty}^\infty \frac{1}{\nu^2 + (\Delta_\phi-\rho)^2} \frac{1}{\Gamma\left( \mathi \nu \right) \Gamma\left( - \mathi \nu \right)} \\
&\qquad\times \Gamma^2\left( \Delta_\phi - \frac{\rho + \mathi \nu}{2} \right) \Gamma^2\left( \Delta_\phi - \frac{\rho - \mathi \nu}{2} \right) \Gamma^2\left( \frac{\rho + \mathi \nu}{2} \right) \Gamma^2\left( \frac{\rho - \mathi \nu}{2} \right) \\
&\qquad\times \left[ \gamma - \psi(\rho) + \psi\left( \frac{\rho + \mathi \nu}{2} \right) + \psi\left( \frac{\rho - \mathi \nu}{2} \right) \right] \total \nu \eqend{.}
\end{splitequation}

To compute the one-loop corrections to the conformal dimension, we also need the three-point function with analytically continued dimensions, which is obtained analogously from the four-point function~\eqref{eq:s_4_tree} by setting $D_3 = D_4 = \Delta_\phi$, summing over $k$ and $l$ and computing the finite part as $P_3, P_4 \to R$, which is again given by the residue of the integrand at $s = 0$. The result can be written in Mellin representation~\eqref{eq:s_n_mellin} with $\Delta_{\phi^2} = 2 \Delta_\phi$ and amplitude
\begin{splitequation}
\label{eq:mellin_tree_phiphiphi2}
&\mathcal{M}^{(0),\Lambda,\Lambda_0}_{\phi_i(D_1) \phi_j(D_2) \phi^2} = \tilde{\lambda}_{ijkk} \frac{\Gamma\left( \frac{D_1+D_2}{2} + \Delta_\phi - \rho \right) \Gamma^2(\Delta_\phi)}{\Gamma\left( 1 + D_1 - \rho \right) \Gamma\left( 1 + D_2 - \rho \right) \Gamma^2\left( 1 + \Delta_\phi - \rho \right) \Gamma(2 \Delta_\phi)} \\
&\quad- \frac{2 \pi^{\rho-1} \ell^{2 \rho-1} \tilde{\lambda}_{ikm} \tilde{\lambda}_{jkm}}{\Gamma\left( 1 + D_1 - \rho \right) \Gamma\left( 1 + D_2 - \rho \right) \Gamma^2\left( 1 + \Delta_\phi - \rho \right)} \\
&\qquad\times \frac{1}{\Gamma\left( \Delta_\phi + \frac{D_1 - D_2}{2} \right) \Gamma\left( \Delta_\phi - \frac{D_1 - D_2}{2} \right) \Gamma\left( \Delta_\phi + \rho - \frac{D_1 + D_2}{2} \right)} \\
&\qquad\times \int_{-\infty}^\infty \frac{1}{\nu^2 + (\Delta_\phi-\rho)^2} \frac{1}{\Gamma\left( \mathi \nu \right) \Gamma\left( - \mathi \nu \right)} \Gamma\left( \frac{\Delta_\phi + D_1 - \rho + \mathi \nu}{2} \right) \Gamma\left( \frac{\Delta_\phi + D_1 - \rho - \mathi \nu}{2} \right) \\
&\qquad\quad\times \Gamma\left( \frac{\Delta_\phi + D_2 - \rho - \mathi \nu}{2} \right) \Gamma\left( \frac{\Delta_\phi + D_2 - \rho + \mathi \nu}{2} \right) \Gamma\left( \frac{\Delta_\phi - D_1 + \rho - \mathi \nu}{2} \right) \\
&\qquad\quad\times \Gamma\left( \frac{\Delta_\phi - D_1 + \rho + \mathi \nu}{2} \right) \Gamma\left( \frac{\Delta_\phi - D_2 + \rho - \mathi \nu}{2} \right) \Gamma\left( \frac{\Delta_\phi - D_2 + \rho + \mathi \nu}{2} \right) \total \nu \eqend{,} \raisetag{12.4em}
\end{splitequation}
where we used the first Barnes lemma~\eqref{eq:barneslemma1} to perform the integral over $t$. Note that while the Mellin amplitude is finite on-shell where $D_1 = D_2 = \Delta_\phi$, the three-point function
\begin{splitequation}
S^{\Lambda,\Lambda_0}_{\phi_i(D_1) \phi_j(D_2) \phi^2}(P_1,P_2,P_3) &= \mathcal{M}^{\Lambda,\Lambda_0}_{\phi_i(D_1) \phi_j(D_2) \phi^2} \\
&\quad\times \Gamma\left( \frac{D_1 + D_2 - 2 \Delta_\phi}{2} \right) \left( - 2 \ell^{-2} P_1 \cdot P_2 \right)^\frac{2 \Delta_\phi - D_1 - D_2}{2} \\
&\quad\times \Gamma\left( \frac{2 \Delta_\phi + D_1 - D_2}{2} \right) \left( - 2 \ell^{-2} P_1 \cdot P_3 \right)^\frac{D_2 - D_1 - 2 \Delta_\phi}{2} \\
&\quad\times \Gamma\left( \frac{2 \Delta_\phi - D_1 + D_2}{2} \right) \left( - 2 \ell^{-2} P_2 \cdot P_3 \right)^\frac{D_1 - D_2 - 2 \Delta_\phi}{2}
\end{splitequation}
itself is not because of the first $\Gamma$ function. For a finite result for the three-point function one would have to first set $D_1 = D_2 = \Delta_\phi$ and afterwards compute the finite part as $P_3, P_4 \to R$, which results in additional logarithms due to the above corrections to the conformal dimension of $\phi^2$.

From the four-point function~\eqref{eq:s_4_tree}, we obtain in this way the three-point function
\begin{splitequation}
&S^{(0),0,\infty}_{\phi_i \phi_j \phi^2}(P_1,P_2,P_3) = \mathcal{P}\!f_{P_4 \to P_3} S^{(0),0,\infty}_{\phi_i \phi_j \phi_k \phi_k}(P_1,P_2,P_3,P_4) \\
&= \left( - 2 \ell^{-2} P_1 \cdot P_3 \right)^{- \Delta_\phi} \left( - 2 \ell^{-2} P_2 \cdot P_3 \right)^{- \Delta_\phi} \int \Gamma^2\left( \frac{\Delta_\phi}{3} + t \right) \Gamma^2\left( \frac{2 \Delta_\phi}{3} - t \right) \\
&\quad\times \Bigg[ \left[ - 2 \gamma - \ln \left( - 2 \ell^{-2} P_1 \cdot P_2 \right) + \ln \left( - 2 \ell^{-2} P_1 \cdot P_3 \right) + \ln \left( - 2 \ell^{-2} P_2 \cdot P_3 \right) - 2 \psi\left( \frac{2 \Delta_\phi}{3} - t \right) \right] \\
&\qquad\times \mathcal{M}^{(0),0,\infty}_{\phi_i \phi_j \phi_k \phi_k}(\delta_{ij}(s=0,t)) + \left. \frac{\partial}{\partial s} \mathcal{M}^{(0),0,\infty}_{\phi_i \phi_j \phi_k \phi_k}(\delta_{ij}(s,t)) \right\rvert_{s=0} \Bigg] \frac{\total t}{2 \pi \mathi} \eqend{,} \raisetag{2.2em}
\end{splitequation}
which has the form of a first-order correction to the three-point function~\eqref{eq:s_3_form} with
\begin{equations}
\begin{split}
\delta c_{\phi_i \phi_j \phi^2} &= \int \Gamma^2\left( \frac{\Delta_\phi}{3} + t \right) \Gamma^2\left( \frac{2 \Delta_\phi}{3} - t \right) \Bigg[ \left[ - 2 \gamma - 2 \psi\left( \frac{2 \Delta_\phi}{3} - t \right) \right] \\
&\quad\times \mathcal{M}^{(0),0,\infty}_{\phi_i \phi_j \phi_k \phi_k}(\delta_{ij}(s=0,t)) + \left. \frac{\partial}{\partial s} \mathcal{M}^{(0),0,\infty}_{\phi_i \phi_j \phi_k \phi_k}(\delta_{ij}(s,t)) \right\rvert_{s=0} \Bigg] \frac{\total t}{2 \pi \mathi} \eqend{,}
\end{split} \\
\delta \Delta_{\phi^2} &= - \frac{2}{c_{\phi_i \phi_j \phi^2}} \int \Gamma^2\left( \frac{\Delta_\phi}{3} + t \right) \Gamma^2\left( \frac{2 \Delta_\phi}{3} - t \right) \mathcal{M}^{(0),0,\infty}_{\phi_i \phi_j \phi_k \phi_k}(\delta_{ij}(s=0,t)) \frac{\total t}{2 \pi \mathi} \eqend{,} \\
\delta \Delta_\phi &= 0 \eqend{,}
\end{equations}
and the correction $\delta \Delta_{\phi^2}$ agrees with the previously derived expression~\eqref{eq:delta_deltaphi2} as it must for consistency. With the explicit form of the four-point on-shell Mellin amplitude~\eqref{eq:mellin_tree_phiphiphiphi} and using the first Barnes lemma~\eqref{eq:barneslemma1} and its corollary~\eqref{eq:barneslemmapoly} to perform the integral over $t$, we obtain
\begin{splitequation}
\label{eq:delta_cphiphiphi2}
\delta c_{\phi_i \phi_j \phi^2} &= - 2 \tilde{\lambda}_{ijkk} \frac{\Gamma^4(\Delta_\phi) \Gamma\left( 2 \Delta_\phi - \rho \right)}{\Gamma^4\left( 1 + \Delta_\phi - \rho \right) \Gamma(2 \Delta_\phi)} \Big[ \gamma + 2 \psi(\Delta_\phi) - \psi(2 \Delta_\phi) \Big] \\
&\quad+ \tilde{\lambda}_{ikm} \tilde{\lambda}_{jkm} \frac{4 \pi^{\rho-1} \ell^{2 \rho-1}}{\Gamma(\rho) \Gamma^4\left( 1 + \Delta_\phi - \rho \right)} \int_{-\infty}^\infty \frac{1}{\nu^2 + (\Delta_\phi-\rho)^2} \frac{1}{\Gamma\left( \mathi \nu \right) \Gamma\left( - \mathi \nu \right)} \\
&\qquad\times \Gamma^2\left( \Delta_\phi - \frac{\rho + \mathi \nu}{2} \right) \Gamma^2\left( \Delta_\phi - \frac{\rho - \mathi \nu}{2} \right) \Gamma^2\left( \frac{\rho + \mathi \nu}{2} \right) \Gamma^2\left( \frac{\rho - \mathi \nu}{2} \right) \\
&\qquad\times \left[ \gamma - \psi(\rho) + \psi\left( \frac{\rho + \mathi \nu}{2} \right) + \psi\left( \frac{\rho - \mathi \nu}{2} \right) \right] \total \nu \eqend{.}
\end{splitequation}
The tree-level correction to the normalised three-point correlator~\eqref{eq:free_theory_gammaphiphiphi2} in the free theory is then obtained as
\begin{splitequation}
\label{eq:delta_gamma_phiphiphi2}
\delta \gamma_{\phi_i \phi_j \phi^2} &= \frac{1}{\sqrt{ c_{\phi\phi}^2 c_{\phi^2 \phi^2} }} \delta c_{\phi_i \phi_j \phi^2} - \frac{\gamma_{\phi_i \phi_j \phi^2}}{2 c_{\phi^2 \phi^2}} \delta c_{\phi^2 \phi^2} \\
&= \frac{4 \pi^{2\rho} \Gamma^2\left( \Delta_\phi - \rho + 1 \right)}{\ell^{2-4\rho} \Gamma^2(\Delta_\phi) \sqrt{ 2 \delta_{kk} }} \left[ \delta c_{\phi_i \phi_j \phi^2} - \frac{\delta_{ij}}{2 \delta_{kk}} \delta c_{\phi^2 \phi^2} \right] \eqend{,}
\end{splitequation}
where the correction $\delta c_{\phi^2 \phi^2}$ is given by~\eqref{eq:delta_c_phi2phi2} and the correction $\delta c_{\phi_i \phi_j \phi^2}$ by~\eqref{eq:delta_cphiphiphi2}, and where we took into account that to this order $\delta c_{\phi\phi} = 0$, i.e., that there are no tree-level corrections to the normalisation of the operator dual to $\phi_i$. Comparing~\eqref{eq:delta_cphiphiphi2} and~\eqref{eq:delta_c_phi2phi2}, we note that $\delta c_{\phi^2 \phi^2} = \delta_{ij} \delta c_{\phi_i \phi_j \phi^2}$, such that the correction~\eqref{eq:delta_gamma_phiphiphi2} is proportional to
\begin{equation}
\left( \delta_{ik} \delta_{jl} - \frac{\delta_{ij} \delta_{kl}}{2 \delta_{mm}} \right) \delta c_{\phi_k \phi_l \phi^2} \eqend{,}
\end{equation}
and depending on the couplings $\tilde{\lambda}_{ijkl}$ and $\tilde{\lambda}_{ijk}$ could even vanish.

\subsection{One loop for the double-trace operator}

Let us determine the one-loop correction to the conformal dimension of $\phi^2$. For this, we need the tree-level four-point correlator of two $\phi^2$ and two single-trace operators $\phi$ with analytically continued dimensions, which we can obtain again from the tree-level flow equation~\eqref{eq:s_n_mellin_flow_exact_tree} for the Mellin amplitude. To keep the computation manageable, we only consider quartic coupling, i.e., set $\tilde{\lambda}_{ijk} = 0$ in the following. Since the three-point functions of two $\phi^2$ and one $\phi$ then vanish, on the right-hand side of the flow equation we only have $\mathcal{F}$ factors~\eqref{eq:mellin_factor_tree} corresponding to three-point Mellin amplitudes of one $\phi^2$ and two $\phi$, which are in turn given by~\eqref{eq:mellin_tree_phiphiphi2}. For the three-point $\mathcal{F}$ factors, we use the result~\eqref{eq:f3_constraints_tree}, \eqref{eq:f3_tree} to obtain
\begin{splitequation}
\label{eq:f3_phi2phiphi}
\mathcal{F}^{(0),\Lambda,\Lambda_0}_{\phi^2 \phi_i(D_1) \phi_m(\rho + \mathi \nu)}(\delta) &= \tilde{\lambda}_{imkk} \frac{\Gamma\left( \Delta_\phi + \frac{D_1-\rho+\mathi \nu}{2} \right) \Gamma\left( \Delta_\phi + \frac{D_1 - \rho - \mathi \nu}{2} \right) \Gamma^2(\Delta_\phi)}{\Gamma\left( 1 + D_1 - \rho \right) \Gamma\left( 1 + \mathi \nu \right) \Gamma^2\left( 1 + \Delta_\phi - \rho \right) \Gamma(2 \Delta_\phi)} \\
&\quad\times \frac{\Gamma\left( \delta - \Delta_\phi - \frac{D_1 - \rho - \mathi \nu}{2} \right)}{\Gamma(\delta)} \eqend{,}
\end{splitequation}
and plugging these into the flow equation~\eqref{eq:s_n_mellin_flow_exact_tree}, integrating over $\Lambda$ and taking the physical limit $\Lambda \to 0$, $\Lambda_0 \to \infty$ it follows that
\begin{splitequation}
\label{eq:mellin_tree_phiphiphi2phi2}
&\mathcal{M}^{(0),0,\infty}_{\phi^2 \phi^2 \phi_i(D_1) \phi_j(D_2)}(\delta_{ij}) = \frac{\tilde{\lambda}_{imkk} \tilde{\lambda}_{jmkk} \pi^{\rho-1} \ell^{2\rho-1} \Gamma^4\left( \frac{\Delta_{\phi^2}}{2} \right)}{2 \Gamma^2(\Delta_{\phi^2}) \Gamma^4\left( 1 + \frac{\Delta_{\phi^2}}{2} - \rho \right) \Gamma\left( 1 + D_1 - \rho \right) \Gamma\left( 1 + D_2 - \rho \right)} \\
&\quad\times \int_{-\infty}^\infty \frac{\Gamma\left( \frac{\Delta_{\phi^2} + D_1-\rho+\mathi \nu}{2} \right) \Gamma\left( \frac{\Delta_{\phi^2} + D_1 - \rho - \mathi \nu}{2} \right) \Gamma\left( \frac{\Delta_{\phi^2} + D_2-\rho-\mathi \nu}{2} \right) \Gamma\left( \frac{\Delta_{\phi^2} + D_2 - \rho + \mathi \nu}{2} \right)}{\Big[ \nu^2 + (\Delta_\phi-\rho)^2 \Big] \Gamma\left( \mathi \nu \right) \Gamma\left( - \mathi \nu \right)} \\
&\qquad\times \Bigg[ \frac{\Gamma\left( \delta_{13} - \frac{\Delta_{\phi^2} + D_1 - \rho - \mathi \nu}{2} \right)}{\Gamma(\delta_{13})} \frac{\Gamma\left( \delta_{24} - \frac{\Delta_{\phi^2} + D_2 - \rho + \mathi \nu}{2} \right)}{\Gamma(\delta_{24})} \\
&\qquad\qquad+ \frac{\Gamma\left( \delta_{14} - \frac{\Delta_{\phi^2} + D_1 - \rho - \mathi \nu}{2} \right)}{\Gamma(\delta_{14})} \frac{\Gamma\left( \delta_{23} - \frac{\Delta_{\phi^2} + D_2 - \rho + \mathi \nu}{2} \right)}{\Gamma(\delta_{23})} + ( \nu \leftrightarrow - \nu ) \Bigg] \total \nu \eqend{.}
\end{splitequation}
The further procedure is by now familiar: we insert this amplitude into the flow equation for the correction to the conformal dimension~\eqref{eq:2pf_flow_delta}, with the $\delta_{ij}$ in the Mellin amplitude given by~\eqref{eq:integral_boundary2_finitepart_deltaij} with $\Delta = 0$ and $\Delta_1 = \Delta_{\phi^2}$, since restricting to one loop amounts to taking tree-level amplitudes and tree-level conformal data on the right-hand side. Integrating over $\Lambda$ and using the first Barnes lemma~\eqref{eq:barneslemma1} to perform the integral over $\delta_{14}$, this gives
\begin{splitequation}
\Delta_{\phi^2}^{(1),0,\infty} &= \Delta_{\phi^2}^{(1),\infty,\infty} - \frac{2 \pi^{2\rho-2} \ell^{4\rho-2} \Gamma\left( \Delta_{\phi^2} - \rho \right)}{c_{\phi^2 \phi^2} \Gamma(\rho) \Gamma(2 \rho - \Delta_{\phi^2})} \frac{\tilde{\lambda}_{mnkk} \tilde{\lambda}_{mnkk} \Gamma^4\left( \frac{\Delta_{\phi^2}}{2} \right)}{\Gamma^2(\Delta_{\phi^2}) \Gamma^4\left( 1 + \frac{\Delta_{\phi^2}}{2} - \rho \right)} \\
&\quad\times \int_{-\infty}^\infty \int_{-\infty}^\infty \frac{1}{\nu^2 + (\Delta_\phi-\rho)^2} \frac{1}{\Gamma\left( \mathi \nu \right) \Gamma\left( - \mathi \nu \right)} \frac{1}{\mu^2 + (\Delta_\phi-\rho)^2} \frac{1}{\Gamma\left( \mathi \mu \right) \Gamma\left( - \mathi \mu \right)} \\
&\qquad\times \Gamma\left( \frac{\Delta_{\phi^2} + \mathi (\mu+\nu)}{2} \right) \Gamma\left( \frac{\Delta_{\phi^2} - \mathi (\mu+\nu)}{2} \right) \Gamma\left( \frac{\Delta_{\phi^2} + \mathi (\mu-\nu)}{2} \right) \\
&\qquad\times \Gamma\left( \frac{\Delta_{\phi^2} - \mathi (\mu-\nu)}{2} \right) \Gamma\left( \rho - \frac{\Delta_{\phi^2} - \mathi (\mu+\nu)}{2} \right) \Gamma\left( \rho - \frac{\Delta_{\phi^2} + \mathi (\mu+\nu)}{2} \right) \\
&\qquad\times \Gamma\left( \rho - \frac{\Delta_{\phi^2} - \mathi (\mu-\nu)}{2} \right) \Gamma\left( \rho - \frac{\Delta_{\phi^2} + \mathi (\mu-\nu)}{2} \right) \total \mu \total \nu \eqend{,} \raisetag{2.2em}
\end{splitequation}
where $\Delta_{\phi^2}^{(1),\infty,\infty}$ is the boundary term of the flow that serves as a counterterm (if required). The integral over $\mu$ and $\nu$ is of the same form as the one we encountered for the single-trace operator~\eqref{eq:oneloop_delta_phi}, except that now the dimension $\Delta_{\phi^2}$ of the external operator $\phi^2$ is different from the dimension $\Delta_\phi$ of the internal one $\phi$. The most important consequence of this is that the integration contour for $\mu$ and $\nu$ cannot be taken to run along the real axis, but must be indented in the complex plane such that the poles of the $\Gamma$ functions with $\pm \mathi \mu$ lie above/below the contour~\cite[Sec.~2.2]{giombisleighttaronna2018}. This gives the proper analytic continuation of the integral (as function of the operator dimensions), but makes numerical evaluation difficult. For $\rho = 1$, we can use the explicit result~\eqref{eq:int_loop_rho1_result}, but since the Witten diagram to which this correction corresponds is UV divergent for $\rho = 1$, the result~\eqref{eq:int_loop_rho1_result} has a pole for $\Delta_{\phi^2} = 2 \Delta_\phi$. Expanding around this pole, the result is given by~\eqref{eq:int_loop_rho1_result2}, such that
\begin{equation}
\label{eq:delta_phi2_corr}
\Delta_{\phi^2}^{(1),0,\infty}(\rho = 1) = \Delta_{\phi^2}^{(1),\infty,\infty} + \frac{8 \pi^2 \ell^2 \tilde{\lambda}_{mnkk} \tilde{\lambda}_{mnkk}}{c_{\phi^2 \phi^2} \left( \Delta_{\phi^2} - 1 \right)^2} \left[ \frac{2}{\Delta_{\phi^2} - 2 \Delta_\phi} - \gamma - \psi\left( \Delta_{\phi^2} - 1 \right) \right] \eqend{,}
\end{equation}
and the divergent term can be subtracted by the counterterm $\Delta_{\phi^2}^{(1),\infty,\infty}$. In~\cite{carmipenedonessilvazhiboedov2021}, loop corrections to the conformal dimension of the set of double-trace operators $\op_{n,\ell}$ were computed; these operators are schematically given by $\phi \left( \partial^2 \right)^n \partial_{\mu_1} \cdots \partial_{\mu_\ell} \phi$ (up to the addition of descendants and the subtraction of traces). However, \cite{carmipenedonessilvazhiboedov2021} only includes operators of spin $\ell \geq 2$, and so we cannot compare their result with~\eqref{eq:delta_phi2_corr}, which is $\op_{0,0}$ and has spin $\ell = 0$.

\section{Discussion}
\label{sec:conclusion}

We have shown how to apply the renormalisation group flow equations to weakly coupled quantum field theories in AdS, and how to derive flow equations for boundary correlators using the split representation of AdS propagators. From this, we derived the flow for the conformal dimensions of operators in the dual CFT, as well as the flow of Mellin amplitudes for higher $n$-point functions. We have then computed the explicit solution of the flow at tree level and one loop for low-order $n$-point functions in full generality, and exemplified the resulting formulas with an $\mathrm{O}(N)$ scalar field model in AdS.

In contrast to the holographic renormalisation group~\cite{deboerverlindeverlinde2000,deboer2000,li2000,heemskerkpolchinski2011,sathiapalansonoda2017}, in which a radial cutoff in the bulk corresponds to a UV cutoff in the CFT which destroys conformal invariance, the boundary correlators are conformally invariant throughout the flow since the bulk propagator with UV and IR cutoffs $\Lambda_0$ and $\Lambda$ is AdS invariant. Nevertheless, to remove UV divergences arising in loop corrections bulk counterterms are still necessary, which arise from the boundary condition of the flow at $\Lambda = \Lambda_0$, and which are in one-to-one correspondence with contact interactions. In Minkowski space, instead of setting the boundary conditions of the flow at the UV scale $\Lambda = \Lambda_0$ in such a way that the UV divergences from loop corrections are canceled, one can also set physical boundary conditions at $\Lambda = 0$ for relevant and marginal operators~\cite{kellerkoppersalmhofer1992} and integrate the flow upwards, which ensures automatic cancellation of the UV divergences. Applied to our case, this would for example mean that one fixes the mass of the bulk AdS scalar field to its tree level value, and therefore automatically cancels all loop corrections to the leading (AdS) boundary behaviour of the two-point function and thus the anomalous dimension of the CFT operator dual to the bulk scalar. Depending on the specific theory, this might or might not be a suitable renormalisation condition. Of course this is just a manifestation of the general issue of scheme dependence of anomalous dimensions, which can only be restricted if the bulk theory has additional symmetries. For the example computations here, we have simply used a variant of analytic regularisation, taking the finite part of the result (and discarding the pole term) as the regulator vanishes. Similar analytic regularisation methods have been used in~\cite{giombisleighttaronna2018}; other methods include an AdS-invariant UV regularisation~\cite{bertansachs2018,bertansachsskvortsov2019}, replacing the chordal distance in the propagator by a non-vanishing function of it.

Once a suitable renormalisation scheme has been chosen, one can study the system of flow equations for the $n$-point functions~\eqref{eq:s_n_flow} or the Mellin amplitudes~\eqref{eq:s_n_mellin_flow_exact} further. In particular, it would be very interesting to use the OPE to relate higher $n$-point functions (or their Mellin amplitudes) to the three-point function or amplitude, and obtain in this way a closed (infinite-dimensional) system for the conformal data of primary operators in the CFT. As explained in the introduction, such a system has been previously derived by Hollands~\cite{hollands2018} for the change of conformal data under a strictly marginal perturbation of the CFT (see also~\cite{behan2018}). However, the natural boundary conditions for this system (a free-field CFT) are degenerate and prevent a numerical solution. On the other hand, the system that one could derive from the flow studied in this article governs large-$N$ corrections in the CFT, which a priori do not need to be obtained from a marginal perturbation, and moreover can have different and possibly non-degenerate boundary conditions.

Lastly, to effectively determine higher loop corrections to the conformal data for higher-twist double-trace operators $\op_{n,\ell}$, one should use a more effective method than explicitly taking derivatives of the four-point function~\eqref{eq:s_4_tree}, followed by the limit where two points coincide, as we did in this article. Of course, it is well known that for the tree-level corrections one can read off the anomalous conformal dimensions by performing the integral over $s$ in the four-point function~\eqref{eq:s_4_tree} as a sum over residues, with each residue corresponding to the contribution of an operator $\op_{n,0}$. However, in the general case it might be more useful to use methods such as conglomeration~\cite{fitzpatrickkaplan2012} or dispersive sum rules~\cite{carmipenedonessilvazhiboedov2021}. As for the connection with the OPE, we leave this and the extension of the present framework to spinning operators to future work.

\begin{acknowledgments}
This work has been funded by the Deutsche Forschungsgemeinschaft (DFG, German Research Foundation) --- project nos. 415803368 and 406116891 within the Research Training Group RTG 2522/1. I thank Martin Ammon and Katharina Wölfl for discussions, António Antunes and Sylvain Fichet for references, and the anonymous referee for useful comments.
\end{acknowledgments}

\appendix

\section{Useful integrals}
\label{sec:appendix}

We determine various integrals that are needed in the paper, integrated over either EAdS or the boundary. The first is an integral appearing in the boundary terms for the flow, which is easily solved in Poincar{\'e} coordinates using the Symanzik result~\cite{symanzik1972}. Since the integral has been computed many times~\cite{penedones2011}, we only give the result (valid for $m \geq 3$):
\begin{splitequation}
\label{eq:integral_symanzik}
&\int \left[ \prod_{k=1}^m \Gamma(\mu_k) \left( - 2 \ell^{-2} P_k \cdot X \right)^{-\mu_k} \right] \total X \\
&\quad= \ell^{d+1} \frac{\pi^\rho}{2} \Gamma\left( \sum_{k=1}^m \frac{\mu_k}{2} - \rho \right) \int \dotsi \int \left[ \prod_{k=1}^m \prod_{l=i+1}^m \Gamma\left( \delta_{kl} \right) \left( - 2 \ell^{-2} P_k \cdot P_l \right)^{- \delta_{kl}} \right] [ \total \delta_{kl} ] \eqend{,}
\end{splitequation}
where the integration measure $[ \total \delta_{kl} ]$ is defined by
\begin{equation}
\label{eq:mellin_integration_measure}
\delta_{kl} = \delta^0_{kl} + \sum_{i=1}^{m(m-3)/2} M_{kl,i} s_i \eqend{,} \quad \left[ \total \delta_{kl} \right] = \prod_{i=1}^{m(m-3)/2} \frac{\total s_i}{2 \pi \mathi} \eqend{,}
\end{equation}
where $\delta^0_{jk}$ is a solution with positive real part of $\sum_{l=1}^{k-1} \delta^0_{lk} + \sum_{l=k+1}^m \delta^0_{kl} = \mu_k$, and where $M_{kl,i}$ fulfills the conditions $\sum_{l=1}^{k-1} M_{lk,i} + \sum_{l=k+1}^m M_{kl,i} = 0$ (such that also $\sum_{l=1}^{k-1} \delta_{lk} + \sum_{l=k+1}^m \delta_{kl} = \mu_k$) and $\abs{ \det M_{kl,i} } = 1$, with only the $M_{2l,i}$, $M_{3l,i}$ and $M_{kl,i}$ for $k,l \geq 4$ being taken into account in the determinant. The integral is absolutely convergent under the conditions $\Re \mu_i < \rho + 1/2$ for all $i$ and $\sum_{k=1}^m \Re \mu_k > \rho + 1/2$, and the integration path is the standard Mellin contour in this case (i.e., such that all arguments of the $\Gamma$ functions have positive real part). Otherwise, the integral is defined by analytic continuation.

For $m = 2$, the corresponding integral is IR-divergent, and we must introduce a cutoff and compute its finite part as the cutoff is removed. Consider thus
\begin{equation}
\label{eq:integral_bulk2_def}
I_{\mu\nu}(P,Q) \equiv \Pf_{\epsilon \to 0} \int \Theta\left( z - \frac{\epsilon}{M} \right) \Gamma(\mu) \left( - 2 \ell^{-2} P \cdot X \right)^{-\mu} \Gamma(\nu) \left( - 2 \ell^{-2} Q \cdot X \right)^{-\nu} \total X \eqend{,}
\end{equation}
where we put a cutoff at $z = \epsilon/M$ in the Poincar{\'e} coordinates~\eqref{eq:coordinates_poincare}, anticipating that $M$ will serve as a renormalisation scale that is determined later. Using the standard integral (valid for $\Re \mu > 0$, $\Re A > 0$)~\cite{dlmf}
\begin{equation}
\label{eq:masterintegral_power}
\Gamma(\mu) A^{-\mu} = \int_0^\infty s^{\mu-1} \mathe^{-A s} \total s \eqend{,}
\end{equation}
we obtain
\begin{splitequation}
I_{\mu\nu}(P,Q) &= \Pf_{\epsilon \to 0} \int \Theta\left( z - \frac{\epsilon}{M} \right) \iint_0^\infty s^{\mu-1} t^{\nu-1} \exp\left[ 2 s \ell^{-2} P \cdot X + 2 t \ell^{-2} Q \cdot X \right] \total s \total t \total X \\
&= \Pf_{\epsilon \to 0} \int \Theta\left( z - \frac{\epsilon}{M} \right) \iint_0^\infty s^{\mu-1} t^{\nu-1} \exp\left[ - (s+t) \frac{z}{\ell} - \frac{s}{\ell z} ( \vec{x} - \vec{p} )^2 - \frac{t}{\ell z} ( \vec{x} - \vec{q} )^2 \right] \\
&\qquad\times \total s \total t \left( \frac{\ell}{z} \right)^{d+1} \total z \total^d \vec{x} \eqend{.}
\end{splitequation}
Shifting $\vec{x} \to \vec{x} + (s \vec{p} + t \vec{q})/(s+t)$, the Gaussian integral over $\vec{x}$ is immediate, and after the rescaling $s \to s t$ also the $t$ integral can be easily done and results in
\begin{splitequation}
I_{\mu\nu}(P,Q) &= \pi^\rho \ell^{3\rho+1} \Gamma(\mu+\nu-\rho) \Pf_{\epsilon \to 0} \int \Theta\left( z - \frac{\epsilon}{M} \right) \int_0^\infty s^{\mu-1} (1+s)^{-\rho} \\
&\hspace{8em}\times \left[ (1+s) \frac{z}{\ell} + \frac{s}{\ell z (1+s)} ( \vec{p} - \vec{q} )^2 \right]^{\rho-\mu-\nu} z^{-\rho-1} \total s \total z \eqend{,}
\end{splitequation}
where for convergence we need to assume that $\Re(\mu+\nu-\rho) > 0$. To disentangle the $s$ and $z$ integrations, we use the Mellin representation (valid for $A,B,\Re s > 0$)
\begin{equation}
\label{eq:sum_mellin_representation}
(A+B)^{-s} = \int_{0 < \Re t < \Re s} \frac{\Gamma(s-t) \Gamma(t)}{\Gamma(s)} A^{-t} B^{t-s} \frac{\total t}{2 \pi \mathi}
\end{equation}
for the factor in brackets, after which the integral over $s$ results in $\Gamma$ functions, and the one over $z$ in a power of $\epsilon$:
\begin{splitequation}
I_{\mu\nu}(P,Q) &= \pi^\rho \ell^{2\rho+\mu+\nu+1} \Pf_{\epsilon \to 0} \int_{0 < \Re t < \min\left[ \Re \mu, \Re \nu, \Re (\mu+\nu-\rho) \right]} [ ( \vec{p} - \vec{q} )^2 ]^{-t} \\
&\hspace{8em}\times \frac{\Gamma(\mu-t) \Gamma(\nu-t) \Gamma(\mu+\nu-\rho-t) \Gamma(t)}{\Gamma(\mu+\nu+1-2t)} \left( \frac{\epsilon}{M} \right)^{2t-\mu-\nu} \frac{\total t}{2 \pi \mathi} \eqend{,}
\end{splitequation}
where for convergence of the integrals we needed to restrict the $t$ integration contour. Since this contour is in fact the standard Mellin--Barnes one, where the arguments of all $\Gamma$ functions have positive real part, one could easily evaluate the integral exactly in terms of hypergeometric functions. However, this is not necessary since it is clear that the finite part of the integral as $\epsilon \to 0$ is given by the residue of the integrand at $t = (\mu+\nu)/2$. Taking w.l.o.g. $\Re \mu \geq \Re \nu$, we thus obtain a non-vanishing result in two cases: one needs either $\Re (\mu-\nu) = 2 n$ with $n \in \mathbb{N}_0$ or $\Re (\mu+\nu) = 2(\rho-n)$ with $n \in \mathbb{N}_0$, $n < \rho/2$ (since we had assumed that $\Re(\mu+\nu-\rho) > 0$). Since we need this integral with $\mu$ and $\nu$ being conformal dimensions, which we assume to be greater than $\rho$, the second case is not relevant. For $\mu = \nu$, the integrand has a double pole at $t = \mu$, and we obtain
\begin{splitequation}
\label{eq:integral_bulk2_result_0}
I_{\mu\mu}(P,Q) &= \pi^\rho \ell^{2\rho+2\mu+1} [ ( \vec{p} - \vec{q} )^2 ]^{-\mu} \Gamma(\mu) \Gamma(\mu-\rho) \left[ \ln ( \vec{p} - \vec{q} )^2 - \psi(\mu) + \psi(\mu-\rho) \right] \\
&= \pi^\rho \ell^{2\rho+1} \Gamma(\mu) \Gamma(\mu-\rho) \left( - 2 \ell^{-2} P \cdot Q \right)^{-\mu} \\
&\quad\times \left[ \ln \left( - 2 \ell^{-2} P \cdot Q \right) + 2 \ln M - \psi(\mu) + \psi(\mu-\rho) \right] \eqend{,}
\end{splitequation}
where $\psi(z) \equiv \Gamma'(z)/\Gamma(z)$ is the logarithmic derivative of the $\Gamma$ function. For $\mu = \nu + 2 n$ with $n \geq 1$, we have instead a simple pole at $t = \nu + n$ and obtain
\begin{equation}
\label{eq:integral_bulk2_result_n}
I_{\nu+2n,\nu}(P,Q) = \frac{(-1)^n}{n} \pi^\rho \ell^{2\rho+1} \left( - 2 \ell^{-2} P \cdot Q \right)^{-\nu-n} \Gamma(\nu+n-\rho) \Gamma(\nu+n) \eqend{.}
\end{equation}

We also need integrals where points are integrated only over the boundary. For the flow of the one-point function, we need
\begin{equation}
\label{eq:integral_boundary1_finitepart_def}
I_{\mu\nu\lambda}(P) \equiv \int \Pf_{Q' \to Q} \left( - 2 \ell^{-2} P \cdot Q \right)^{-\mu} \left( - 2 \ell^{-2} P \cdot Q' \right)^{-\nu} \left( - 2 \ell^{-2} Q \cdot Q' \right)^{-\lambda} \total Q \eqend{.}
\end{equation}
Let us define $Q'$ as the embedding of the boundary vector $\vec{q} + \vec{k}$, such that we need the finite part as $\vec{k} \to 0$. Shifting $\vec{q} \to \vec{q} + \vec{p}$, we compute
\begin{splitequation}
I_{\mu\nu\lambda}(P) &= \ell^{2(\mu+\nu+\lambda)} \Pf_{\vec{k} \to 0} \left[ ( \vec{k}^2 )^{-\lambda} \int ( \vec{q}^2 )^{-\mu} \left[ (\vec{q} + \vec{k})^2 \right]^{-\nu} \total^d \vec{q} \right] \\
&= \ell^{2(\mu+\nu+\lambda)} \Pf_{\vec{k} \to 0} \left[ \frac{( \vec{k}^2 )^{-\lambda}}{\Gamma(\mu) \Gamma(\nu)} \int \int_0^\infty \int_0^\infty s^{\mu-1} t^{\nu-1} \exp\left[ - \vec{q}^2 s - (\vec{q} + \vec{k})^2 t \right] \total s \total t \total^d \vec{q} \right]
\end{splitequation}
using again the standard integral~\eqref{eq:masterintegral_power}. Shifting $\vec{q} \to \vec{q} - \vec{k} \, t/(s+t)$ and performing the Gaussian integral over $\vec{q}$, we obtain
\begin{equation}
I_{\mu\nu\lambda}(P) = \pi^\rho \ell^{2(\mu+\nu+\lambda)} \Pf_{\vec{k} \to 0} \left[ \frac{( \vec{k}^2 )^{-\lambda}}{\Gamma(\mu) \Gamma(\nu)} \int_0^\infty \int_0^\infty s^{\mu-1} t^{\nu-1} (s+t)^{-\rho} \exp\left( - \frac{s t}{s+t} \vec{k}^2 \right) \total s \total t \right] \eqend{.}
\end{equation}
The rescaling $s \to s t$ allows to perform the integral over $t$ with the result
\begin{equation}
I_{\mu\nu\lambda}(P) = \pi^\rho \ell^{2(\mu+\nu+\lambda)} \Pf_{\vec{k} \to 0} \left[ \frac{( \vec{k}^2 )^{-\lambda-\mu-\nu+\rho}}{\Gamma(\mu) \Gamma(\nu)} \int_0^\infty s^{-\nu+\rho-1} (1+s)^{\mu+\nu-2\rho} \total s \right] \eqend{,}
\end{equation}
valid for $\Re(\mu+\nu-\rho) > 0$. We see that a non-vanishing finite part exists only for $\lambda+\mu+\nu = \rho$, which is given by
\begin{equation}
\label{eq:integral_boundary1_finitepart_result}
I_{\mu\nu,\rho-\mu-\nu}(P) = \pi^\rho \ell^d \frac{\Gamma(\rho-\mu) \Gamma(\rho-\nu)}{\Gamma(\mu) \Gamma(\nu) \Gamma(2\rho-\mu-\nu)} \eqend{,}
\end{equation}
and the convergence conditions (also for the integration over $s$) can be rephrased as the statement that the argument of all $\Gamma$ functions has positive real part.

For the flow of the two-point function, we have to compute instead
\begin{splitequation}
\label{eq:integral_boundary2_finitepart_def}
I^{\Lambda,\Lambda_0,\nu}_{A_1 A_2 B}(P_1,P_2) &\equiv \int \Pf_{Q' \to Q} \int M^{\Lambda,\Lambda_0}_{A_1 A_2 B(\rho+\mathi \nu) B(\rho-\mathi \nu)}(\delta_{ij}) \left[ \prod_{1 \leq i < j \leq 4} \Gamma(\delta_{ij}) \right] \left( - 2 \ell^{-2} P_1 \cdot P_2 \right)^{-\delta_{12}} \\
&\qquad\times \left( - 2 \ell^{-2} P_1 \cdot Q \right)^{-\delta_{13}} \left( - 2 \ell^{-2} P_1 \cdot Q' \right)^{-\delta_{14}} \left( - 2 \ell^{-2} P_2 \cdot Q \right)^{-\delta_{23}} \\
&\qquad\times \left( - 2 \ell^{-2} P_2 \cdot Q' \right)^{-\delta_{24}} \left( - 2 \ell^{-2} Q \cdot Q' \right)^{-\delta_{34}} [ \total \delta_{ij} ] \total Q \eqend{,}
\end{splitequation}
where the integration measure is the one of equation~\eqref{eq:mellin_integration_measure}, with the $\mu_k$ given by the conformal dimensions of the operators $A_k$, or $\rho \pm \mathi \nu$ for the analytically continued ones. Taking again $Q'$ to be the embedding of the boundary vector $\vec{q} + \vec{k}$ and performing an appropriate shift of $\vec{q}$, we compute using the integral~\eqref{eq:masterintegral_power}
\begin{splitequation}
I^{\Lambda,\Lambda_0,\nu}_{A_1 A_2 B}(P_1,P_2) &= \Pf_{\vec{k} \to 0} \int M^{\Lambda,\Lambda_0}_{A_1 A_2 B(\rho+\mathi \nu) B(\rho-\mathi \nu)}(\delta_{ij}) \left( - 2 \ell^{-2} P_1 \cdot P_2 \right)^{-\delta_{12}} \ell^{2 (\delta_{13} + \delta_{14} + \delta_{23} + \delta_{24} + \delta_{34})} \\
&\qquad\times ( \vec{k}^2 )^{-\delta_{34}} \int_0^\infty\dotsi\int_0^\infty t_1^{\delta_{13}-1} t_2^{\delta_{14}-1} t_3^{\delta_{23}-1} t_4^{\delta_{24}-1} \int \mathe^{ - (t_1 + t_2 + t_3 + t_4) \vec{q}^2 } \total^d \vec{q} \\
&\qquad\times \exp\bigg[ - \frac{(t_2 + t_4) (t_1 + t_3)}{t_1 + t_2 + t_3 + t_4} \vec{k}^2 + 2 \frac{t_2 t_3 - t_1 t_4}{t_1 + t_2 + t_3 + t_4} ( \vec{p}_1 - \vec{p}_2 ) \vec{k} \\
&\qquad\qquad\quad- \frac{(t_3 + t_4) (t_1 + t_2)}{t_1 + t_2 + t_3 + t_4} (\vec{p}_1 - \vec{p}_2)^2 \bigg] \total t_1 \total t_2 \total t_3 \total t_4 \, \Gamma(\delta_{12}) \Gamma(\delta_{34}) [ \total \delta_{ij} ] \eqend{.}
\end{splitequation}
The Gaussian integral over $\vec{q}$ is immediate, and after the rescalings $t_i \to t_i t_4$ for $i = 1,2,3$ we can also perform the integral over $t_4$ with the result
\begin{splitequation}
I^{\Lambda,\Lambda_0,\nu}_{A_1 A_2 B}(P_1,P_2) &= \pi^\rho \Pf_{\vec{k} \to 0} \int M^{\Lambda,\Lambda_0}_{A_1 A_2 B(\rho+\mathi \nu) B(\rho-\mathi \nu)}(\delta_{ij}) \left( - 2 \ell^{-2} P_1 \cdot P_2 \right)^{-\delta_{12}} \Gamma(\delta_{12}) \Gamma(\delta_{34}) \\
&\qquad\times \ell^{2 (\delta_{13} + \delta_{14} + \delta_{23} + \delta_{24} + \delta_{34})} \Gamma(\delta_{13}+\delta_{14}+\delta_{23}+\delta_{24}-\rho) ( \vec{k}^2 )^{-\delta_{34}} \\
&\qquad\times \iiint_0^\infty t_1^{\delta_{13}-1} t_2^{\delta_{14}-1} t_3^{\delta_{23}-1} (1 + t_1 + t_2 + t_3)^{\delta_{13}+\delta_{14}+\delta_{23}+\delta_{24}-2\rho} \\
&\qquad\times \bigg[ (1 + t_2) (t_1 + t_3) \vec{k}^2 - 2 (t_2 t_3 - t_1) ( \vec{p}_1 - \vec{p}_2 ) \vec{k} \\
&\qquad\qquad+ (1 + t_3) (t_1 + t_2) (\vec{p}_1 - \vec{p}_2)^2 \bigg]^{-(\delta_{13}+\delta_{14}+\delta_{23}+\delta_{24}-\rho)} \total t_1 \total t_2 \total t_3 [ \total \delta_{ij} ] \eqend{.}
\end{splitequation}
After the rescalings $t_2 \to t_1 t_2$ and $t_3 \to t_1 t_3$ we obtain
\begin{splitequation}
\label{eq:integral_boundary2_finitepart_1}
I^{\Lambda,\Lambda_0,\nu}_{A_1 A_2 B}(P_1,P_2) &= \pi^\rho \Pf_{\vec{k} \to 0} \int M^{\Lambda,\Lambda_0}_{A_1 A_2 B(\rho+\mathi \nu) B(\rho-\mathi \nu)}(\delta_{ij}) \left( - 2 \ell^{-2} P_1 \cdot P_2 \right)^{-\delta_{12}} \Gamma(\delta_{12}) \Gamma(\delta_{34}) \\
&\qquad\times \ell^{2 (\delta_{13} + \delta_{14} + \delta_{23} + \delta_{24} + \delta_{34})} \Gamma(\delta_{13}+\delta_{14}+\delta_{23}+\delta_{24}-\rho) ( \vec{k}^2 )^{-\delta_{34}} \\
&\qquad\times \iiint_0^\infty t_1^{\rho-\delta_{24}-1} t_2^{\delta_{14}-1} t_3^{\delta_{23}-1} (1 + t_1 + t_1 t_2 + t_1 t_3)^{\delta_{13}+\delta_{14}+\delta_{23}+\delta_{24}-2\rho} \\
&\qquad\quad\times \bigg[ (1 + t_1 t_2) (1 + t_3) \vec{k}^2 - 2 (t_1 t_2 t_3 - 1) ( \vec{p}_1 - \vec{p}_2 ) \vec{k} \\
&\qquad\qquad+ (1 + t_1 t_3) (1 + t_2) (\vec{p}_1 - \vec{p}_2)^2 \bigg]^{-(\delta_{13}+\delta_{14}+\delta_{23}+\delta_{24}-\rho)} \total t_1 \total t_2 \total t_3 [ \total \delta_{ij} ] \eqend{,}
\end{splitequation}
and since the last term in brackets is bounded for small $t_i$ even as $\vec{k} \to 0$, we can take the limit $\vec{k} \to 0$ there. The finite part of the result as $\vec{k} \to 0$ is then clearly given by the residue of the integrand at $\delta_{34} = 0$, for which we need to have a more detailed look at the integration measure. If we choose $\delta_{34}$ and $\delta_{14}$ as the two independent integration variables, the others are determined from equation~\eqref{eq:mellin_integration_measure} to be (with $\Delta_i \equiv \Delta_{A_i}^{\Lambda,\Lambda_0}$)
\begin{equations}[eq:integral_boundary2_finitepart_deltaij]
\delta_{12} &= \delta_{34} - \rho + \Delta_1 - \Delta \eqend{,} \\
\delta_{13} &= - \delta_{14} - \delta_{34} + \rho + \Delta \eqend{,} \\
\delta_{23} &= \delta_{14} + \mathi \nu - \Delta \eqend{,} \\
\delta_{24} &= - \delta_{14} - \delta_{34} + \rho - \mathi \nu \eqend{,}
\end{equations}
where we set $\Delta \equiv (\Delta_1 - \Delta_2)/2$. The positivity conditions give the integration contours $\Delta < \Re \delta_{14} < \Delta_1 - \Delta$ and $\Delta + \rho - \Delta_1 < \Re \delta_{34} < \rho - \Delta$, taking w.l.o.g. $\Delta \geq 0$, and we assume that $\Delta_1$ and $\Delta_2$ are such that these can be fulfilled, in particular $\Delta_i \geq \rho$. Inserting~\eqref{eq:integral_boundary2_finitepart_deltaij} into the expression~\eqref{eq:integral_boundary2_finitepart_1} and rescaling $t_3 \to t_3/t_1$, it follows that
\begin{splitequation}
I^{\Lambda,\Lambda_0,\nu}_{A_1 A_2 B}(P_1,P_2) &= \pi^\rho \Pf_{\vec{k} \to 0} \int M^{\Lambda,\Lambda_0}_{A_1 A_2 B(\rho+\mathi \nu) B(\rho-\mathi \nu)}(\delta_{ij}) \left( - 2 \ell^{-2} P_1 \cdot P_2 \right)^{\delta_{34} - \Delta_1 + \Delta} \ell^{d + 2 \delta_{34}} \\
&\qquad\times \Gamma(\delta_{34} - \rho + \Delta_1 - \Delta) \Gamma(\rho - 2 \delta_{34}) \Gamma(\delta_{34}) ( \vec{k}^2 )^{-\delta_{34}} \\
&\qquad\times \iiint_0^\infty t_1^{\delta_{34} + \Delta - 1} t_2^{\delta_{14}-1} t_3^{\delta_{14} + \mathi \nu - \Delta-1} (1 + t_1 + t_1 t_2 + t_3)^{- 2 \delta_{34}} \\
&\qquad\qquad\times (1 + t_3)^{2 \delta_{34} - \rho} (1 + t_2)^{2 \delta_{34} - \rho} \total t_1 \total t_2 \total t_3 \frac{\total \delta_{34}}{2 \pi \mathi} \frac{\total \delta_{14}}{2 \pi \mathi} \eqend{.}
\end{splitequation}
However, since the $t_1$ integral is divergent when $\delta_{34} = 0$, we first have to perform the $t_i$ integrals explicitly before taking the residue. For this, we use the Mellin representation~\eqref{eq:sum_mellin_representation} for the factor $(1 + t_1 + t_1 t_2 + t_3)^{- 2 \delta_{34}}$ with $A = t_3$ to disentangle the $t_i$ integrals. Performing them in the order $t_3,t_1,t_2$ they are elementary and result in $\Gamma$ functions, and we obtain
\begin{splitequation}
I^{\Lambda,\Lambda_0,\nu}_{A_1 A_2 B}(P_1,P_2) &= \pi^{\rho + \frac{1}{2}} \Pf_{\vec{k} \to 0} \int M^{\Lambda,\Lambda_0}_{A_1 A_2 B(\rho+\mathi \nu) B(\rho-\mathi \nu)}(\delta_{ij}) \left( - 2 \ell^{-2} P_1 \cdot P_2 \right)^{\delta_{34} - \Delta_1 + \Delta} \\
&\qquad\times 2^{1-2\delta_{34}} \ell^{d + 2 \delta_{34}} ( \vec{k}^2 )^{-\delta_{34}} \frac{\Gamma(\Delta + \delta_{34})}{\Gamma\left( \delta_{34} + \frac{1}{2} \right)} \frac{\Gamma(\delta_{34} - \rho + \Delta_1 - \Delta)}{\Gamma(\Delta - \delta_{34} + \rho)} \\
&\qquad\times \int_{0 < \Re z < 2 \Re \delta_{34}} \Gamma(z + \Delta - \delta_{14} - 2 \delta_{34} - \mathi \nu + \rho) \Gamma(\delta_{34} - \Delta - z) \\
&\qquad\quad\times \Gamma(\delta_{14} - z + \mathi \nu - \Delta) \Gamma(z) \frac{\total z}{2 \pi \mathi} \, \Gamma(\Delta - \delta_{14} - \delta_{34} + \rho) \Gamma(\delta_{14}) \frac{\total \delta_{34}}{2 \pi \mathi} \frac{\total \delta_{14}}{2 \pi \mathi} \eqend{,}
\end{splitequation}
where we have used some standard $\Gamma$ function identities~\cite{dlmf} to simplify the result. The integral over $z$ can be done using the first Barnes lemma~\eqref{eq:barneslemma1}, since the integration contour is in fact the standard Mellin--Barnes one if we take into account the integration contours for $\delta_{14}$ and $\delta_{34}$ determined after equation~\eqref{eq:integral_boundary2_finitepart_deltaij}. We thus have
\begin{splitequation}
I^{\Lambda,\Lambda_0,\nu}_{A_1 A_2 B}(P_1,P_2) &= \pi^{\rho + \frac{1}{2}} \Pf_{\vec{k} \to 0} \int M^{\Lambda,\Lambda_0}_{A_1 A_2 B(\rho+\mathi \nu) B(\rho-\mathi \nu)}(\delta_{ij}) \left( - 2 \ell^{-2} P_1 \cdot P_2 \right)^{\delta_{34} - \Delta_1 + \Delta} \\
&\qquad\times 2^{1-2\delta_{34}} \ell^{d + 2 \delta_{34}} ( \vec{k}^2 )^{-\delta_{34}} \frac{\Gamma(\Delta + \delta_{34})}{\Gamma\left( \delta_{34} + \frac{1}{2} \right)} \frac{\Gamma(\delta_{34} - \rho + \Delta_1 - \Delta)}{\Gamma(\Delta - \delta_{34} + \rho)} \\
&\qquad\times \frac{\Gamma(\delta_{34} - \Delta) \Gamma(\delta_{14} - \Delta + \mathi \nu) \Gamma(\rho - 2 \delta_{34}) \Gamma(\rho - \mathi \nu - \delta_{14} - \delta_{34})}{\Gamma(\rho - \Delta - \delta_{34})} \\
&\qquad\times \Gamma(\Delta - \delta_{14} - \delta_{34} + \rho) \Gamma(\delta_{14}) \frac{\total \delta_{34}}{2 \pi \mathi} \frac{\total \delta_{14}}{2 \pi \mathi} \eqend{,}
\end{splitequation}
which actually does \emph{not} have a pole at $\delta_{34} = 0$ anymore --- except if $\Delta \in \mathbb{N}_0$.\footnote{We assume here that the Mellin amplitude is regular at $\delta_{34} = 0$, which is the case for the amplitudes constructed from weakly coupled field theories in EAdS.} For $\Delta = 0$ and $\Delta_1 > \rho$, we have a double pole, and obtain
\begin{splitequation}
\label{eq:integral_boundary2_finitepart_result_0}
I^{\Lambda,\Lambda_0,\nu}_{A_1 A_2 B}(P_1,P_2) &= \frac{2 \pi^\rho \ell^d \Gamma(\Delta_1 - \rho)}{\Gamma(\rho)} \left( - 2 \ell^{-2} P_1 \cdot P_2 \right)^{- \Delta_1} \int \Gamma(\delta_{14}) \Gamma(\rho - \delta_{14}) \\
&\quad\times \Gamma(\delta_{14} + \mathi \nu) \Gamma(\rho - \delta_{14} - \mathi \nu) \bigg[ \left. \partial_{\delta_{34}} M^{\Lambda,\Lambda_0}_{A_1 A_2 B(\rho+\mathi \nu) B(\rho-\mathi \nu)}(\delta_{ij}) \right\rvert_{\delta_{34} = 0} \\
&\qquad+ \left. M^{\Lambda,\Lambda_0}_{A_1 A_2 B(\rho+\mathi \nu) B(\rho-\mathi \nu)}(\delta_{ij}) \right\rvert_{\delta_{34} = 0} \Big[ \ln \left( - 2 P_1 \cdot P_2 \right) - \gamma + \psi(\Delta_1 - \rho) \\
&\qquad\qquad- \psi(\rho - \delta_{14} - \mathi \nu) - \psi(\rho - \delta_{14}) \Big] \bigg] \frac{\total \delta_{14}}{2 \pi \mathi} \eqend{,}
\end{splitequation}
where the $\delta_{ij}$ in the Mellin amplitude are given by~\eqref{eq:integral_boundary2_finitepart_deltaij} with $\Delta = 0$. On the other hand, for $\Delta = n > 0$, we have a simple pole and obtain
\begin{splitequation}
\label{eq:integral_boundary2_finitepart_result_n}
I^{\Lambda,\Lambda_0,\nu}_{A_1 A_2 B}(P_1,P_2) &= \frac{2 (-1)^n \pi^\rho \ell^d}{n} \frac{\Gamma(\Delta_1 - \rho - n) \Gamma(\rho) }{\Gamma(n + \rho) \Gamma(\rho - n)} \left( - 2 \ell^{-2} P_1 \cdot P_2 \right)^{- \Delta_1 + n} \\
&\quad\times \int \left. M^{\Lambda,\Lambda_0}_{A_1 A_2 B(\rho+\mathi \nu) B(\rho-\mathi \nu)}(\delta_{ij}) \right\rvert_{\delta_{34} = 0} \\
&\qquad\times \Gamma(\delta_{14} - n + \mathi \nu) \Gamma(\rho - \mathi \nu - \delta_{14}) \Gamma(n - \delta_{14} + \rho) \Gamma(\delta_{14}) \frac{\total \delta_{14}}{2 \pi \mathi} \eqend{,}
\end{splitequation}
where the $\delta_{ij}$ in the Mellin amplitude are given by~\eqref{eq:integral_boundary2_finitepart_deltaij} with $\Delta = n$.

\subsection{Evaluating a Mellin--Barnes integral}

We consider the double integral
\begin{splitequation}
\label{eq:int_loop_rho}
I(\Delta,\Delta_1,\Delta_2,\rho) &\equiv \int_{-\infty}^\infty \int_{-\infty}^\infty \Gamma\left( \frac{\Delta - \mathi (\mu-\nu)}{2} \right) \Gamma\left( \frac{\Delta + \mathi (\mu-\nu)}{2} \right) \Gamma\left( \rho - \frac{\Delta - \mathi (\mu-\nu)}{2} \right) \\
&\quad\times \Gamma\left( \rho - \frac{\Delta + \mathi (\mu-\nu)}{2} \right) \Gamma\left( \rho - \frac{\Delta + \mathi (\mu+\nu)}{2} \right) \Gamma\left( \rho - \frac{\Delta - \mathi (\mu+\nu)}{2} \right) \\
&\quad\times \frac{\Gamma\left( \frac{\Delta + \mathi (\mu+\nu)}{2} \right)}{\Gamma\left( \mathi \nu \right) \Gamma\left( \mathi \mu \right)} \frac{\Gamma\left( \frac{\Delta - \mathi (\mu+\nu)}{2} \right)}{\Gamma\left( - \mathi \nu \right) \Gamma\left( - \mathi \mu \right)} \frac{1}{[ \nu^2 + (\Delta_1-\rho)^2 ] [ \mu^2 + (\Delta_2-\rho)^2 ]} \total \mu \total \nu \eqend{,}
\end{splitequation}
which is absolutely convergent for $\rho < 1$ and $\rho < \{ \Delta, \Delta_1, \Delta_2 \} < 2 \rho$, conditionally convergent for $\rho = 1$ and the same restriction on the $\Delta_i$, and for other values of the parameters must be defined by analytic continuation. As shown in~\cite{giombisleighttaronna2018,carmidipietrokomatsu2019}, it can be exactly evaluated in terms of $\psi$ functions for $\rho = 1$, and we present here a simple derivation of the result.

Setting $\rho = 1$, the integral simplifies to
\begin{splitequation}
I(\Delta,\Delta_1,\Delta_2,1) &= \int_{-\infty}^\infty \int_{-\infty}^\infty \frac{\pi^2}{\sin\left[ \pi \frac{\Delta - \mathi (\mu-\nu)}{2} \right] \sin\left[ \pi \frac{\Delta + \mathi (\mu-\nu)}{2} \right] \sin\left[ \pi \frac{\Delta + \mathi (\mu+\nu)}{2} \right] \sin\left[ \pi \frac{\Delta - \mathi (\mu+\nu)}{2} \right]} \\
&\quad\times \frac{\mu \nu \sinh(\pi \mu) \sinh(\pi \nu)}{[ \nu^2 + (\Delta_1-1)^2 ] [ \mu^2 + (\Delta_2-1)^2 ]} \total \mu \total \nu \eqend{,}
\end{splitequation}
and we evaluate the $\mu$ integral using Cauchy's theorem, by summing residues of poles in the upper half-plane. There is a single pole at $\mu = \mathi (\Delta_2-1)$ which gives a contribution
\begin{splitequation}
I^{(1)}(\Delta,\Delta_1,\Delta_2,1) &= \int_{-\infty}^\infty \frac{\pi^2}{\cos\left[ \pi \frac{\Delta + \Delta_2 + \mathi \nu}{2} \right] \cos\left[ \pi \frac{\Delta - \Delta_2 - \mathi \nu}{2} \right] \cos\left[ \pi \frac{\Delta - \Delta_2 + \mathi \nu}{2} \right] \cos\left[ \pi \frac{\Delta + \Delta_2 - \mathi \nu}{2} \right]} \\
&\quad\times \frac{\nu \sinh(\pi \nu)}{[ \nu^2 + (\Delta_1-1)^2 ]} \total \nu \, \pi \sin(\pi \Delta_2) \eqend{,}
\end{splitequation}
and four infinite series at $\mu = \pm \nu + \mathi \Delta + 2 \mathi k$ and $\mu = \pm \nu + \mathi (2-\Delta) + 2 \mathi k$ which give a contribution
\begin{splitequation}
I^{(2)}(\Delta,\Delta_1,\Delta_2,1) &= \frac{4 \mathi \pi^2}{\sin(\pi\Delta)} \int_{-\infty}^\infty \sum_{k=0}^\infty \Bigg[ \frac{\Delta - \mathi \nu + 2 k}{(\Delta_2-1)^2 - ( \Delta-\mathi\nu + 2 k )^2} - \frac{\Delta + \mathi \nu + 2 k}{(\Delta_2-1)^2 - ( \Delta+\mathi\nu + 2 k )^2} \\
&\qquad+ \frac{2-\Delta + \mathi \nu + 2 k}{(\Delta_2-1)^2 - ( 2-\Delta+\mathi \nu + 2 k )^2} - \frac{2-\Delta - \mathi \nu + 2 k}{(\Delta_2-1)^2 - ( 2-\Delta-\mathi\nu + 2 k )^2} \\
&\qquad\Bigg] \frac{\nu}{[ \nu^2 + (\Delta_1-1)^2 ]} \total \nu \\
&= \frac{\mathi \pi^2}{\sin(\pi\Delta)} \int_{-\infty}^\infty \Bigg[ \psi\left( \frac{3 - \Delta - \Delta_2 + \mathi \nu}{2} \right) - \psi\left( \frac{3 - \Delta - \Delta_2 - \mathi \nu}{2} \right) \\
&\qquad+ \psi\left( \frac{\Delta + \Delta_2 - 1 - \mathi \nu}{2} \right) - \psi\left( \frac{\Delta + \Delta_2 - 1 + \mathi \nu}{2} \right) \\
&\qquad+ \psi\left( \frac{1 + \Delta - \Delta_2 - \mathi \nu}{2} \right) - \psi\left( \frac{1 + \Delta - \Delta_2 + \mathi \nu}{2} \right) \\
&\qquad+ \psi\left( \frac{1 - \Delta + \Delta_2 + \mathi \nu}{2} \right) - \psi\left( \frac{1 - \Delta + \Delta_2 - \mathi \nu}{2} \right) \Bigg] \frac{\nu}{[ \nu^2 + (\Delta_1-1)^2 ]} \total \nu \eqend{.}
\end{splitequation}
The $\nu$ integrals are also done by summing residues of the poles in the upper half-plane, and again there is a single pole at $\nu = \mathi (\Delta_1-1)$ and four infinite series at $\nu = \mathi (3 - \Delta - \Delta_2) + 2 \mathi k$, $\nu = \mathi (\Delta + \Delta_2 - 1) + 2 \mathi k$, $\nu = \mathi (1 + \Delta - \Delta_2) + 2 \mathi k$ and $\nu = \mathi (1 - \Delta + \Delta_2) + 2 \mathi k$ for both $I^{(1)}$ and $I^{(2)}$. We thus obtain
\begin{splitequation}
&I^{(1)}(\Delta,\Delta_1,\Delta_2,1) = \frac{\pi^4 \sin(\pi \Delta_1) \sin(\pi \Delta_2)}{\sin\left[ \pi \frac{\Delta + \Delta_2 - \Delta_1}{2} \right] \sin\left[ \pi \frac{\Delta - \Delta_2 + \Delta_1}{2} \right] \sin\left[ \pi \frac{\Delta - \Delta_2 - \Delta_1}{2} \right] \sin\left[ \pi \frac{\Delta + \Delta_2 + \Delta_1}{2} \right]} \\
&\qquad- \frac{4 \pi^3}{\sin(\pi \Delta)} \sum_{k=0}^\infty \Bigg[ \frac{3 - \Delta - \Delta_2 + 2 k}{(\Delta_1-1)^2 - (3 - \Delta - \Delta_2 + 2 k)^2} + \frac{\Delta + \Delta_2 - 1 + 2 k}{(\Delta_1-1)^2 - (\Delta + \Delta_2 - 1 + 2 k)^2} \\
&\qquad\qquad- \frac{1 + \Delta - \Delta_2 + 2 k}{(\Delta_1-1)^2 - (1 + \Delta - \Delta_2 + 2 k)^2} - \frac{1 - \Delta + \Delta_2 + 2 k}{(\Delta_1-1)^2 - (1 - \Delta + \Delta_2 + 2 k)^2} \Bigg] \\
&= \frac{\pi^4 \sin(\pi \Delta_1) \sin(\pi \Delta_2)}{\sin\left[ \pi \frac{\Delta + \Delta_2 - \Delta_1}{2} \right] \sin\left[ \pi \frac{\Delta - \Delta_2 + \Delta_1}{2} \right] \sin\left[ \pi \frac{\Delta - \Delta_2 - \Delta_1}{2} \right] \sin\left[ \pi \frac{\Delta + \Delta_2 + \Delta_1}{2} \right]} \\
&\quad- \frac{\pi^3}{\sin(\pi \Delta)} \Bigg[ \psi\left( \frac{4 - \Delta - \Delta_1 - \Delta_2}{2} \right) - \psi\left( \frac{2 + \Delta - \Delta_1 - \Delta_2}{2} \right) + \psi\left( \frac{2 - \Delta + \Delta_1 - \Delta_2}{2} \right) \\
&\qquad- \psi\left( \frac{\Delta + \Delta_1 - \Delta_2}{2} \right) - \psi\left( \frac{2 - \Delta - \Delta_1 + \Delta_2}{2} \right) + \psi\left( \frac{\Delta - \Delta_1 + \Delta_2}{2} \right) \\
&\qquad- \psi\left( \frac{- \Delta + \Delta_1 + \Delta_2}{2} \right) + \psi\left( \frac{- 2 + \Delta + \Delta_1 + \Delta_2}{2} \right) \Bigg]
\end{splitequation}
and
\begin{splitequation}
I^{(2)}(\Delta,\Delta_1,\Delta_2,1) &= - \frac{\pi^3}{\sin(\pi\Delta)} \Bigg[ \psi\left( \frac{4 - \Delta - \Delta_1 - \Delta_2}{2} \right) - \psi\left( \frac{2 + \Delta - \Delta_1 - \Delta_2}{2} \right) - \psi\left( \frac{2 - \Delta + \Delta_1 - \Delta_2}{2} \right) \\
&\qquad+ \psi\left( \frac{\Delta + \Delta_1 - \Delta_2}{2} \right) + \psi\left( \frac{2 - \Delta - \Delta_1 + \Delta_2}{2} \right) - \psi\left( \frac{\Delta - \Delta_1 + \Delta_2}{2} \right) \\
&\qquad- \psi\left( \frac{- \Delta + \Delta_1 + \Delta_2}{2} \right) + \psi\left( \frac{- 2 + \Delta + \Delta_1 + \Delta_2}{2} \right) \Bigg] \\
&\quad- \frac{4 \pi^3}{\sin(\pi\Delta)} \sum_{k=0}^\infty \Bigg[ \frac{3 - \Delta - \Delta_2 + 2 k}{(2k + 4 - \Delta - \Delta_1 - \Delta_2) (2k + 2 - \Delta + \Delta_1 - \Delta_2)} \\
&\qquad\quad- \frac{\Delta + \Delta_2 - 1 + 2 k}{(2k + \Delta - \Delta_1 + \Delta_2) (2k - 2 + \Delta + \Delta_1 + \Delta_2)} \\
&\qquad\quad- \frac{1 + \Delta - \Delta_2 + 2 k}{(2k + 2 + \Delta - \Delta_1 - \Delta_2) (2k + \Delta + \Delta_1 - \Delta_2)} \\
&\qquad\quad+ \frac{1 - \Delta + \Delta_2 + 2 k}{(2k + 2 - \Delta - \Delta_1 + \Delta_2) (2k - \Delta + \Delta_1 + \Delta_2)} \Bigg] \\
&= \frac{2 \pi^3}{\sin(\pi\Delta)} \Bigg[ \psi\left( \frac{2 - \Delta + \Delta_1 - \Delta_2}{2} \right) - \psi\left( \frac{\Delta + \Delta_1 - \Delta_2}{2} \right) \\
&\qquad\qquad+ \psi\left( \frac{- \Delta + \Delta_1 + \Delta_2}{2} \right) - \psi\left( \frac{- 2 + \Delta + \Delta_1 + \Delta_2}{2} \right) \Bigg] \eqend{.}
\end{splitequation}
Summing both and using the reflection formula
\begin{equation}
\psi(1-z) = \psi(z) + \pi \cot(\pi z)
\end{equation}
for the $\psi$ function as well as trigonometric identities, this results in
\begin{splitequation}
\label{eq:int_loop_rho1_result}
I(\Delta,\Delta_1,\Delta_2,1) &= \frac{4 \pi^4}{\cos[\pi(\Delta_1+\Delta_2)] - \cos(\pi\Delta)} \\
&\quad- \frac{2 \pi^3}{\sin(\pi \Delta)} \Bigg[ \psi\left( \frac{4 - \Delta - \Delta_1 - \Delta_2}{2} \right) - \psi\left( \frac{2 + \Delta - \Delta_1 - \Delta_2}{2} \right) \\
&\qquad\qquad- \psi\left( \frac{- \Delta + \Delta_1 + \Delta_2}{2} \right) + \psi\left( \frac{- 2 + \Delta + \Delta_1 + \Delta_2}{2} \right) \Bigg] \eqend{,}
\end{splitequation}
which agrees (taking into account the overall normalisation) with~\cite[Eq.~(2.45)]{giombisleighttaronna2018}.

As $\Delta \to \Delta_1 + \Delta_2$, we obtain a divergent result, which tells us that in this case the integration contour gets pinched between two single poles. Performing the integral over $\mu$, one sees that this occurs in the $\nu$ integration for the pole at $\nu = - \mathi (\Delta_1-\rho)$, which has to lie below the integration contour, and the one at $\nu = - \mathi (\Delta-\Delta_2-\rho) + 2 \mathi k$ for $k = 0$, which has to lie above the contour. Instead of re-doing the complete computation however, we can just expand the result~\eqref{eq:int_loop_rho1_result} around $\Delta = \Delta_1 + \Delta_2$, which gives
\begin{equation}
\label{eq:int_loop_rho1_result2}
I(\Delta_1+\Delta_2,\Delta_1,\Delta_2,1) \approx \frac{4 \pi^3}{\sin(\pi\Delta)} \Bigg[ \frac{2}{\Delta - \Delta_1 - \Delta_2} - \gamma - \psi(\Delta-1) \Bigg] \eqend{.}
\end{equation}

\addcontentsline{toc}{section}{References}
\bibliography{literature}

\end{document}